\documentclass[a4paper,11pt]{article}
\pdfoutput=1 
\usepackage{jheppub} 
\usepackage[utf8]{inputenc} 
\usepackage[T1]{fontenc}
\usepackage{ae}
\usepackage{aecompl}

\newcommand{\ads}{\mathrm{AdS}}
\newcommand{\ds}{\mathrm{dS}}
\newcommand{\ess}{\mathbb{S}}

\newcommand{\adsts}{{\ads \times \ess}}

\numberwithin{equation}{section}

\newenvironment{eqaed}
    {\begin{equation}
    \begin{aligned}
    }
    { 
    \end{aligned}
    \end{equation}
    }
    
\usepackage{xcolor} 
\usepackage[framemethod=default]{mdframed}
\newmdenv[skipabove=10pt,
skipbelow=7pt,
rightline=false,
leftline=true,
topline=false,
bottomline=false,
linecolor=black,
backgroundcolor=black!5,
innerleftmargin=4pt,
innerrightmargin=0pt,
innertopmargin=0pt,
leftmargin=2pt,
rightmargin=0pt,
linewidth=2pt,
innerbottommargin=0pt]{lbBox}
\newenvironment{importantbox}{\begin{lbBox}\vspace{1 mm}
	} {\vspace{1.5 mm}\end{lbBox}}

\title{\boldmath de Sitter in non-supersymmetric string theories: no-go theorems and brane-worlds}

\author[a]{Ivano Basile,}
\author[b]{Stefano Lanza}

\affiliation[a]{Scuola Normale Superiore and I.N.F.N.\\Piazza dei Cavalieri 7, 56126, Pisa, Italy}
\affiliation[b]{Jefferson Physical Laboratory, Harvard University\\Cambridge, MA 02138, USA}

\emailAdd{ivano.basile@sns.it}
\emailAdd{slanza@g.harvard.edu}

\abstract{We study de Sitter configurations in ten-dimensional string models where supersymmetry is either absent or broken at the string scale. To this end, we derive expressions for the cosmological constant in general warped flux compactifications with localized sources, which yield no-go theorems that extend previous works on supersymmetric cases. We frame our results within a dimensional reduction and connect them to a number of Swampland conjectures, corroborating them further in the absence of supersymmetry. Furthermore, we construct a top-down string embedding of de Sitter brane-world cosmologies within unstable anti-de Sitter landscapes, providing a concrete realization of a recently revisited proposal.}

\begin{document} 
\maketitle
\flushbottom

\section{Introduction}

Despite the numerous successes of string theory, its connection to realistic phenomenology remains a remarkably subtle challenge. The theory appears to contain all the ingredients necessary to concoct standard-like models with the inclusion of dark energy, but upon supersymmetry breaking most of the computational power is typically lost due to uncontrolled back-reactions. As a result, the very existence of de Sitter ($\ds$) landscapes and of similarly desirable constructions is still unsettled, despite a long and meticulous scrutiny. Most prominently, KKLT-type settings~\cite{Kachru:2003aw} with anti-brane uplifts entail a number of subtleties, and a complete ten-dimensional picture is still lacking at present. On the other hand, $\ds$ solutions built out of purely classical ingredients within a supergravity approximation appear to necessarily contain uncontrolled regimes in the vicinity of orientifold planes~\cite{Danielsson:2009ff, Cordova:2018dbb, Blaback:2019zig, Cordova:2019cvf}. We shall not attempt to provide a comprehensive account of this extensive subject, since our focus in this paper will be on higher-dimensional approaches~\cite{Koerber:2007xk, Moritz:2017xto, Kallosh:2018nrk, Bena:2018fqc, Gautason:2018gln,Danielsson:2018ztv, Hamada:2019ack, Carta:2019rhx, Gautason:2019jwq} and, in particular, on the search for new solutions.

This state of affairs provided fertile ground for the development of the `Swampland program'~\cite{Vafa:2005ui}\footnote{See~\cite{Brennan:2017rbf,Palti:2019pca} for reviews.}, whose ultimate aim is to identify a set of criteria that consistent effective field theories (EFTs) coupled to gravity ought to satisfy. In this context, one can try to frame the apparent absence of $\ds$ solutions as a distinguishing feature of UV-complete models, rather than a mere technical obstacle to model building. Along these lines, the `de Sitter conjecture'~\cite{Obied:2018sgi} states that any EFT coupled to gravity stemming from a UV-complete model cannot accommodate $\ds$ minima. This conjecture is partly corroborated by no-go theorems~\cite{Maldacena:2000mw,Giddings:2001yu,Hertzberg:2007wc} that forbid classical $\ds$ vacua in supergravity, and thus in supersymmetric compactifications of string theory or M-theory. However, evidence for the conjecture in full-fledged non-supersymmetric settings is still lacking at present.

Motivated by these issues, in this paper we consider the non-supersymmetric string models in ten dimensions whose perturbative spectra are devoid of tachyons. In particular, we focus on the $SO(16) \times SO(16)$ heterotic model of~\cite{AlvarezGaume:1986jb, Dixon:1986iz} and on two orientifold models: the $U(32)$ type $0'\text{B}$ model of~\cite{Sagnotti:1995ga, Sagnotti:1996qj} and the $USp(32)$ model of~\cite{Sugimoto:1999tx}, in which supersymmetry is non-linearly realized~\cite{Antoniadis:1999xk, Angelantonj:1999jh, Aldazabal:1999jr, Angelantonj:1999ms} via `brane supersymmetry breaking' (BSB). The low-energy effective actions that describe these models involve exponential dilaton potentials generated by gravitational tadpoles, and the issue at stake is whether the ingredients provided by string-scale supersymmetry breaking can allow for $\ds$ configurations. While a number of parallels between lower-dimensional anti-brane uplifts and the ten-dimensional BSB scenario appear encouraging to this effect, as we shall see shortly the presence of exponential potentials does not ameliorate the situation, insofar as (warped) flux compactifications are concerned. On the other hand, as we shall explain in Section~\ref{sec:braneworld}, the very presence of exponential potentials allows for intriguing brane-world scenarios within the $\ads$ landscapes discussed in~\cite{Antonelli:2019nar}, whose non-perturbative instabilities play a crucial rôle in this respect.

This paper is structured as follows. After a brief overview of the relevant string models with broken, or without, supersymmetry in Section~\ref{sec:non-susy_strings}, we begin our investigation in Section~\ref{sec:FR} studying Freund-Rubin compactifications, which turn out to be either excluded or unstable, consistently with the results of~\cite{Montero:2020rpl}. Then, in Section~\ref{sec:no-go}, we proceed to study warped compactifications with fluxes threading cycles of general internal manifolds, along the lines of~\cite{Giddings:2001yu}, and we obtain conditions that fix the (sign of the) resulting cosmological constant in terms of the model parameters, generalizing the results of~\cite{Maldacena:2000mw} to models with supersymmetry-breaking exponential potentials. Furthermore, in Section~\ref{sec:no-go_sources} we include the contribution of localized sources, which leads to a generalized expression for the cosmological constant. The resulting sign cannot be fixed \textit{a priori} in the entire space of parameters, but one can derive sufficient conditions that exclude $\ds$ solutions for certain ranges of parameters. In Section~\ref{sec:red_pot} we connect our results to a lower-dimensional description, showing that even in bottom-up models where $\ds$ Freund-Rubin solutions are allowed they are unstable. In Section~\ref{sec:tcc_dsc} we employ the lower-dimensional formulation to discuss how our results relate to recent Swampland conjectures~\cite{Obied:2018sgi,Bedroya:2019snp,Lanza:2020qmt}, showing that the ratio $\frac{|{\nabla \mathcal{V}}|}{\mathcal{V}}$ is bounded from below whenever the effective potential $\mathcal{V} > 0$. Finally, in Section~\ref{sec:braneworld} we review a recently revisited proposal~\cite{Banerjee:2018qey, Banerjee:2019fzz, Banerjee:2020wix} which rests on the observation that branes nucleating amidst $\ads \to \ads$ transitions host $\ds$ geometries on their world-volumes, and we embed a construction of this type in the non-supersymmetric string models that we consider, building on the results of~\cite{Antonelli:2019nar}. We conclude in Section~\ref{sec:conclusions} with some closing remarks. The paper contains two appendices. In Appendix~\ref{app:no-go_proof} we provide details of the derivation of the no-go results in Section~\ref{sec:no-go}, while in Appendix~\ref{app:reduced_pot} we discuss in detail the computation of the effective potential in the dimensional reduction discussed in Section~\ref{sec:red_pot}.

\section{Non-supersymmetric string models}
\label{sec:non-susy_strings}

In this section we introduce the ten-dimensional non-supersymmetric string models that we shall consider in the remainder of this paper. They comprise two orientifold models, namely the $USp(32)$ model of~\cite{Sugimoto:1999tx} and the $U(32)$ type $0'$B model of~\cite{Sagnotti:1995ga, Sagnotti:1996qj}, and the $SO(16) \times SO(16)$ heterotic model of~\cite{AlvarezGaume:1986jb, Dixon:1986iz}. While the latter two models feature non-supersymmetric perturbative spectra with no tachyons, the $USp(32)$ model is particularly intriguing, since via `brane supersymmetry breaking' (BSB) it realizes supersymmetry non-linearly in the open sector~\cite{Antoniadis:1999xk, Angelantonj:1999jh, Aldazabal:1999jr, Angelantonj:1999ms}. These models can be described in terms of vacuum amplitudes, whose modular properties encode perturbative spectra in (combinations of) characters $\left( O_{2n} \, , V_{2n} \, , S_{2n} \, , C_{2n} \right)$ of the level-one affine $\mathfrak{so}(2n)$ algebra. For a review of this formalism and of related constructions, see~\cite{Dudas:2000bn, Angelantonj:2002ct, Mourad:2017rrl}.

\subsection{The orientifold models}\label{sec:orientifolds}

In order to introduce the orientifold models at stake\footnote{The original works on orientifolds can be found in~\cite{Sagnotti:1987tw, Pradisi:1988xd, Horava:1989vt, Horava:1989ga, Bianchi:1990yu, Bianchi:1990tb, Bianchi:1991eu, Sagnotti:1992qw}.}, let us recall that in the more familiar case of the type I superstring the perturbative spectrum is encoded in the torus amplitude
\begin{eqaed}\label{eq:torus_amp_i}
    \mathcal{T}_{\text{I}} = \frac{1}{2} \int_{\mathcal{F}} \frac{d^2\tau}{\tau_2^6} \, \frac{\left(V_8 - S_8 \right) \overline{\left(V_8 - S_8 \right)}}{|\eta(\tau)|^{16}} \, ,
\end{eqaed}
which is half of the corresponding amplitude in the type IIB superstring, together with the amplitudes associated to the Klein bottle, the annulus and the Möbius strip, which read
\begin{eqaed}\label{eq:klein_annulus_mobius_mod}
	\mathcal{K} & = \frac{1}{2} \int_0^\infty \frac{d\tau_2}{\tau_2^6} \, \frac{\left( V_8 - S_8 \right) \!\left(2i \tau_2\right)}{\eta^8\!\left(2i \tau_2 \right)} \, , \\
	\mathcal{A} & = \frac{N^2}{2} \int_0^\infty \frac{d\tau_2}{\tau_2^6} \, \frac{\left( V_8 - S_8 \right) \!\left(\frac{i\tau_2}{2}\right)}{\eta^8\!\left(\frac{i\tau_2}{2} \right)} \, , \\
	\mathcal{M} & = \frac{\varepsilon \, N}{2} \int_0^\infty \frac{d\tau_2}{\tau_2^6} \, \frac{\left( \widehat{V}_8 - \widehat{S}_8 \right) \!\left(\frac{i\tau_2}{2} + \frac{1}{2}\right)}{\widehat{\eta}^8\!\left(\frac{i\tau_2}{2} + \frac{1}{2} \right)} \, .
\end{eqaed}
These amplitudes feature (loop-channel) UV divergences which can be ascribed to tadpoles in the NS-NS and R-R sectors, whose cancellation requires
\begin{eqaed}\label{eq:tadpole_super_mod}
	N = 32 \, , \qquad \varepsilon = - 1 \, ,
\end{eqaed}
selecting the $SO(32)$ superstring. The $USp(32)$ model of~\cite{Sugimoto:1999tx} can be obtained from the type IIB superstring introducing an $\text{O}9_+$-plane with positive tension and charge, preserving the R-R tadpole cancellation while generating a non-vanishing NS-NS tadpole, thus breaking supersymmetry at the string scale. This is reflected by a sign change in the Möbius strip amplitude, so that now
\begin{eqaed}\label{eq:mobius_bsb_mod}
	\mathcal{M}_{\text{BSB}} = \frac{\varepsilon \, N}{2} \int_0^\infty \frac{d\tau_2}{\tau_2^6} \, \frac{\left( \widehat{V}_8 + \widehat{S}_8 \right) \!\left(\frac{i\tau_2}{2} + \frac{1}{2}\right)}{\widehat{\eta}^8\!\left(\frac{i\tau_2}{2} + \frac{1}{2} \right)} \, .
\end{eqaed}
The resulting R-R tadpole cancellation condition requires that $\varepsilon = 1$ and $N = 32$, \textit{i.e.} a $USp(32)$ gauge group. However, one is now left with a NS-NS tadpole, and thus at low energies a runaway exponential potential of the type\footnote{For a more detailed analysis of the low-energy physics of the BSB model, see~\cite{Dudas:2000nv, Pradisi:2001yv}.}
\begin{eqaed}\label{eq:runaway_potential_string_frame}
	T \int d^{10} x \, \sqrt{- g_{\rm s}} \, e^{- \phi}
\end{eqaed}
emerges in the string frame, while its Einstein-frame counterpart is
\begin{eqaed}\label{eq:runaway_potential_einstein_frame}
	T \int d^{10} x \, \sqrt{- g} \, e^{\gamma \phi} \, , \qquad \gamma = \frac{3}{2} \, .
\end{eqaed}
Exponential potentials of the type of eq.~\eqref{eq:runaway_potential_einstein_frame} are smoking guns of string-scale supersymmetry breaking, and in order to balance their runaway effects in a controlled fashion we shall introduce fluxes.

The $U(32)$ type $0'\text{B}$ model arises via an orientifold projection of the type $0\text{B}$ model, described by the torus amplitude
\begin{eqaed}\label{eq:0_mod}
	\mathcal{T}_{\text{0B}} = \int_{\mathcal{F}} \frac{d^2\tau}{\tau_2^6} \, \frac{O_8 \, \overline{O_8} + V_8 \, \overline{V_8} + S_8 \, \overline{S_8} + C_8 \, \overline{C_8}}{|\eta(\tau)|^{16}} \, ,
\end{eqaed}
which entails adding to (half of) it the contributions
\begin{eqaed}\label{eq:0'b_annulus_mobius_mod}
	\mathcal{K}_{0'\text{B}} & = \frac{1}{2} \int_0^\infty \frac{d\tau_2}{\tau_2^6} \left( - \, O_8 + V_8 + S_8 - C_8 \right) \, , \\
	\mathcal{A}_{0'\text{B}} & \int_0^\infty \frac{d\tau_2}{\tau_2^6} \, n \, \overline{n} \, V_8 - \, \frac{n^2 + \overline{n}^2}{2} \, C_8 \, , \\
	\mathcal{M}_{0'\text{B}} & = \int_0^\infty \frac{d\tau_2}{\tau_2^6} \, \frac{n + \overline{n}}{2} \, \widehat{C}_8 \, ,
\end{eqaed}
where the complex `eigencharges' $n = \overline{n}$ pertain to unitary groups $U(n)$, and tadpole cancellation fixes $n = 32$. As in the case of the $USp(32)$ model, this model admits a space-time description in terms of orientifold planes, now with vanishing tension, and the low-energy physics of both non-supersymmetric orientifold models can be captured by the exponential potential of eq.~\eqref{eq:runaway_potential_einstein_frame}. In addition to these orientifold models, the low-energy description can also encompass the non-supersymmetric heterotic model, which we shall now discuss in detail, with a simple replacement of numerical coefficients in the effective action.

\subsection{The heterotic model}\label{sec:heterotic}

In the heterotic case, in order to break supersymmetry via a tachyon-free projection, one can start from the torus amplitude of the $E_8 \times E_8$ superstring, which reads
\begin{eqaed}\label{eq:torus_amp_h}
    \mathcal{T}_{\text{HE}} = \int_{\mathcal{F}} \frac{d^2\tau}{\tau_2^6} \, \frac{\left(V_8 - S_8 \right) \overline{\left( O_{16} + S_{16} \right)}^2}{|\eta(\tau)|^{16}} \, ,
\end{eqaed}
and project onto the states with even total fermion number\footnote{In contrast, projecting onto the states with even right-moving fermion number leads one to the T-dual $SO(32)$ heterotic superstring.}. This amounts to adding to (half of) the amplitude of eq.~\eqref{eq:torus_amp_h} its images under $S$ and $T$ modular transformations in such a way that the resulting total amplitude is modular invariant. The result is
\begin{eqaed}\label{eq:h_soxso_mod}
	\mathcal{T}_{SO(16) \times SO(16)} = \int_{\mathcal{F}} \frac{d^2\tau}{\tau_2^6} \frac{1}{|\eta(\tau)|^{16}} \, & \bigg[  O_8 \, \overline{\left( V_{16} \, C_{16} + C_{16} \, V_{16} \right)} \\
	& + V_8 \, \overline{\left( O_{16} \, O_{16} + S_{16} \, S_{16} \right)} \\
	& - S_8 \, \overline{\left( O_{16} \, S_{16} + S_{16} \, O_{16} \right)} \\
	& - C_8 \, \overline{\left( V_{16} \, V_{16} + C_{16} \, C_{16} \right)} \bigg] \, .
\end{eqaed}
Level matching purges tachyons from the spectrum, but the vacuum energy does not vanish\footnote{In some orbifold models, it is possible to obtain suppressed or vanishing leading contributions to the cosmological constant~\cite{Dienes:1990ij,Dienes:1990qh,Kachru:1998hd, Angelantonj:2004cm, Abel:2017rch}.}, since it is not protected by supersymmetry. Up to a volume prefactor, its value is given by eq.~\eqref{eq:h_soxso_mod}, and, since the resulting string-scale vacuum energy couples with the gravitational sector in a universal fashion\footnote{At the level of the space-time effective action, the vacuum energy contributes to the string-frame cosmological constant. In the Einstein frame, it corresponds to a runaway exponential potential for the dilaton.}, its presence also entails a gravitational tadpole, and thus a runaway exponential potential for the dilaton. In the Einstein frame, it takes the form
\begin{eqaed}\label{eq:runaway_potential_het}
	T \int d^{10}x \, \sqrt{- g} \, e^{\gamma \phi} \, , \qquad \gamma = \frac{5}{2} \, , 
\end{eqaed}
and thus the effect of the gravitational tadpoles on the low-energy physics of both the orientifold models and the $SO(16) \times SO(16)$ heterotic model can be accounted for with the same type of exponential dilaton potential.

\subsection{The low-energy description}

The string models introduced in the preceding sections can be described, at low energies, by effective actions of the type
\begin{equation}\label{eq:action}
    S = \frac{1}{16\pi G_D}\int d^D x \, \sqrt{-g} \, \left( R - \frac{4}{D-2}\, (\partial \phi)^2 - T \, e^{\gamma \phi} - \frac{f(\phi)}{2(p+2)!}\, H_{p+2}^2 \right) \, ,
    \end{equation}
following the notation of~\cite{Antonelli:2019nar} where
\begin{eqaed}\label{eq:f_coupling}
    f(\phi) \equiv e^{\alpha \phi} \, ,
\end{eqaed}
$T = \mathcal{O}\!\left(\frac{1}{\alpha'}\right)$ is the (fixed) supersymmetry-breaking gravitational tadpole and the form field is taken in the \textit{electric frame} where $\alpha > 0$. In this frame, the orientifold models are described by
\begin{equation}\label{eq:bsb_electric_params}
    D = 10 \, , \quad p = 1 \, , \quad \gamma = \frac{3}{2} \, , \quad \alpha = 1 \, ,
\end{equation}
while for the heterotic model the electric frame, described by
\begin{equation}\label{eq:het_magnetic_params}
    D = 10 \, , \quad p = 5 \, , \quad \gamma = \frac{5}{2} \, , \quad \alpha = 1 \, ,
\end{equation}
arises from the original magnetic frame via duality.

The equations of motion stemming from the action of eq.~\eqref{eq:action} are
\begin{subequations}\label{eq:eoms}
	\begin{align}
		R_{MN} & = \widetilde{T}_{MN} \, , \label{eq:eoms_R} \\
		\Box \, \phi - V'(\phi) - \frac{f'(\phi)}{2(p+2)!} \, H_{p+2}^2 & = 0 \, , \label{eq:eoms_phi} \\
		d  \star (f(\phi) \, H_{p+2}) & = 0 \label{eq:eoms_H} \, ,
	\end{align}
\end{subequations}
where the trace-reversed stress-energy tensor is
\begin{eqaed}\label{eq:stress_tensor}
    \widetilde{T}_{MN} & = \frac{4}{D-2} \, \partial_M \phi \partial_N \phi + \frac{f(\phi)}{2(p+1)!} \, (H_{p+2}^2)_{MN} \\
    &\quad\, + \frac{g_{MN}}{D-2} \left( V - \frac{p+1}{2(p+2)!} \, f(\phi) \, H_{p+2}^2 \right) \, .
\end{eqaed}
In the following the shall also need the parameters in the magnetic frame. Dualizing, for the orientifold models one finds
\begin{equation}\label{eq:bsb_magnetic_params}
    D = 10 \, , \quad p = 5 \, , \quad \gamma = \frac{3}{2} \, , \quad \alpha = -1 \, ,
\end{equation}
while for the heterotic model one recovers
\begin{equation}\label{eq:het_electric_params}
    D = 10 \, , \quad p = 1 \, , \quad \gamma = \frac{5}{2} \, , \quad \alpha = -1 \, ,
\end{equation}
where the Kalb-Ramond field strength appears as a $3$-form.

\section{Freund-Rubin compactifications}
\label{sec:FR}

In this section we initiate our search for possible $\ds$ vacua in non-supersymmetric string models, starting from Freund-Rubin compactifications. Let us remark that, in the presence of exponential potentials, the dilaton, whose VEV $\phi_0$ defines the string coupling $g_{\rm s} \equiv e^{\phi_0}$, is to be stabilized by (large) fluxes in order that the solutions be perturbative globally~\cite{Mourad:2016xbk}. To wit, the ten-dimensional low-energy description of eq.~\eqref{eq:action} does not admit flux-less vacua where the dilaton is stabilized. Therefore, maximally symmetric space-times may only arise from special compactifications, where the internal manifold is supported by fluxes. In this fashion, let us consider unwarped products of a $d$-dimensional, non-compact manifold $X$ and a $q$-dimensional, compact manifold $Y$. The Lorentzian space-time $X$ is considered \emph{external}, while $Y$ is the \emph{internal} Riemannian manifold. The ansatz for the metric reads
\begin{eqaed}
	\label{eq:gen_ds2_ansatz}
	d  s^2 & = L^2 \, d  s_X^2 + R^2 \, d  s_Y^2 \, , 
\end{eqaed}
with $L$ and $R$ the curvature radii of $X$ and $Y$ respectively. We require that both $X$ and $Y$ be maximally symmetric\footnote{As remarked in~\cite{Antonelli:2019nar}, our considerations apply to any Einstein internal manifold.}, and in the ensuing discussion we shall not specify the curvature of either. Namely, $X$ can be either ${\rm AdS}_{d}$, $\mathbb{M}_{d}$ or ${\rm dS}_{d}$, while $Y$ can be either a sphere $\mathbb{S}^q$ or a hyperbolic plane $\mathbb{H}^q$. We now look for solutions to eqs.~\eqref{eq:eoms} with constant dilaton $\phi = \phi_0$, generalizing the ones of~\cite{Antonelli:2019nar} to arbitrary (signs of the) curvatures. As emphasized above, from eq.~\eqref{eq:eoms_phi} one can readily deduce that, in order to stabilize the dilaton to a constant, a non-trivial $(p+2)$-form flux $H_{p+2}$ ought to be included. In \emph{Freund-Rubin} compactifications, a single flux threads either $X$ or $Y$, corresponding to the electric or magnetic frame respectively. In the former case, $H_{p+2}$ threads the whole space-time $X$, whose dimension is thus fixed to $d = p+2$. Therefore,
\begin{eqaed}
\label{eq:gen_Hel_ansatz}
	H_{p+2} & = c \, d {\rm vol}_{X} \,, 
\end{eqaed}
with $d {\rm vol}_{X}$ the volume form of $X$. Here $c$ is determined by the quantization condition
\begin{equation}\label{eq:gen_electric_flux}
n = \frac{1}{\Omega_Y}\int_{Y} f \star H_{p+2} = c \, f \, R^{q} \, ,
\end{equation}
where $n$ is quantized. Then, eq.~\eqref{eq:eoms_phi} yields
\begin{equation}
\label{eq:gen_eom_dilaton_el}
n^2 = 2 R^{2q} \frac{f^2 V'}{f'},
\end{equation}
while separating eq.~\eqref{eq:eoms_R} into its external and internal components using eq.~\eqref{eq:gen_ds2_ansatz}, one finds
\begin{eqaed}\label{eq:gen_eom_gravity_el}
	\sigma_{X} \,\frac{p+1}{L^2} \, (D-2) & = - \, (q-1) \, \frac{n^2}{2 f R^{2q}} \ +  V  \,,
	\\
	\sigma_{Y} \,\frac{q-1}{R^2} \, (D-2) & = (p+1) \, \frac{n^2}{2 f R^{2q}} \ + V  \,,
\end{eqaed}
where $\sigma_{X,Y} = +1$ if $X$ or $Y$ is elliptical and $\sigma_{X,Y} = -1$ if it is hyperbolic. 

On the other hand, eqs.~\eqref{eq:eoms} may be also solved considering a magnetic flux which threads the internal space $Y$. The corresponding ansatz for $H_{p+2}$ is
\begin{eqaed}
	\label{eq:gen_Hans}
	H_{p+2} & = c \, d {\rm vol}_{Y} \,,
\end{eqaed}
with $p+2 = D-d$ and with $d {\rm vol}_{Y}$ the volume form of $Y$. The quantization condition now reads
\begin{equation}\label{eq:gen_magnetic_flux}
n = \frac{1}{\Omega_Y}\int_{Y} H_{p+2} = c \, R^{p+2} \, ,
\end{equation}
and substituting in eq.~\eqref{eq:eoms_phi} leads to
\begin{equation}\label{eq:gen_eom_dilaton_mag}
n^2 = - \, 2 \, R^{2(p+2)} \, \frac{V'}{f'}\,.
\end{equation}
Eq.~\eqref{eq:eoms_R} now takes the form
\begin{eqaed}
	\label{eq:gen_eom_gravity_mag}
	\sigma_{X} \,\frac{D-p-3}{L^2} \, (D-2) & = - \, (p+1) \, \frac{ n^2 f }{2 R^{2(p+2)}} \, +  V  \,,
	\\
	\sigma_{Y} \,\frac{p+1}{R^2} \, (D-2) & = (D-p-3) \, \frac{n^2 f}{2 R^{2(p+2)}} \, + V  \,,
\end{eqaed}
which are simply the electromagnetic dual of eq.~\eqref{eq:gen_eom_gravity_el}.

Clearly, Freund-Rubin compactifications are allowed if and only if eqs.~\eqref{eq:gen_eom_dilaton_el} and \eqref{eq:gen_eom_gravity_el}, for an electric flux, or eqs.~\eqref{eq:gen_eom_dilaton_mag} and \eqref{eq:gen_eom_gravity_mag}, for a magnetic flux, admit positive solutions for the string coupling $g_{\rm s} = e^{\phi_0}$ and the curvature radii $R$ and $L$. In this regard, eqs.~\eqref{eq:gen_eom_dilaton_el} and \eqref{eq:gen_eom_dilaton_mag} provide important constraints: indeed, an electric flux requires that
\begin{equation}
	{\rm sgn} \, \alpha = {\rm sgn} \,\gamma\,.
\end{equation}
Hence, only the orientifold models, described by eq.~\eqref{eq:bsb_electric_params}, afford solutions of this type with an electric flux. Conversely, a magnetic flux requires that
\begin{equation}
{\rm sgn} \,\alpha = - \,{\rm sgn}\, \gamma\,,
\end{equation}
which is the case for the heterotic model, described by eq.~\eqref{eq:het_magnetic_params}. Furthermore, eqs.~\eqref{eq:gen_eom_gravity_el} and~\eqref{eq:gen_eom_gravity_mag} imply that $\sigma_Y = +1$, \textit{i.e.} the internal manifold $Y = \mathbb{S}^q$.

Given these preliminary constraints, in the following we shall explore which space-time geometries are allowed out of ${\rm AdS}_{d}$, $\mathbb{M}_{d}$ or ${\rm dS}_{d}$.

\subsection{AdS solutions}
\label{sec:FR_AdS}

The ${\rm AdS}_d$ solutions of~\cite{Antonelli:2019nar} can be recovered setting $\sigma_X= -1$ and $\sigma_Y = 1$, and they are perturbative for large fluxes whenever the constraints are satisfied. In the string models described in Section~\ref{sec:non-susy_strings}, in the electric frame $p=1$ for the orientifold models, while $p=5$ for the heterotic model. Hence, the orientifold models allow only ${\rm AdS}_3 \times \mathbb{S}^7$ solutions of this type, and in this case
\begin{eqaed}
	\label{eq:caseI_elbsb_sol}
	g_{\rm s} &=  2^{\frac74} \times 3 n^{-\frac14} T^{\frac34} \,,
	\\
	R^2 &=  2^{-\frac58} \times 3^{-\frac12} n^{\frac38} T^{\frac18} \,,
	\\
	L^2 &= \frac{R^2}{6}\,.
\end{eqaed}
Conversely, the $SO(16) \times SO(16)$ heterotic model admits only ${\rm AdS}_7 \times \mathbb{S}^3$ solutions of this type, with a magnetic flux threading the internal $\mathbb{S}^3$. Solving the equations of motion, one finds
\begin{eqaed}
	\label{eq:caseI_maghet_sol}
	g_{\rm s} &=  5^{\frac14}  n^{-\frac12} T^{-\frac12} \,,
	\\
	R^2 &=  5^{-\frac58}  n^{\frac54} T^{\frac14} \,,
	\\
	L^2 &= 12 \, R^2\,.
\end{eqaed}
As a final remark, in both cases the solution is completely specified by the flux parameter $n$. For large values of $n$ the string coupling is small, while the curvature radii $R$ and $L$ are large, and thus both curvature and string loop corrections are expected to be negligible. Moreover, the curvature radii scale with the same power of $n$, and thus these solutions are not scale separated.

\subsection{The obstructions to de Sitter and Minkowski solutions}

Let us now seek solutions with $\ds$ or Minkowski space-times. For an external ${\rm dS}_{d}$ space-time, $\sigma_X =1$. The internal manifold $Y$ is necessarily a sphere $\mathbb{S}^q$, as in~\cite{Montero:2020rpl}, or an Einstein manifold of positive curvature. As for the $\ads$ case, one can consider either ${\rm dS}_{3} \times \mathbb{S}^7$ solutions with an electric flux in the orientifold models or ${\rm dS}_{7} \times \mathbb{S}^3$ solutions with a magnetic flux in the heterotic model. However, no such solutions exist: the flux quantization conditions do not exclude them at the outset, but eqs.~\eqref{eq:gen_eom_gravity_el} and \eqref{eq:gen_eom_gravity_mag} do not admit solutions of this type with positive curvature radii $R$ and $L$.

Moreover, for different reasons, one cannot generically find Minkowski solutions. This case would correspond to the limit $L \to \infty$ in eq.~\eqref{eq:gen_ds2_ansatz}. However, as is evident from eqs.~\eqref{eq:gen_eom_gravity_el} and \eqref{eq:gen_eom_gravity_mag}, in this limit a solution can only exist if the contribution of the dilaton potential is exactly canceled by the flux contribution. Alternatively, in the absence of fluxes, one could conceive an asymptotically Minkwoski vacuum with $\phi \to -\infty$, but the considerations in~\cite{Antonelli:2019nar} exclude this scenario.

\section{A no-go theorem for dS and Minkowski solutions}
\label{sec:no-go}

The difficulties in finding $\ds$ solutions encountered in the preceding section can be put on more general and firmer grounds. As is known from supersymmetric compactifications~\cite{Gibbons:1984kp,Dasgupta:1999ss,Maldacena:2000mw,Giddings:2001yu,Green:2011cn}, reducing ten or eleven-dimensional supergravity theories over a compact manifold imposes stringent, global constraints on the lower-dimensional theory\footnote{Similar conclusions can be reached studying the Raychaudhuri equation in higher-dimensional settings~\cite{Das:2019vnx}, or employing a world-sheet analysis as in~\cite{Kutasov:2015eba}.}. In particular, the value of the cosmological constant in the reduced theory is restricted. 

To wit, in~\cite{Maldacena:2000mw} it was demonstrated that compactifying fairly generic $D$-dimensional theories of gravity over a compact and non-singular $(D-d)$-dimensional manifold necessarily leads to a $d$-dimensional theory with a strictly negative cosmological constant. However, the proof in~\cite{Maldacena:2000mw} relies on the assumption that the potential in the $D$-dimensional theory not be positive definite. This is ostensibly in contrast with the non-supersymmetric string models described by actions of the form of eq.~\eqref{eq:action}, where a positive definite potential strikingly appears in the ten-dimensional low-energy description. On the one hand, this might suggest that the no-go theorems forbidding dS or Minkowski compactifications could be evaded by the non-supersymmetric models presented in Section~\ref{sec:non-susy_strings}; on the other hand, the obstructions to compactifications of this type that we have found in the preceding section compel one to seek proper and more general justifications.

Indeed, let us compactify the ten-dimensional non-supersymmetric string models, specified by the general action in eq.~\eqref{eq:action}, down to $d < D$ space-time dimensions, and let $q \equiv D - d$. We consider the metric ansatz
\begin{equation}
\label{no-go_metr_ans}
d s_{10}^2  = e^{- \frac{2q}{d-2}C(y)} \, \widehat g_{\mu\nu} (x) d  x^\mu d  x^\nu+ e^{2C(y)} \,  \widetilde g_{ab} (y) d  y^a d  y^b\, ,
\end{equation}
where $x^\mu$, $\mu = 0,1,\ldots,d-1$ denote the external coordinates and $y^a$,  $a=1,\ldots,q$ denote the internal coordinates. Then, retracing the same arguments of~\cite{Maldacena:2000mw}, we arrive at the formulation of the following no-go result:

\begin{importantbox}
{\bf dS and Minkowski no-go theorem for non-supersymmetric string theories:}\\ \emph{Consider a (warped) compactification of the ten-dimensional non-supersymmetric string models of Section~\ref{sec:non-susy_strings}, described at low energies by the action in eq.~\eqref{eq:action}, over a closed, compact manifold $Y$, with ${\rm dim}\, Y > 2$. The internal manifold $Y$ is threaded by a magnetic $(p+2)$-form flux $H_{p+2}$ spanning an arbitrary cycle of dimension $p+2 \leq q$.  Then, whenever
\begin{equation}
\label{eq:no-go}
\frac{\alpha}{\gamma}+ (p+1)  > 0 \, ,
\end{equation}
no compactifications to either $d$-dimensional Minkowski or dS space-times are allowed.}
\end{importantbox}

This result generalizes the no-go theorem for Freund-Rubin compactifications, which was also discussed in~\cite{Montero:2020rpl} and follows from the general expressions in~\cite{Antonelli:2019nar}, and the proof, which proceeds along the lines of~\cite{Maldacena:2000mw,Giddings:2001yu}, is given in Appendix~\ref{app:no-go_proof}. Below we shall limit ourselves to comment on the implications of this result. The inequality~\eqref{eq:no-go} is to be interpreted as a constraint on the parameters entering the action in eq.~\eqref{eq:action} in order \emph{not} to admit dS or Minkowski compactifications. Indeed, \eqref{eq:no-go} does not entirely exclude dS or Minkowski compactifications, which might be realized when the inequality in eq.~\eqref{eq:no-go} is violated. In order for this to happen, since $p>0$, it is necessary that
\begin{equation}
{\rm sgn} \,\alpha = {\rm sgn} \, \frac{f'}{f} = - \, {\rm sgn} \,\frac{V'}{V}  = - \, {\rm sgn} \,\gamma\,.
\end{equation}
For instance, let us consider the non-supersymmetric string models introduced in Section~\ref{sec:non-susy_strings}, which feature a 3-form flux or a 7-form flux. In particular, the BSB and type $0'$B orientifold models, specified by the parameters in eq.~\eqref{eq:bsb_electric_params}, and the heterotic model, specified by those in eq.~\eqref{eq:het_magnetic_params}, have $\gamma>0$ and $\alpha>0$. As such, they cannot allow for $\ds$ or Minkowski vacua of this type. On the other hand, dualizing the orientifold models the relevant parameters are encoded in eq.~\eqref{eq:bsb_magnetic_params}, while dualizing the heterotic model the relevant parameters are encoded in eq.~\eqref{eq:het_electric_params}. In this case $\gamma>0$, while $\alpha<0$. Nevertheless, \eqref{eq:no-go} holds, and one is thus led to the conclusion that, in the non-supersymmetric string models under consideration, dS and Minkowski compactifications are not allowed. Therefore, one may refine the above no-go theorem by specializing to the UV-complete models examined in Section~\ref{sec:non-susy_strings}: 

\begin{importantbox}
{\bf dS and Minkowski no-go theorem for orientifold and heterotic models with broken supersymmetry:} \emph{The non-supersymmetric BSB orientifold model, the type $0'$B orientifold model and the $SO(16)\times SO(16)$ heterotic model do not admit (warped) compactifications to $d>2$ dimensional dS or Minkowski space-times.}
\end{importantbox}

As a final remark, as was observed in~\cite{Maldacena:2000mw}, the no-go theorems stated above strictly hold for \emph{static} compactifications, and with warp factors depending only on the internal coordinates. In other words, time-dependent dS solutions might still be viable in principle, albeit explicit constructions appear quite challenging. 

\subsection{Including space-time filling sources}
\label{sec:no-go_sources}

A possible way to evade the no-go theorem relies on the inclusion of localized sources, which may introduce singularities in the internal manifold $Y$. Specifically, \emph{localized} objects are intended as objects which are not resolved and whose world-volumes are described by $\delta$-functions.

Let us consider a single such localized object in the low-energy action of eq.~\eqref{eq:action}. It spans a $(p+1)$-dimensional hypersurface which is parametrized by the world-volume coordinates $\xi^i$, $i=0,\ldots,p$. In the ambient $D$-dimensional space-time, it spans a hypersurface that is specified by the embedding $\xi^i \mapsto x^M(\xi)$. Its dynamics and coupling to bulk fields are encoded in an action of the form
\begin{eqaed}
	\label{eq:sources_Sloc}
	 S_{\rm loc} = - \,\sigma_{\tau} \int  d^{p+1}\xi\, \sqrt{-h} \,\mathcal{T}_p(\phi) + q_{p} \int C_{p+1}\,.
\end{eqaed}
In the first term, $h \equiv {\rm det}\, h_{ij}$, where $h_{ij}$ is the pullback
\begin{eqaed}\label{eq:sources_hij}
	h_{ij} = g_{MN} \frac{\partial x^M}{\partial \xi^i} \frac{\partial x^N}{\partial \xi^j}
\end{eqaed}
of the space-time metric $g_{MN}$ to the world-volume. For convenience, we shall work in the \emph{static gauge}, in which the world-volume coordinates $\xi^i$ coincide with the first $p+1$ space-time coordinates $x^M$,  
\begin{eqaed}\label{eq:sources_xpar}
	x^i = \xi^i \, .
\end{eqaed}
We shall further denote the residual space-time coordinates, which are transverse to the object, as $x_\perp^K$, with $K=p+1,\ldots,D$. As a further simplifying assumption, we shall also assume that the object be \emph{static}, namely
\begin{eqaed}\label{eq:sources_xperp}
	x_{\perp}^{K} = z^{K}_0
\end{eqaed}
for constant $z^{K}_0$, which implies that $h_{ij} = g_{ij}$. In eq.~\eqref{eq:sources_Sloc}, the positive-definite tension $\mathcal{T}_ p$ is allowed to depend on the dilaton, the sole bulk scalar field. Echoing the behavior of fundamental branes, we shall set
\begin{eqaed}\label{eq:sources_tau}
	\mathcal{T}_p = |\tau_p| \, e^{\sigma \phi}\,,
\end{eqaed}
with $\sigma$ and $\tau_p$ constants. Furthermore, $\sigma_{\tau} = \pm 1$ in eq.~\eqref{eq:sources_Sloc} may eventually account for sources with negative tension, such as orientifold planes. In addition, we have assumed the object to have charge $q$ under the $(p+1)$-form field. The full action describing the coupling of the source to the bulk is then
\begin{eqaed}\label{eq:sources_Stot}
	S = S_{\rm bulk} + S_{\rm loc}\,,
\end{eqaed}
where the bulk contribution arises from eq.~\eqref{eq:action}.

An argument analogous to the preceding one then leads to the following, extended no-go theorem

\begin{importantbox}
	{\bf dS and Minkowski no-go theorem for non-supersymmetric string theories with space-time filling sources:}
	\\
	\emph{Consider a compactification of the ten-dimensional non supersymmetric string models of Section~\ref{sec:non-susy_strings}, described by the effective action of eq.~\eqref{eq:action}, over a closed, compact manifold $Y$, with ${\rm dim}\, Y > 2$. Assume the presence of a single space-time-filling source, which is localized in $Y$ and described by the action in eq.~\eqref{eq:sources_Sloc}. Furthermore, the internal manifold $Y$ is threaded by a magnetic $(p+2)$-form flux $H_{p+2}$ spanning an arbitrary cycle of dimension $p+2 \leq q$. Then, if both
	\begin{equation}
	\label{eq:no-go_gen_sources}
	\frac{\alpha}{\gamma}+ (p+1)  > 0 \quad {\rm and} \quad \sigma_\tau \left(p -7 -\frac{2 \, \sigma}{\gamma}  \right)< 0\,,
	\end{equation}
	no compactifications to either $d$-dimensional Minkowski or dS space-times are allowed.}
\end{importantbox}

The proof can be found in Appendix~\ref{app:no-go_proof}. As in the source-less case, the inequalities in eq.~\eqref{eq:no-go_gen_sources} are to be intended as constraints on the parameters in the bulk action of eq.~\eqref{eq:action} and the source-dilaton coupling in eq.~\eqref{eq:sources_Sloc} which exclude dS or Minkowski solutions. In particular, if
\begin{equation}
	\sigma_\tau \left(p -7 -\frac{2 \sigma}{\gamma}  \right) > 0
\end{equation}
the non-supersymmetric string models introduced in Section~\ref{sec:non-susy_strings} might \textit{a priori} admit dS or Minkowski compactifications. The contribution of additional sources can be included without further difficulties.

As an additional remark, it is worthwhile mentioning that the inclusion of localized objects may be compulsory if (generalized) global symmetries~\cite{Gaiotto:2014kfa} are to be avoided. In fact, if models described by actions of eq.~\eqref{eq:action} ought to be promoted to quantum gravity models, no global symmetries can be present\footnote{For arguments to this effect, see~\cite{Banks:2010zn}.}. In particular, the action of eq.~\eqref{eq:action} exhibits two global symmetries: a $(p+1)$-form symmetry shifting the gauge field by a flat connection, namely $B_{p+1} \to B_{p+1} + \Lambda_{p+1}$, and its magnetic counterpart. Whenever no other mechanisms are available, including localized sources such as those described by eq.~\eqref{eq:sources_Sloc} explicitly breaks these symmetries~\cite{Montero:2017yja,McNamara:2019rup}.

\section{Vacua of the lower-dimensional theory and perturbative instabilities}
\label{sec:red_pot}

In the preceding section we have provided a top-down argument against the existence of dS solutions for UV-complete non-supersymmetric string models. However, one can reach similar conclusions via purely lower-dimensional arguments. To this effect, in this section we shall prove that, for arbitrary values of the gauge and dilaton couplings, bottom-up non-supersymmetric models with exponential potentials that afford dS compactifications necessarily develop perturbative instabilities.

To begin with, let us compute the relevant effective action of the $d$-dimensional reduction of eq.~\eqref{eq:action} over a $(10-d)$-dimensional manifold. In addition to our preceding considerations, we shall also include the radion field, a universal modulus that parametrizes the volume of the internal manifold. In detail, let us consider the metric ansatz
\begin{equation}
\label{eq:bound_ds2}
d s^2 = e^{2{B} \rho(x)} \widehat{d s}^2_X + e^{2 {A} \rho(x)} \widetilde{d s}_Y^2 \, ,
\end{equation}
where the parameters ${A}$ and ${B}$, given by
\begin{equation}
\label{eq:bound_alphabeta}
{A} = - \, \sqrt{ \frac{d-2}{16(10-d)}} \,, \qquad {B} = \sqrt{ \frac{10-d}{16(d-2)}}\,,
\end{equation}
have been chosen in order that the $d$-dimensional action be expressed in the Einstein frame and with canonically normalized kinetic terms. Furthermore, we shall assume the presence of a magnetic $(10-d)$-form flux, which threads the internal manifold $Y$ and is quantized according to
\begin{equation}
\label{eq:bound_Hn}
\frac{1}{\widetilde{\rm vol}_Y} \int_{Y} H_{d_Y} = n\,.
\end{equation}
As explained in detail in Appendix~\ref{app:reduced_pot}, using the metric ansatz in eq.~\eqref{eq:eft_ds2} the ten-dimensional action of eq.~\eqref{eq:action} leads to the reduced action
\begin{equation}
\label{eq:bound_redS}
\begin{aligned}
S = \int_X d^dx\, \sqrt{|\widehat g_X|} \left(\widehat R- \frac12 (\widehat\partial \rho)^2 -  \frac12 (\widehat\partial \phi)^2 - \mathcal{V}(\phi,\rho)\right) \, ,
\end{aligned}
\end{equation}
where the effective potential $\mathcal{V}(\phi,\rho)$ for the radion and the dilaton takes the form
\begin{equation}
\label{eq:bound_redpot}
\begin{aligned}
\mathcal{V}(\phi,\rho) &= e^{2 {B}\rho}V(\phi) + \frac12 f n^2  \sigma_Y   e^{2 {B} \rho (d-1)}- r \, e^{2({B}-{A})\rho}\, .
\end{aligned}    
\end{equation}
While we have derived eq.~\eqref{eq:bound_redpot} from unwarped compactifications, one can carry out an analogous computation in the warped case, and our ensuing discussion is unaffected by this generalization.\footnote{Let us remark that the action of eq.~\eqref{eq:bound_redS} may not describe a proper EFT in general, since we have not included all the geometry-dependent moduli. In addition, in the absence of scale separation one ought to include higher Kaluza-Klein modes, which were studied in detail in~\cite{Basile:2018irz} for the $\ads$ solutions discussed in Section~\ref{sec:FR_AdS}.}

In addition to the tadpole contribution proportional to $V(\phi)$, the potential in eq.~\eqref{eq:bound_redpot} includes the contribution
\begin{equation}
\label{eq:bound_redpotflux}
    \mathcal{V}_{\rm flux} = \frac12 f n^2  \sigma_Y   e^{2 {B} \rho (d-1)} \, ,
\end{equation}
arising from the magnetic flux, with $\sigma_{Y}=+1$ ($\sigma_{Y}=-1$) if $Y$ is elliptical (hyperbolic), and a contribution from the internal curvature, with 
\begin{eqaed}\label{eq:curvature_r}
    r \equiv \frac{1}{\widetilde{\rm vol}_Y} \int_Y \sqrt{\widetilde g_Y}  \widetilde R(y) \, .
\end{eqaed}
Concretely, for a maximally symmetric internal space, $r= (10-d)(9-d) \, \sigma_Y$. As we have discussed in Section~\ref{sec:FR}, the choice $\sigma_Y = -1$ does not lead to Freund-Rubin solutions, and therefore in the following we shall set $\sigma_Y = +1$. Moreover, let us recall that, for the non-supersymmetric models at stake, the gauge kinetic function $f$ and the potential $V$ take the form
\begin{equation}
    f(\phi) = e^{\alpha\phi}\,,\qquad V(\phi) = T e^{\gamma \phi}\,,
\end{equation}
with $\alpha$ and $\gamma$ real parameters. In this case, the potential in eq.~\eqref{eq:bound_redpot} can be recast in terms of its derivatives according
\begin{equation}
\label{eq:bound_redpotb}
\boxed{\begin{aligned}
	\mathcal{V} (\phi,\rho) &= \frac{{A}}{\gamma ({A}-{B})}\, \partial_{\phi} \mathcal{V} + \frac{1}{2({B} - {A})}\,\partial_{\rho} \mathcal{V}  - \frac{d-2}{8} \left(p+1 +\frac{\alpha}{\gamma} \right) \mathcal{V}_{\rm flux}
	\\
	&= \frac{{A} + {B} (d-2)}{\alpha ({B}-{A})} \,\partial_{\phi} \mathcal{V} + \frac{1}{2({B} - {A})}\,\partial_{\rho} \mathcal{V}  + \frac{d-2}{8} \left( (p+1) \frac{\gamma}{\alpha}  +1\right)  e^{2 {B}\rho}V(\phi)\,.
\end{aligned}}
\end{equation}
The above useful form of the lower-dimensional potential allows a systematic study of the vacua and of their perturbative stability. Indeed, as one can readily observe from eq.~\eqref{eq:bound_redpotb}, extremizing the reduced potential with respect to the dilaton and the radion yields the local extremum
\begin{equation}
\label{eq:bound_redpot_min}
 \mathcal{V}_{\rm ext} = - \, \frac{d-2}{16} \,n^2\, e^{2(d-1){B}\rho + \alpha \phi} \left(p+1  + \frac{\alpha}{\gamma} \right) \, ,
\end{equation}
whose sign depends on the parameters $\gamma$ and $\alpha$ mirroring the inequality of eq.~\eqref{eq:no-go}.

In the following, after re-examining the vacua in the non-supersymmetric models introduced in Section~\ref{sec:non-susy_strings}, we shall comment on the stability of dS vacua in bottom-up models.

\subsection{The BSB and type \texorpdfstring{$0'$}{0'}B orientifold models}

\begin{center}
	\begin{figure}[ht]
		\centering
		\includegraphics[width=7.0cm]{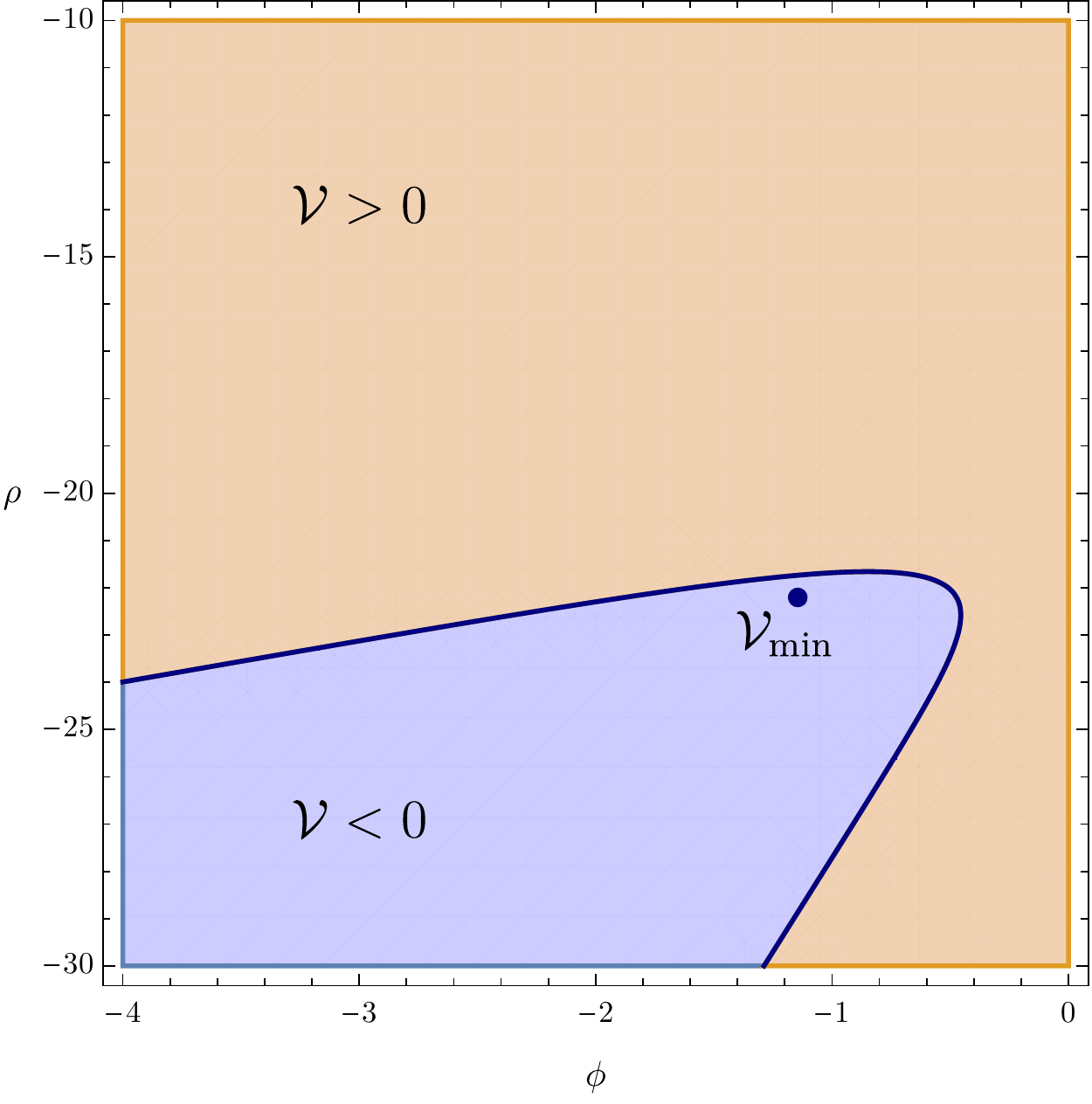}\hspace{0.5cm}
		\includegraphics[width=7.0cm]{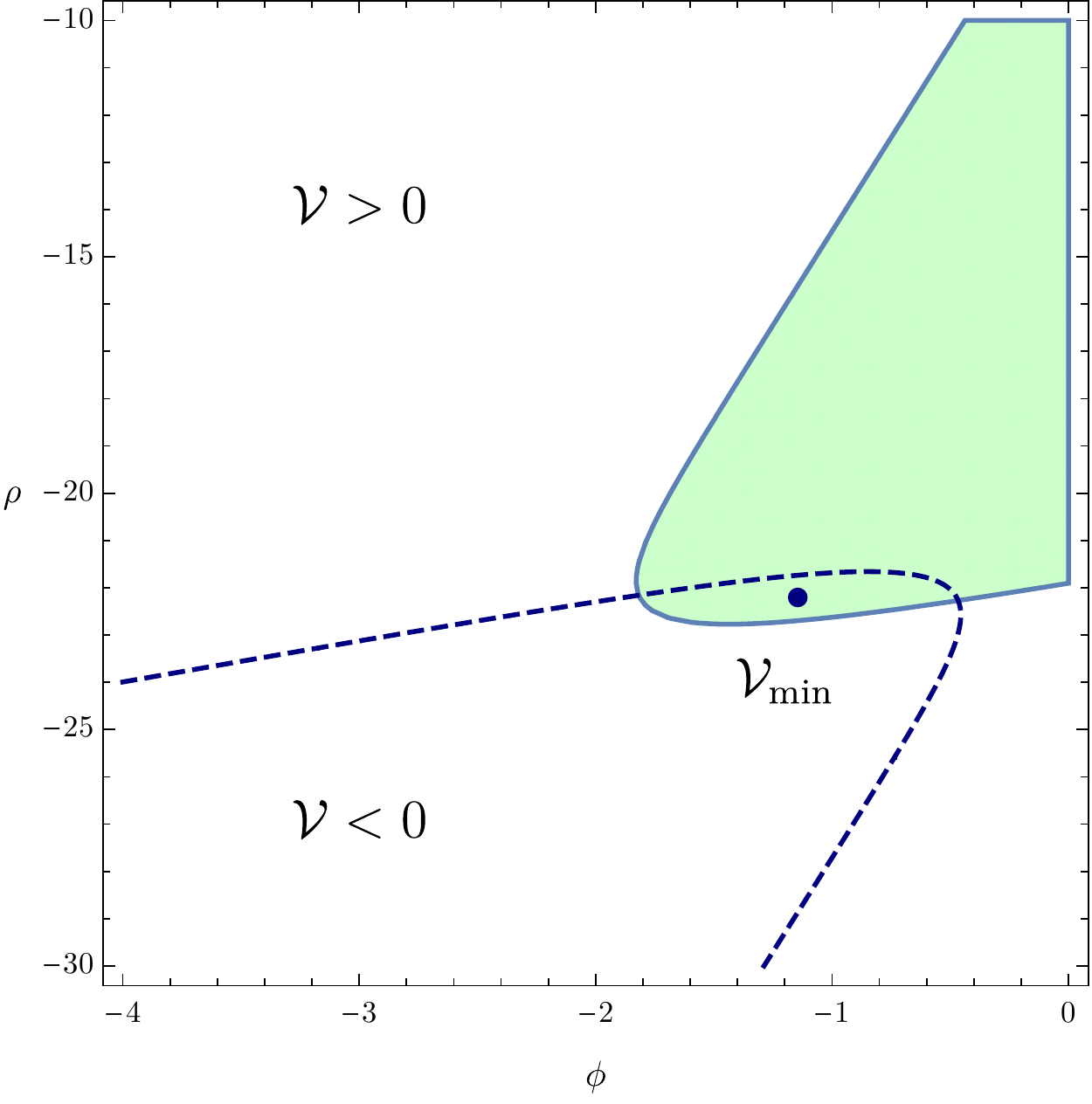}
		\caption{Plots of the regions where the potential in eq.~\eqref{eq:bound_bsb_redpot} is positive and negative definite (left) and the region of stability, where the Hessian is positive definite (right), for the BSB and type $0'$B orientifold models. We have chosen $n=10^6$ and $\mathcal{V}$ is expressed in units of $T$. \label{Fig:V_bsb}}
	\end{figure}
\end{center}

The compactification of the BSB and type $0'$B orientifold models, described by the parameters in eq.~\eqref{eq:bsb_electric_params}, on a seven-dimensional manifold leads to the effective potential
\begin{equation}
\label{eq:bound_bsb_redpot}
\begin{aligned}
\mathcal{V}(\phi,\rho) &= e^{\frac{\sqrt{7}}2\rho+\frac32 \phi} T+ \frac12 e^{\sqrt{7}\rho-\phi} n^2  \sigma_Y   - r \, e^{\frac{4}{\sqrt{7}}\rho}
\end{aligned}    
\end{equation}
for the radion and the dilaton. As we have discussed in Section~\ref{sec:FR_AdS}, there is a single AdS minimum, with
\begin{equation}
e^{\phi} =  2^{\frac74} \times 3 n^{-\frac14} T^{\frac34}\,,\qquad e^{-\frac{\rho}{2\sqrt{7}}} = 2^{-\frac58} 3^{-\frac12} n^{\frac38} T^{\frac1{8}} \, ,
\end{equation}
consistently with eq.~\eqref{eq:caseI_elbsb_sol}, and no dS or Minkowski solutions. At the AdS minimum, the masses of dilaton and radion fluctuations are
\begin{equation}
\{m^2\} = \frac{62208}{n^3 \, T} \, \{1,3\} \, ,
\end{equation}
the eigenvalues of the Hessian matrix of the potential in eq.~\eqref{eq:bound_bsb_redpot}. This minimum is depicted in Fig.~\ref{Fig:V_bsb}, along with the sign of the potential and its region of stability in the $(\phi,\rho)$-plane, where the Hessian is positive definite.

\subsection{The \texorpdfstring{$SO(16)\times SO(16)$}{SO(16) x SO(16)} heterotic model}

\begin{center}
	\begin{figure}[ht]
		\centering
			\includegraphics[width=7.0cm]{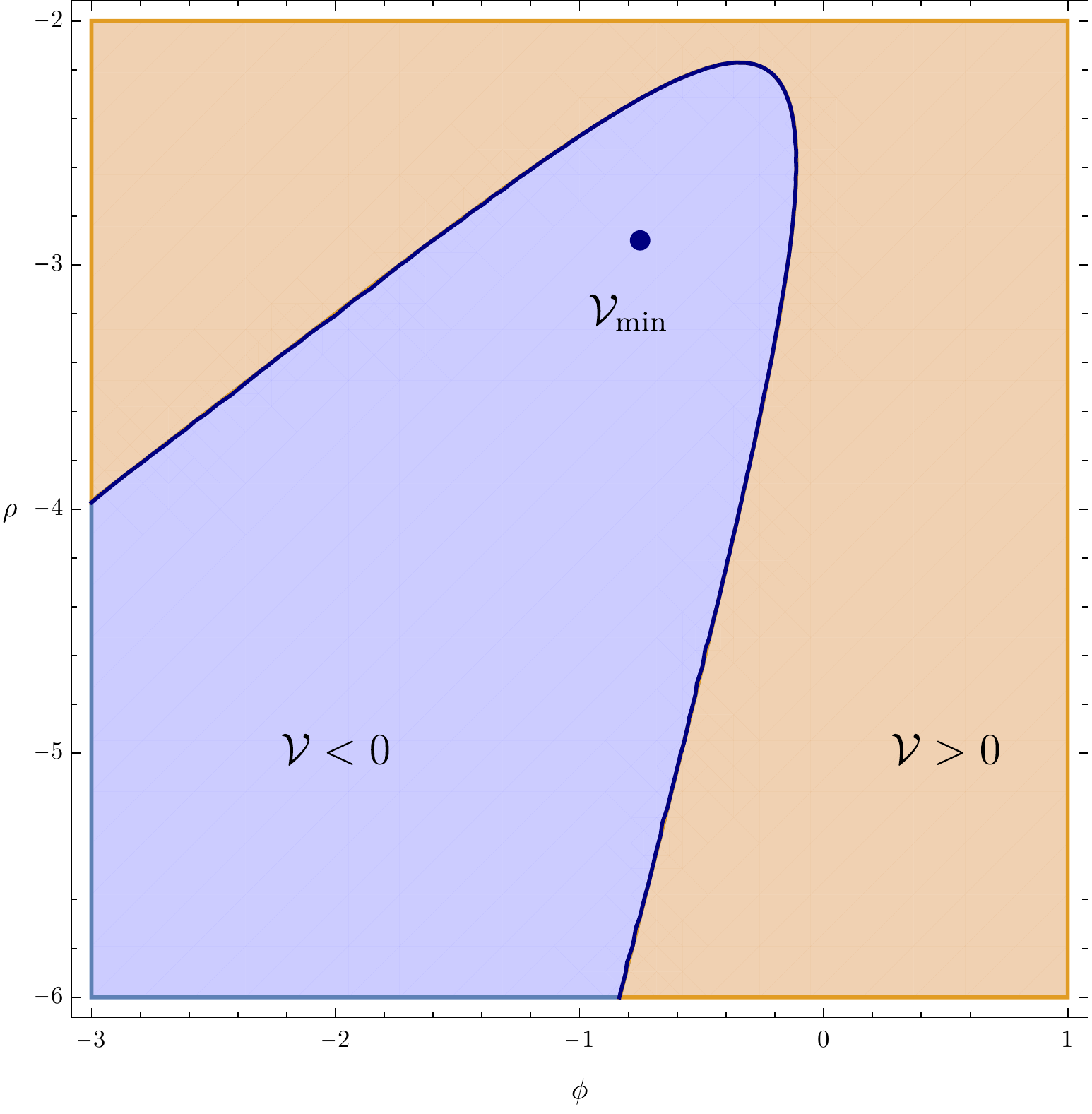}\hspace{0.5cm}
			\includegraphics[width=7.0cm]{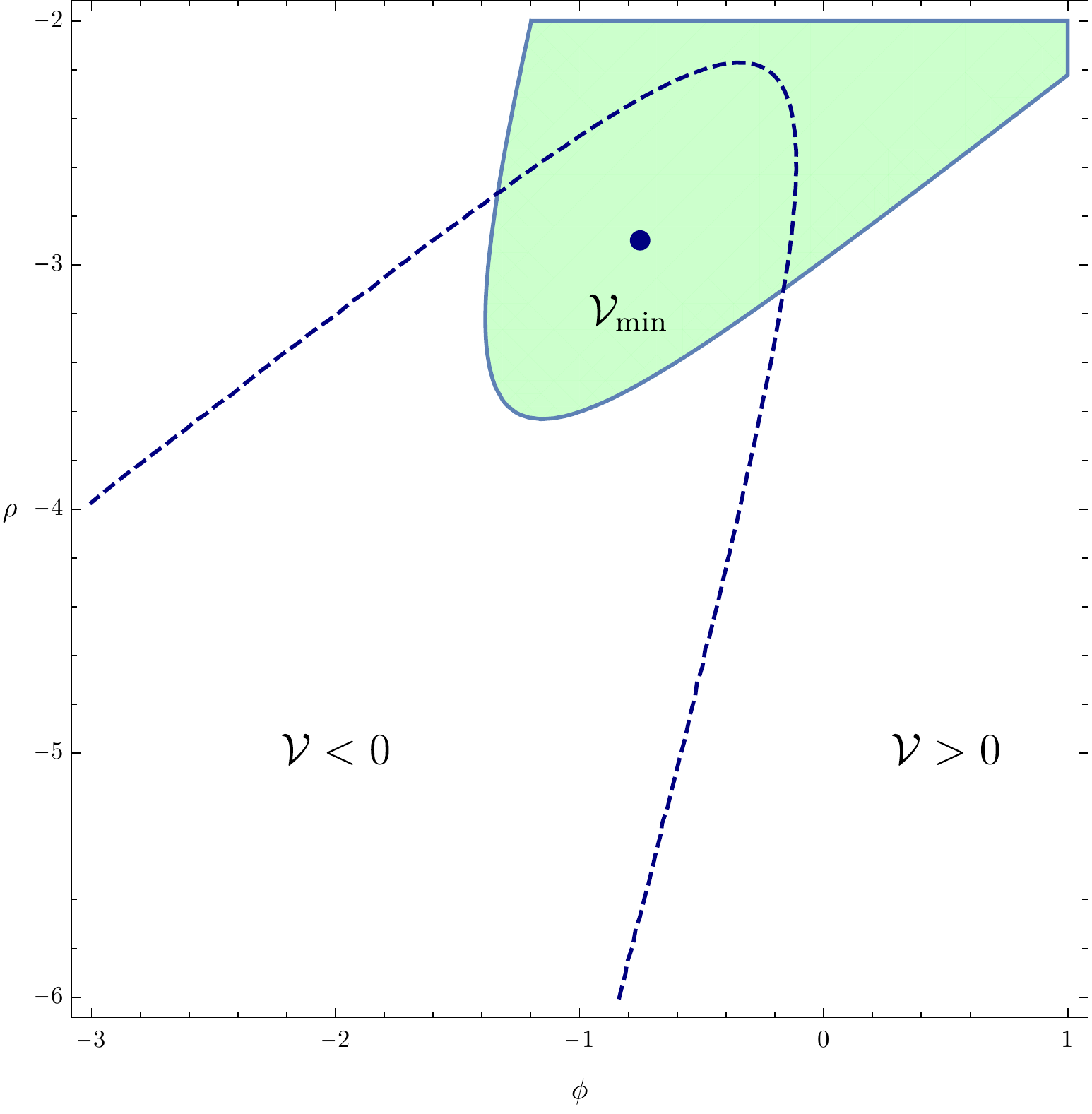}
		\caption{Plots of the regions where the potential in eq.~\eqref{eq:bound_het_redpot} is positive and negative definite (left) and the region of stability, where the hessian is positive definite (right), for the $SO(16)\times SO(16)$ heterotic model. We have chosen $n=10$ and $\mathcal{V}$ is expressed in units of $T$. \label{Fig:V_het}}
	\end{figure}
\end{center}

Let us now turn to the $SO(16)\times SO(16)$ heterotic model, described by the parameters in eq.~\eqref{eq:het_magnetic_params}, compactified over a three-dimensional manifold. The radion-dilaton potential in the reduced theory then reads
\begin{equation}
\label{eq:bound_het_redpot}
\begin{aligned}
\mathcal{V}(\phi,\rho) &= e^{\frac12 \sqrt{\frac35}\rho+\frac52 \phi} T+ \frac12 e^{3 \sqrt{\frac35}\rho-\phi} n^2  \sigma_Y   - r \, e^{-\frac{4}{\sqrt{15}}\rho}\, .
\end{aligned}    
\end{equation}
Analogously to the orientifold models, one can be show that it admits a single AdS minimum with
\begin{equation}
	e^{\phi} =  5^{\frac14}  n^{-\frac12} T^{-\frac12}\,,\qquad e^{\frac{\rho}{2\sqrt{15} }} = 5^{\frac18} n^{-\frac14} T^{-\frac1{20}} \, ,
\end{equation}
in agreement with eq.~\eqref{eq:caseI_maghet_sol}. The resulting masses of dilaton and radion flucuations, the eigenvalues of the Hessian matrix of the potential at the minimum, are
\begin{equation}
\{m^2\} = \frac{10}{n^2 \, T^{\frac25}} \, \{4 - \sqrt{6}, 4+\sqrt{6}\} \, .
\end{equation}
The sign of the potential, along with its region of stability in the $(\phi,\rho)$-plane, is depicted in Fig.~\ref{Fig:V_het}.

\subsection{dS vacua and instabilities}

As we have remarked above, and as expected from the no-go theorem introduced in Section~\ref{sec:no-go}, the $SO(16)\times SO(16)$ heterotic model, the BSB model and type $0'$B model admit only a single, AdS minimum, which however can develop perturbative instabilities in higher scalar Kaluza-Klein sectors, depending on the choice of internal manifold~\cite{Basile:2018irz}. However, for general values of the parameters $\gamma$ and $\alpha$, the potential in eq.~\eqref{eq:bound_redpot} may afford dS extrema: in particular, from eq.~\eqref{eq:bound_redpot_min} one can conclude that they exist whenever
\begin{equation}
\label{eq:bound_dSres}
    \frac{\alpha}{\gamma} + p+1 < 0\,,
\end{equation}
in compliance with the no-go theorem of eq.~\eqref{eq:no-go}. Conversely, requiring that the potential \eqref{eq:bound_redpot} cannot accommodate dS extrema recovers eq.~\eqref{eq:no-go} from a bottom-up perspective, which resonates with the analysis of~\cite{Montero:2020rpl}. However, it is worth noting that the top-down proof of the no-go theorem in Section~\ref{sec:no-go} is more general, since it does not rest on any hypothesis on the structure of the moduli space. 

\begin{center}
	\begin{figure}[ht]
		\centering
			\includegraphics[width=7.0cm]{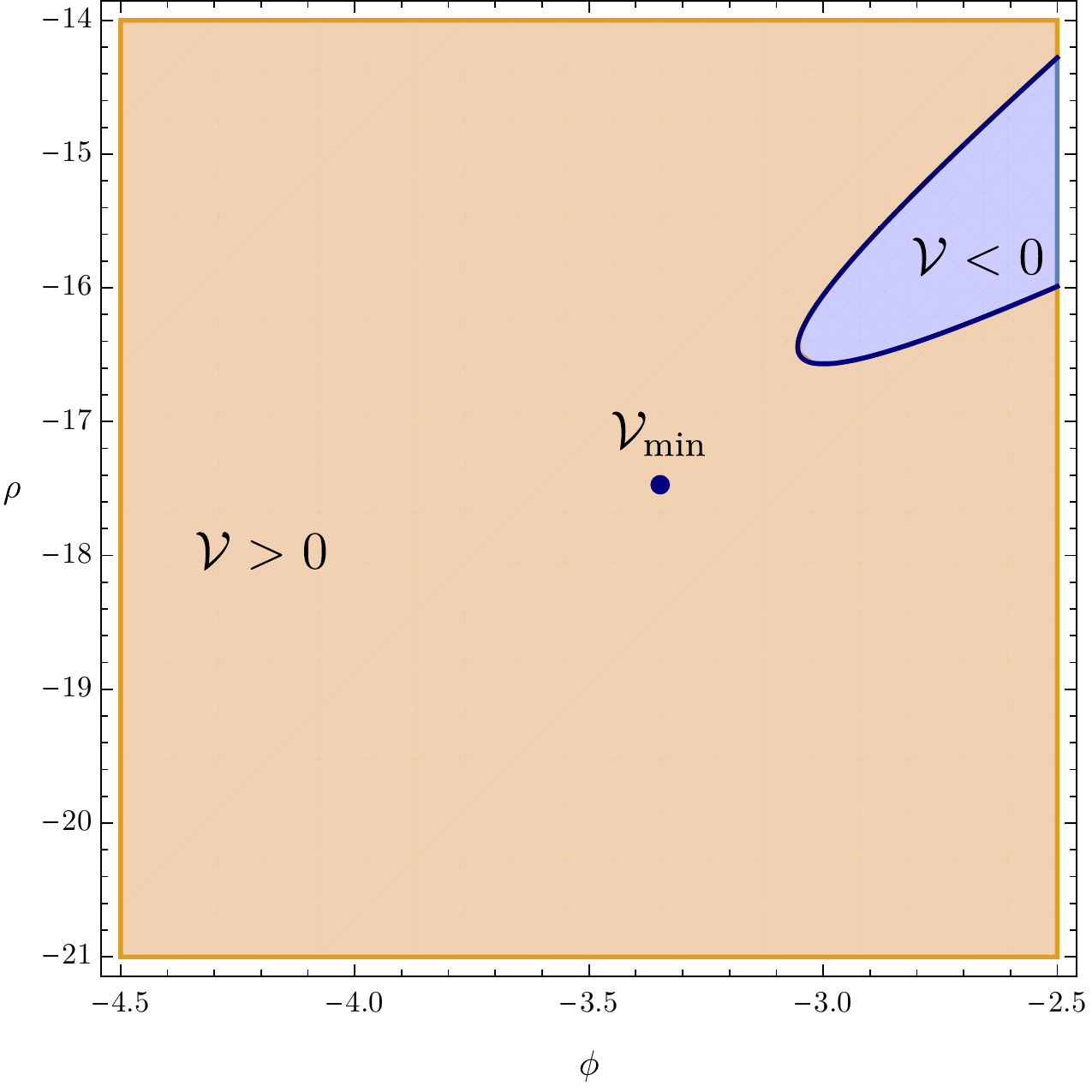}\hspace{0.5cm}
			\includegraphics[width=7.0cm]{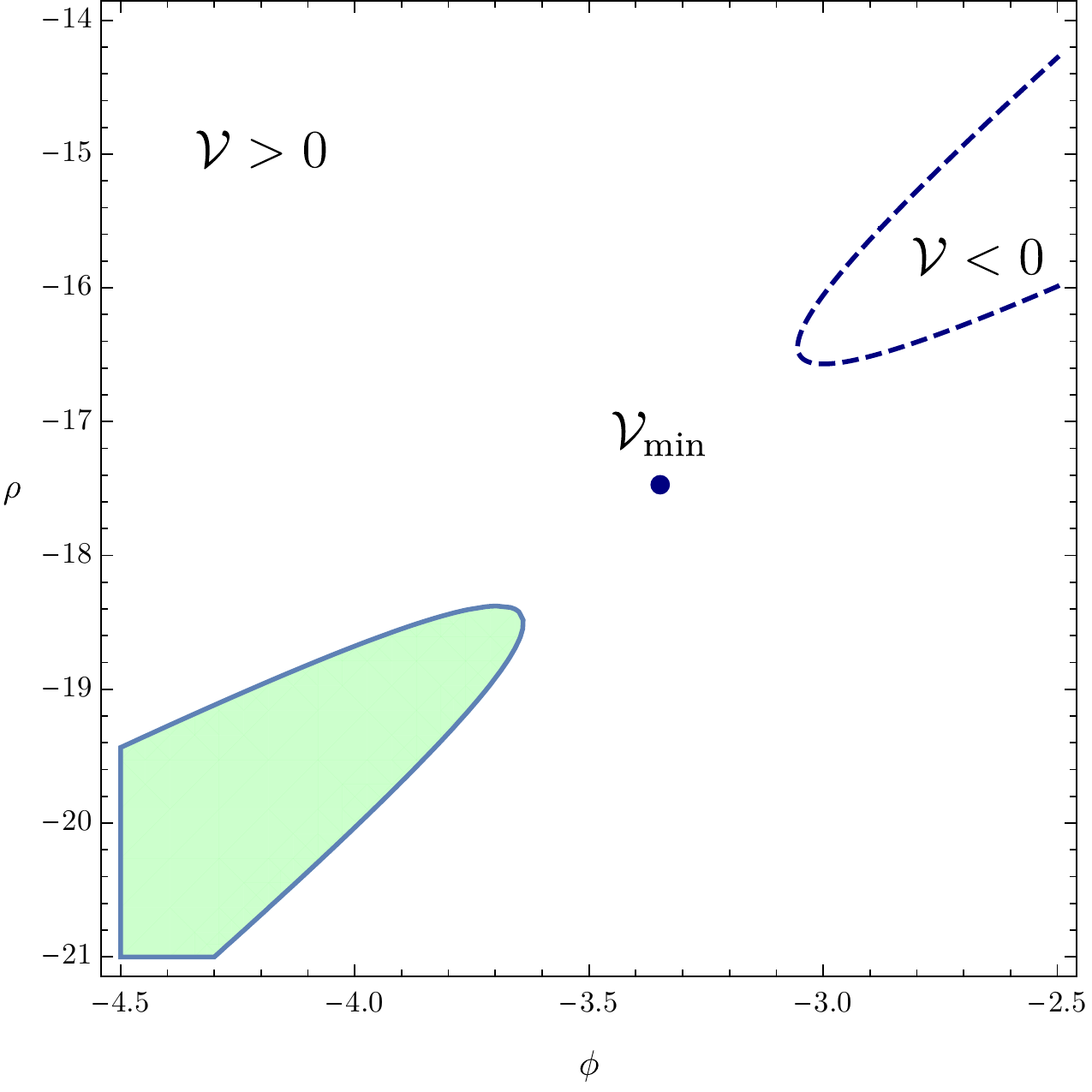}
		\caption{An example of four-dimensional dS vacuum, where we have highlighted of the sign of the potential and its region of stability. We have chosen $\alpha = -4$ and $\gamma=\frac12$.\label{Fig:V_dS}}
	\end{figure}
\end{center}

Although dS vacua are allowed in this case, it turns out that they are necessarily unstable, as in~\cite{Montero:2020rpl}. Indeed, at the extremum the $\partial^2_\phi \mathcal{V}$ and the determinant of the Hessian matrix take the form
\begin{equation}
	\begin{aligned}
	\partial^2_\phi \mathcal{V} |_{\rm ext} &= \frac12 \, e^{2{B}(d-1)\rho + \alpha\phi} \, \alpha \, n^2 \, (\alpha -\gamma)\,,
	\\
	\det \mathrm{Hess}(\mathcal{V}) |_{\rm ext} &= \frac{n^4 \, \alpha }{16} \, e^{4{B} \rho + 2\alpha \phi} \left[(\alpha - (d-1) \gamma)\left(p+1 + \frac{\alpha}{\gamma}\right)\right]\,,
	\end{aligned}
\end{equation}
and, for a dS extremum, eq.~\eqref{eq:bound_dSres} implies that ${\rm sgn}\, \alpha = -\, {\rm sgn}\,  \gamma$. Hence, whenever eq.~\eqref{eq:bound_dSres} holds, either $\partial^2_\phi \mathcal{V} |_{\rm ext}$ or the determinant of the Hessian matrix is negative. An example of perturbatively unstable dS solution is depicted in Fig.~\ref{Fig:V_dS}.

\section{Swampland conjectures and non-supersymmetric string theories}

In this section we frame our preceding considerations in the context of the swampland. Consider a class of lower-dimensional theories which couple gravity to some dynamical fields. Generically, it is expected that only a portion of them constitute the \emph{landscape} of theories originating from a higher dimensional theory, for which string theory ought to provide a UV completion. The remaining theories are said to belong to the \emph{swampland}, namely the set of EFTs which, although apparently consistent from a lower-dimensional perspective, cannot be completed by string theory in the UV. The aim of the Swampland program\footnote{For reviews, see~\cite{Brennan:2017rbf,Palti:2019pca}.} is to identify, within a bottom-up framework, criteria that separate the landscape from the swampland.

Concretely, let us again consider the class of $d$-dimensional theories described by an action of the form
\begin{equation}
\label{eq:swamp_redS}
\begin{aligned}
S = \int_X d^dx \,\sqrt{|\widehat g_X|} \left(\widehat R- \frac12 (\widehat\partial \rho)^2 -  \frac12 (\widehat\partial \phi)^2 - \mathcal{V}(\phi,\rho)\right) \, ,
\end{aligned}
\end{equation}
which couple gravity to two real scalar fields $\phi$ and $\rho$, subjected to the potential
\begin{equation}
\label{eq:swamp_redpot}
\begin{aligned}
\mathcal{V}(\phi,\rho) &= T \,e^{2 {B}\rho + \gamma \phi} + \frac12 \,n^2   \,e^{2 {B} \rho (d-1) + \alpha\phi}- r \, e^{2({B}-{A})\rho}\, .
\end{aligned}    
\end{equation}
In this section we regard these models from a bottom-up perspective, so that ${A}$, ${B}$, $\gamma$ and $\alpha$ are free parameters, but our preceding considerations imply that eq.~\eqref{eq:swamp_redS} can arise reducing the higher-dimensional effective action of eq.~\eqref{eq:action} on a $(D-d)$-dimensional manifold. In this context, $\phi$ and $\rho$ play a specific r\^ole: the VEV of $\phi$ defines the string coupling $g_{\rm s} = e^\phi$, while the radion $\rho$ is the universal volume modulus. Moreover, one can expect that eq.~\eqref{eq:swamp_redS} is endowed with a proper string theory origin only for some specific values of the parameters $\gamma$ and $\alpha$, for instance those in eqs.~\eqref{eq:bsb_magnetic_params} and~\eqref{eq:bsb_electric_params}. In other words, the parameters in eqs.~\eqref{eq:bsb_magnetic_params} and~\eqref{eq:bsb_electric_params} build a landscape of models that can be completed to non-supersymmetric string theories.

Here, however, we would like to pursue a different approach with respect to the preceding sections. In compliance with the Swampland program, we shall take eq.~\eqref{eq:swamp_redS} as a \emph{starting point}, momentarily foregoing its UV origin. We shall then investigate whether general features can distinguish consistent models exclusively on lower-dimensional grounds, testing to what extent the models characterized by the action in eq.~\eqref{eq:swamp_redS} satisfy the proposed Swampland conjectures. On the one hand, we shall discuss how our findings in the preceding sections resonate with the Swampland program and, on the other hand, we shall provide further non-trivial evidence for some recently proposed Swampland conjectures in non-supersymmetric settings\footnote{Recent efforts in this respect have addressed the Weak Gravity conjecture in Scherk-Schwarz compactifications~\cite{Bonnefoy:2018tcp, Bonnefoy:2020fwt}.}.

\subsection{The de Sitter conjecture and the Transplanckian Censorship conjecture}
\label{sec:tcc_dsc}

The de Sitter conjecture~\cite{Obied:2018sgi} excludes (perturbative) dS vacua in any EFT consistent with quantum gravity. Consider an EFT with some real scalar fields $\phi^i$, described by the action
\begin{equation}
\label{eq:swamp_no-dS_S}
	S = \int_X d^dx \,\sqrt{|g_X|} \left(R -\frac12 G_{ij} (\phi) \, \partial \phi^i \, \partial \phi^j - \mathcal{V}(\phi,\rho)\right) \, ,
\end{equation}
with $G_{ij}(\phi)$ the field-space metric and $\mathcal{V}(\phi)$ the scalar potential. The dS conjecture asserts that the slope of the potential is bounded from below according to
\begin{equation}
\label{eq:swamp_no-dS}
	|\nabla \mathcal{V}| \geq c\, \mathcal{V} \, ,
\end{equation}
where $|\nabla \mathcal{V}| \equiv \sqrt{G^{ij}(\phi) \, \partial_i \mathcal{V} \, \partial_j \mathcal{V}}$, $\partial_i \mathcal{V} \equiv \frac{\partial}{\partial \phi^i}\mathcal{V}$ and $G^{ij}$ denotes the inverse of $G_{ij}$. In the original incarnantion of the de Sitter conjecture~\cite{Obied:2018sgi}, $c$ was left as an unspecified $\mathcal{O}(1)$ parameter. Since its formulation, the de Sitter conjecture has been subjected to further refinements, most notably~\cite{Garg:2018reu,Ooguri:2018wrx} (see also~\cite{Dvali:2018fqu,Andriot:2018wzk}). In particular, in four-dimensional Calabi-Yau compactifications of string theory or M-theory, a no-go theorem was proposed in~\cite{Grimm:2019ixq} to the effect that, asymptotically in field space, there is \emph{no dS critical point} near any two-moduli parametrically controlled limit.

Although originally the parameter $c$ entering eq.~\eqref{eq:swamp_no-dS} was not specified, a proper estimate of its value is of utmost importance for inflationary scenarios~\cite{Obied:2018sgi,Bedroya:2019snp,Bedroya:2019tba}\footnote{Single-field inflation appears to provide observational constraints on $c$. See, for instance,~\cite{Kinney:2018nny}.}. The issue of determining the constant $c$ was later addressed by another conjecture: the Transplanckian Censorship conjecture (TCC)~\cite{Bedroya:2019snp} asserts that, in a $d$-dimensional theory consistent with quantum gravity, \emph{asymptotically} in field space, the slope of the potential for the scalar field is bounded from below according to
\begin{equation}
\label{eq:swamp_TCC}
    \frac{|\nabla \mathcal{V}|}{\mathcal{V}} \Bigg|_{\rm asymp} \geq \frac{2}{\sqrt{(d-1)(d-2)}}\,.
\end{equation}
Clearly, this conjecture is less powerful than the original de Sitter conjecture~\cite{Obied:2018sgi} since it holds only in the asymptotic regions of the field space, where the theory is expected to be weakly coupled and thus more reliable. Nevertheless, in contrast to eq.~\eqref{eq:swamp_no-dS}, eq.~\eqref{eq:swamp_TCC} yields a concrete lower bound on the slope of the potential and, thus, on the parameter $c$. In the following we shall investigate to what extent the de Sitter conjecture and the TCC are satisfied by the non-supersymmetric string models of Section~\ref{sec:non-susy_strings}, for which we shall provide explicit lower bounds for the parameter $c$, relating them with the predictions of the TCC.

As a preliminary check, it is straightforward to see that the ten-dimensional models specified by eq.~\eqref{eq:action} satisfy the de Sitter conjecture: indeed, the ten-dimensional scalar potential $\mathcal{V} (\phi) = T \, e^{\gamma \phi}$ depends solely on the dilaton, and thus
\begin{equation}
    \frac{|\nabla \mathcal{V}|}{\mathcal{V}} = \frac{|\partial_\phi V |}{V} = |\gamma|\,.
\end{equation}
In this case one can therefore identify the parameter $c$ in eq.~\eqref{eq:swamp_no-dS} with $|\gamma|$. For generic models, \textit{i.e.} for arbitrary values of $\gamma$, the TCC bound of eq.~\eqref{eq:swamp_TCC} is not necessarily satisfied, and in particular any ten-dimensional exponential potential which ought to be consistent with the TCC is to satisfy
\begin{equation}
\label{eq:swamp_10Dgamma}
    |\gamma | \geq \frac{1}{3\sqrt{2}}\,.
\end{equation}
Clearly, the orientifold models and the heterotic model of Section~\ref{sec:non-susy_strings}, specified by the parameters in eq.~\eqref{eq:runaway_potential_einstein_frame} and eq.~\eqref{eq:runaway_potential_het} respectively, satisfy the inequality in eq.~\eqref{eq:swamp_10Dgamma}.

However, in dimensions $d<10$, due to additional contributions to the scalar potential it is less trivial to show to what extent the dS conjecture and the TCC bound are satisfied. To this end, in order to obtain a lower bound on the parameter $c$ in eq.~\eqref{eq:swamp_no-dS}, we shall proceed as in~\cite{Grimm:2019ixq}. Let us assume that there exists a positive constant $\widetilde c$ such that, given an $N$-dimensional vector $u_i$
\begin{equation}
	\label{eq:swamp_diseq0}
	 {\widetilde c}^{-2} \geq u_i \, G^{ij} \, u_j \,.
\end{equation}
Then, the inequality
\begin{equation}
\label{eq:swamp_diseq}
|\nabla \mathcal{V}| \geq  {\widetilde c}\,  |\nabla \mathcal{V}| \left(u_i \, G^{ij} \, u_j\right)^{\frac12}
\end{equation}
holds, and applying the Cauchy-Schwarz inequality one is led to
\begin{equation}
\label{eq:swamp_CS}
|\nabla \mathcal{V}| \geq  {\widetilde c}\,  |\nabla \mathcal{V}| \left(u_i G^{ij} u_j \right)^{\frac12} \geq {\widetilde c} \, u_i \, G^{ij} \, \partial_j \mathcal{V}\,.
\end{equation}
Eq.~\eqref{eq:swamp_CS} provides a lower bound for $|\nabla \mathcal{V}|$, but it is not yet in the form of eq.~\eqref{eq:swamp_no-dS} as required by the de Sitter conjecture. For the moment, let us assume that eq.~\eqref{eq:no-go} holds, which is indeed the case for the string models of Section~\ref{sec:non-susy_strings}. Recalling that the potential in eq.~\eqref{eq:swamp_redpot} can be recast in the form of eq.~\eqref{eq:bound_redpotb}, one finds the inequality
\begin{equation}
\label{eq:swamp_ineqV_a}
\mathcal{V} - \frac{{A}}{\gamma ({A}-{B})}\, \partial_{\phi} \mathcal{V} - \frac{1}{2({B} - {A})}\,\partial_{\rho} \mathcal{V}  \leq 0 \, ,
\end{equation}
or alternatively
\begin{equation}
\label{eq:swamp_ineqV_b}
\mathcal{V} -\frac{{A} + {B} (d-2)}{\alpha ({B}-{A})} \,\partial_{\phi} \mathcal{V} - \frac{1}{2({B} - {A})}\,\partial_{\rho} \mathcal{V}  \leq 0 \, ,
\end{equation}
using the first and the second line of eq.~\eqref{eq:bound_redpotb} respectively\footnote{Inequalities similar to eqs.~\eqref{eq:swamp_ineqV_a} and~\eqref{eq:swamp_ineqV_b} are commonly satisfied by potentials stemming from supersymmetric compactifications string theory or M-theory. This feature was originally employed in~\cite{Hertzberg:2007wc} to derive a no-go theorem that excludes $\ds$ solutions in type IIA compactifications from a lower-dimensional perspective. See also~\cite{Andriot:2018ept,Andriot:2018wzk,Andriot:2019wrs} for analogous results in more general settings.
}. Let us first consider the relation in eq.~\eqref{eq:swamp_ineqV_a}. Choosing
\begin{equation}
\label{eq:swamp_ineqV_u1}
	u_\phi = \frac{{A}}{\gamma ({A}-{B})} \,,\qquad u_\rho = \frac{1}{2({B} - {A})}\,,
\end{equation}
one obtains the following chain of inequalities:
\begin{equation}
\label{eq:swamp_CSa}
|\nabla \mathcal{V}| \geq  {\widetilde c}\,  |\nabla \mathcal{V}| \left(u_i \, G^{ij} \,u_j \right)^{\frac12} \geq {\widetilde c} \, u_i \, G^{ij} \, \partial_j \mathcal{V} \geq {\widetilde c} \, \mathcal{V}\,.
\end{equation}
Hence, since in our case the field-space metric is constant, the maximal $c$ that delivers the bound in eq.~\eqref{eq:swamp_no-dS} is
\begin{equation}
c^{(1)} \equiv  {\rm sup}\, \widetilde c = \frac{1}{\sqrt{u_i \, G^{ij} \, u_j}} = \frac{2 |\gamma| ({B} -{A})}{\sqrt{4 {A}^2 + \gamma^2}} \,.
\end{equation}
On the other hand, starting instead from eq.~\eqref{eq:swamp_ineqV_b} and proceeding as above, one would arrive at
\begin{equation}
c^{(2)} = \frac{2 |\alpha | ({B}- {A})}{\sqrt{4 ({A} + (d-2) {B})^2+ \alpha^2}} \, .
\end{equation}
Therefore we may conclude that, generically, the parameter $c$ appearing in eq.~\eqref{eq:swamp_no-dS} is given by
\begin{equation}
\label{eq:swamp_cbound}
c = {\rm max} \{c^{(1)}, c^{(2)} \} \, .
\end{equation}
Let us stress that this estimate for $c$ is quite general, and relies only on the assumption that eq.~\eqref{eq:no-go} holds. However, it is important to recognize that eq.~\eqref{eq:swamp_cbound} typically delivers only a lower bound on the possible values of $c$ such that eq.~\eqref{eq:swamp_no-dS} is satisfied. 
\begin{center}
	\begin{figure}[ht]
		\centering
			\includegraphics[width=6.0cm]{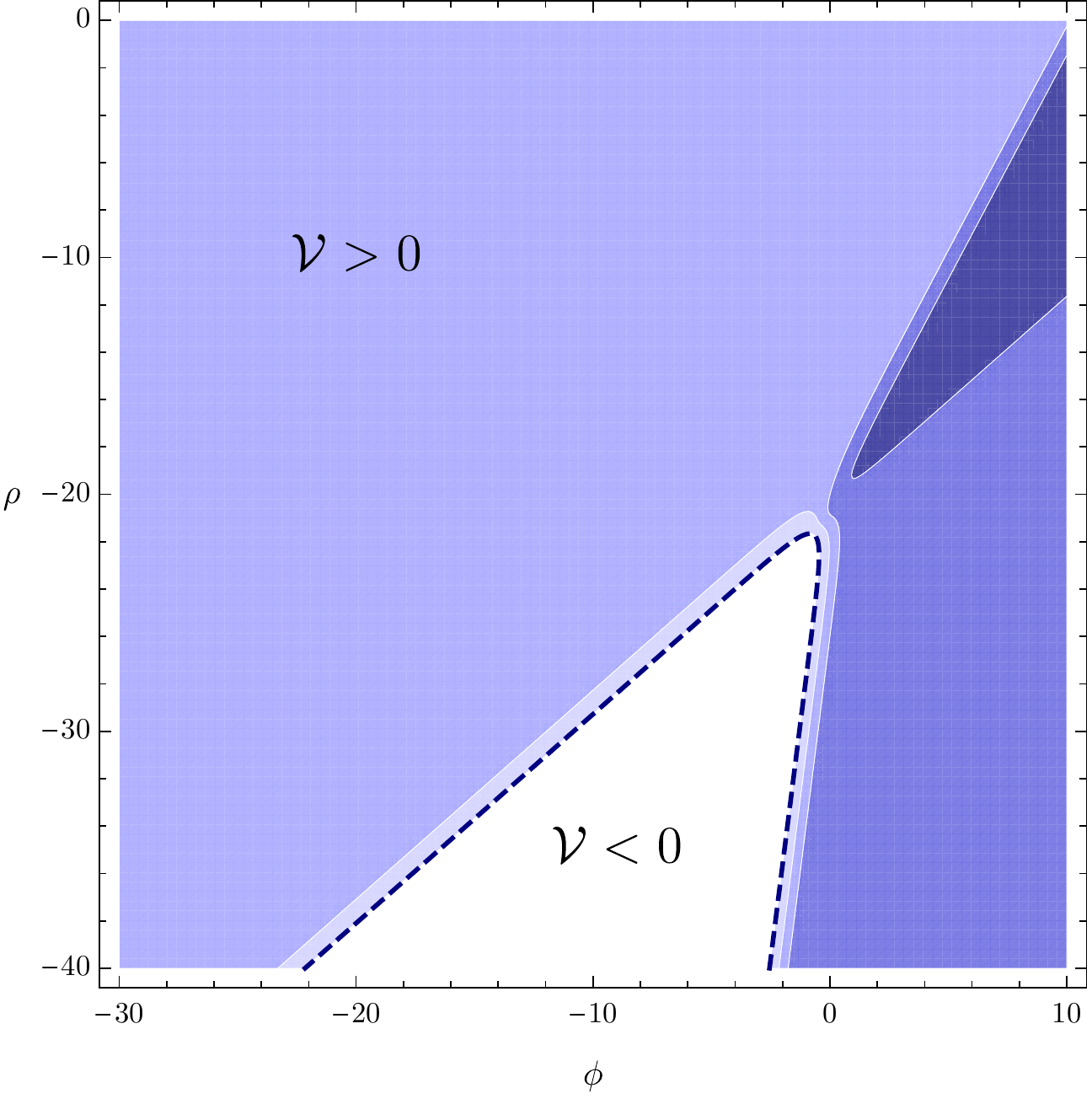}\hspace{0.5cm}
			\includegraphics[width=8.0cm]{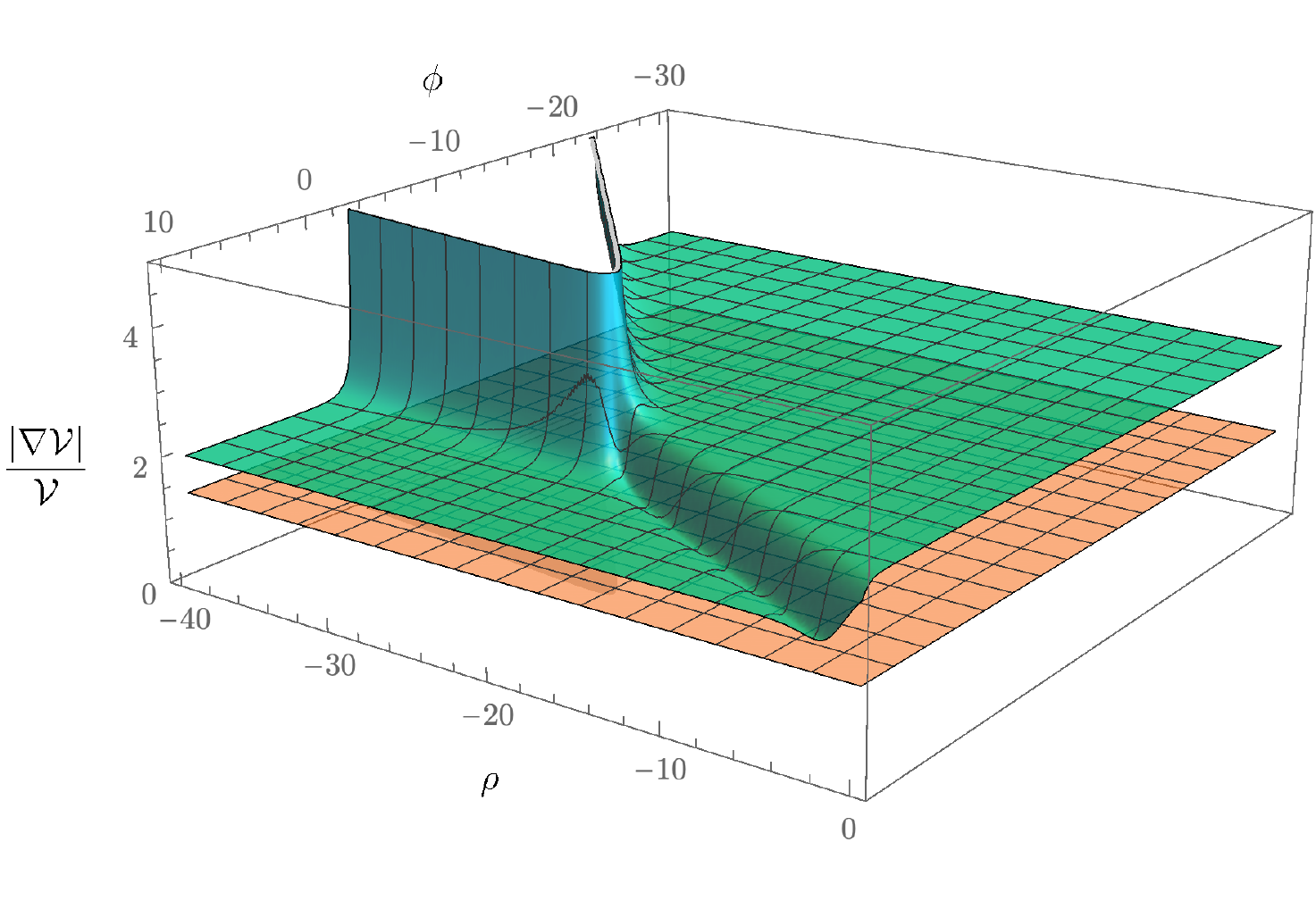}
		\caption{On the left, in blue, the four regions of field space where $1.5 \leq \frac{|\nabla \mathcal{V}|}{\mathcal{V}} < 2$, $2 \leq \frac{|\nabla \mathcal{V}|}{\mathcal{V}} < 2.5$, $2.5 \leq \frac{|\nabla \mathcal{V}|}{\mathcal{V}} < 3.5$ and  $\frac{|\nabla \mathcal{V}|}{\mathcal{V}} \geq 3.5$ are depicted for the BSB model and the type $0'$B model, described by the parameters in eq.~\eqref{eq:bsb_magnetic_params}. Lighter colors correspond to greater values of $\frac{|\nabla \mathcal{V}|}{\mathcal{V}}$. On the right, a plot of $\frac{|\nabla \mathcal{V}|}{\mathcal{V}}$ in the region where the potential is positive definite, compared to the constant value predicted by the TCC in eq.~\eqref{eq:swamp_TCC} (orange). The parameters take the same values as in Fig.~\ref{Fig:V_bsb}. \label{Fig:R_bsb}}
	\end{figure}
\end{center}

For concreteness, let consider the orientifold models specified by the parameters in eq.~\eqref{eq:bsb_magnetic_params}. One finds $c^{(1)} =1.5$, $c^{(2)}= 1$, obtaining $c = 1.5$. However, a numerical computation leads to the stronger estimation for $c$
\begin{equation}
    c_{\rm orientifold} \gtrsim 1.871 \,.
\end{equation}
Remarkably, as depicted in Fig.~\ref{Fig:R_bsb}, this holds within the whole $(\phi,\rho)$-plane, including the regions where the lower-dimensional description of eq.~\eqref{eq:swamp_redS} is not expected to be reliable. Thus, also the TCC in eq.~\eqref{eq:swamp_TCC} is realized. This proves that \emph{non-supersymmetric three-dimensional compactifications originating from the BSB model and the type $0'$B model are consistent with both eq.~\eqref{eq:swamp_no-dS} and eq.~\eqref{eq:swamp_TCC}.} The procedure that we have outlined yields similar bounds for warped compactifications in general dimensions.

\begin{center}
	\begin{figure}[ht]
		\centering
			\includegraphics[width=6.0cm]{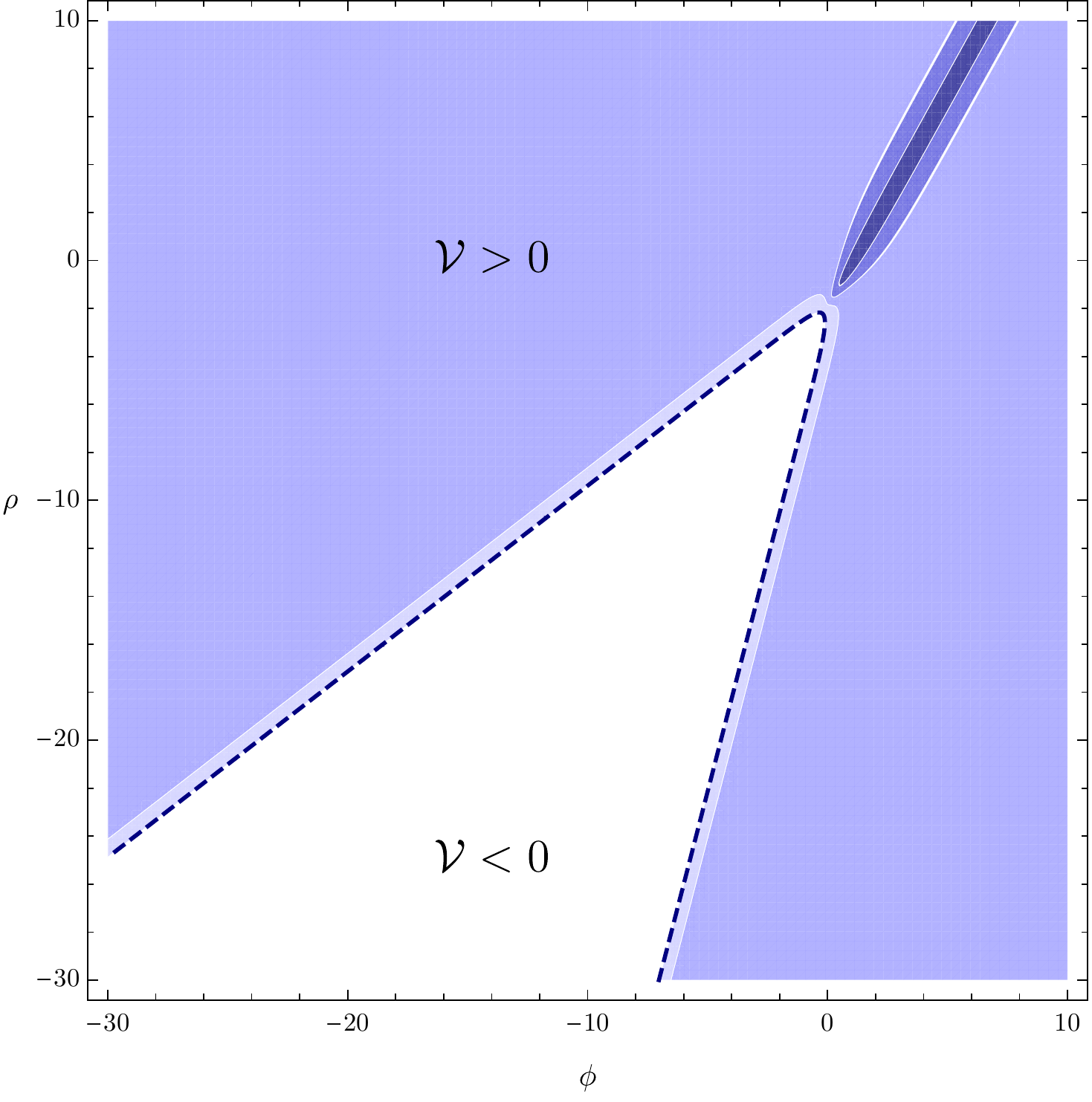}\hspace{0.5cm}
			\includegraphics[width=8.0cm]{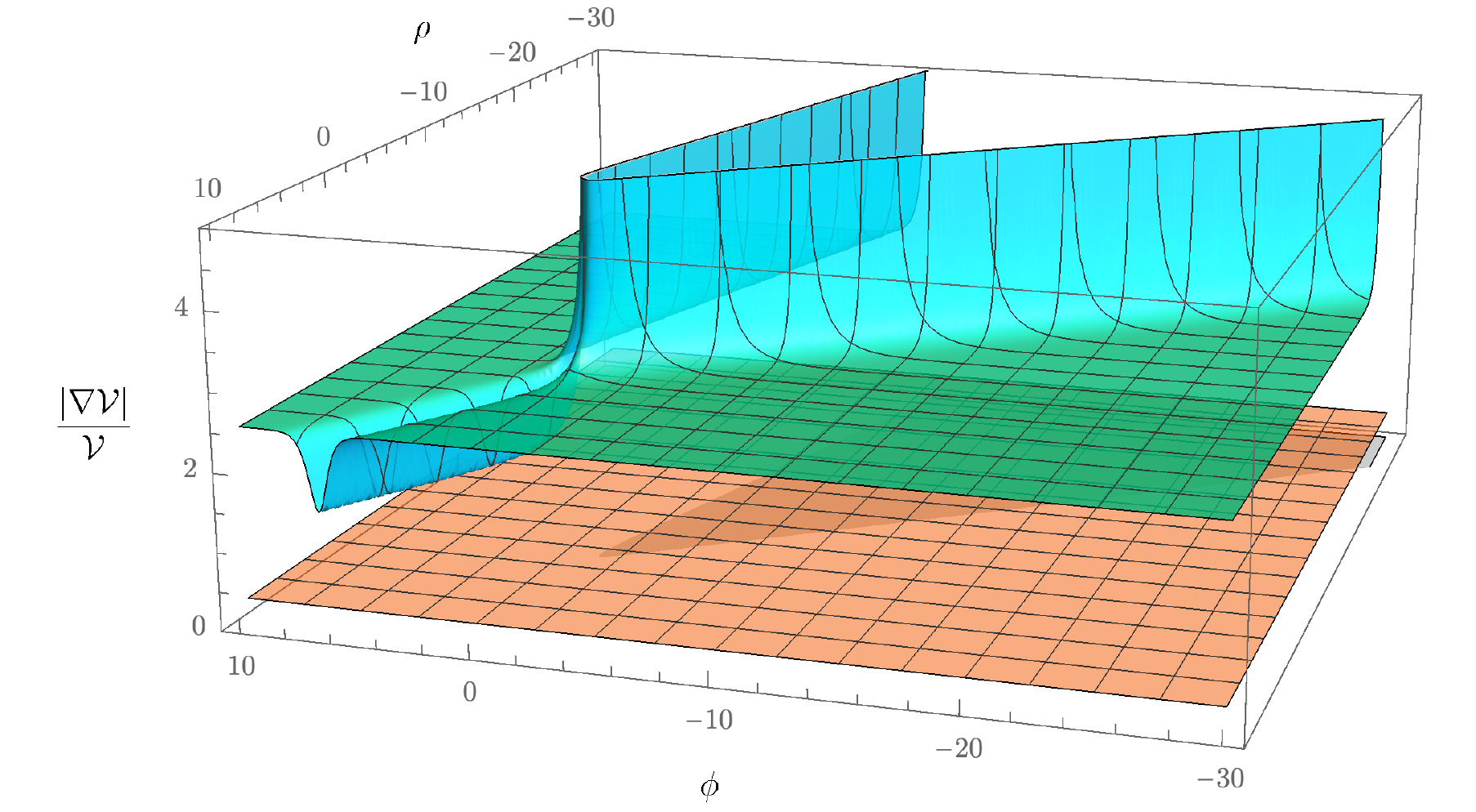}
		\caption{On the left, in blue, the four regions of field space where $1.5 \leq \frac{|\nabla \mathcal{V}|}{\mathcal{V}} < 2$, $2 \leq \frac{|\nabla \mathcal{V}|}{\mathcal{V}} < 2.5$, $2.5 \leq \frac{|\nabla \mathcal{V}|}{\mathcal{V}} < 3.5$ and  $\frac{|\nabla \mathcal{V}|}{\mathcal{V}} \geq 3.5$ are depicted for the heterotic model, specified by the parameters in eq.~\eqref{eq:het_magnetic_params}. Lighter colors correspond to greater values of $\frac{|\nabla \mathcal{V}|}{\mathcal{V}}$. On the right, a plot of $\frac{|\nabla \mathcal{V}|}{\mathcal{V}}$ in the region where the potential is positive definite, compared to the constant value predicted by the TCC in eq.~\eqref{eq:swamp_TCC} (orange). The parameters take the same values as in Fig.~\ref{Fig:V_het}. \label{Fig:R_het}}
	\end{figure}
\end{center}

Also the heterotic model, with the scalar potential as in eq.~\eqref{eq:bound_het_redpot}, satisfies the inequalities. In this case, $c^{(1)} =1$ and $c^{(2)} \simeq 0.632$, from which one would conclude that $c = 1$. However, this is a weak estimate: as depicted in Fig.~\ref{Fig:R_bsb}, the ratio $|\nabla \mathcal{V}|/\mathcal{V}$ can be numerically bounded from below by
\begin{equation}
    c_{\rm heterotic} \gtrsim 1.549 \,.
\end{equation}
Hence, \emph{seven-dimensional non-supersymmetric compactifications originating from the $SO(16)\times SO(16)$ heterotic model are consistent with both eq.~\eqref{eq:swamp_no-dS} and eq.~\eqref{eq:swamp_TCC}.} Once more, the procedure that we have outlined yields similar bounds for warped compactifications in general dimensions.

\subsection{The de Sitter conjecture and the Weak Gravity conjecture for membranes}

As put forward in~\cite{Lanza:2020qmt}, extended objects can be useful to study Swampland conjectures, and their properties can facilitate the development of a \emph{web} among the proposed conjectures. In this regard membranes, namely objects of codimension one, are helpful to constrain the effective potential.

In order to apply this idea to the present context, let us momentarily assume that the potential in the $d$-dimensional theory arises solely from the flux contribution of eq.~\eqref{eq:bound_redpotflux}, $\mathcal{V}(\phi,\rho)  = \mathcal{V}_{\rm flux}$. Crucially, the background flux $n$ can be regarded as \emph{dual} to a $(d-1)$-form field $B_{d-1}$. In fact, in $d$-dimensions $(d-1)$-form fields carry no propagating degrees of freedom, and thus can be effortlessly integrated out. This procedure generates a potential that is characterized by a constant~\cite{Brown:1987dd,Brown:1988kg,Bousso:2000xa,Lanza:2019nfa}.
On the other hand, membranes are the objects which electrically couple to $(d-1)$-forms, and thus source the corresponding fluxes. Indeed, as we shall now discuss in detail, $\mathcal{V}_{\rm flux}$ can be entirely generated by a single membrane, whose charge corresponds to the flux parameter $n$ of the background. To this end, let us consider the action
\begin{equation}
\begin{aligned}
\label{eq:swamp_mem_S}
	S &= \int_X d^d x\, \sqrt{-\widehat g_X} \left(R -\frac12 (\partial \phi)^2  -\frac12 (\partial \rho)^2 - \frac1{2\, d!}  F H_d^2  \right)
	\\
	&\quad\, +  \frac{1}{(d-1)!} \int_X d^dx\,  \partial_\mu \left(\sqrt{-\widehat g_X}  e^{-\alpha \phi - 2{B} (d-1) \rho}  H^{\mu \mu_2 \ldots \mu_d} B_{\mu_2 \ldots \mu_d} \right) 
	\\
	&\quad\, -\int_{\mathcal{M}} d^{d-1}\xi \, \sqrt{-h}\, \mathcal{T} + \nu \int_{\mathcal{M}} B_{d-1} \, .
\end{aligned}
\end{equation}
In the first line, aside from the contributions from gravity and the kinetic terms for the scalar fields, we have included the kinetic term of a $(d-1)$-form field, with $H_d = d B_{d-1}$. For instance, in the orientifold models the form field arises from the R-R sector, while for the heterotic model the form field is the magnetic dual of the Kalb-Ramond field $B_2$. Furthermore, we have introduced the coupling function
\begin{equation}
\label{eq:swamp_mem_F}
	F(\phi,\rho) =  e^{-\alpha \phi - 2{B} (d-1) \rho} \,.
\end{equation}
The second line of eq.~\eqref{eq:swamp_mem_S} contains a boundary term. It is necessary in order to formulate a well-posed variational problem, which requires unconstrained variations $\delta B_{d-1}$ of the $(d-1)$-form gauge field on the boundary. Finally, in the last line of eq.~\eqref{eq:swamp_mem_S} we have included the contribution of a fundamental membrane, spanning the world-volume $\mathcal{M}$ which we assume to be defined by $x^d =0$. The tension $\mathcal{T}$ of the membrane may depend on both $\phi$ and $\rho$; finally, and the last term expresses the electric coupling of the membrane to $B_{d-1}$. The charge $\nu$ corresponds to the background flux parameter $n$, and they coincide in units of a suitable fundamental charge.

\begin{center}
	\begin{figure}[t]
		\centering
		\includegraphics[width=7cm]{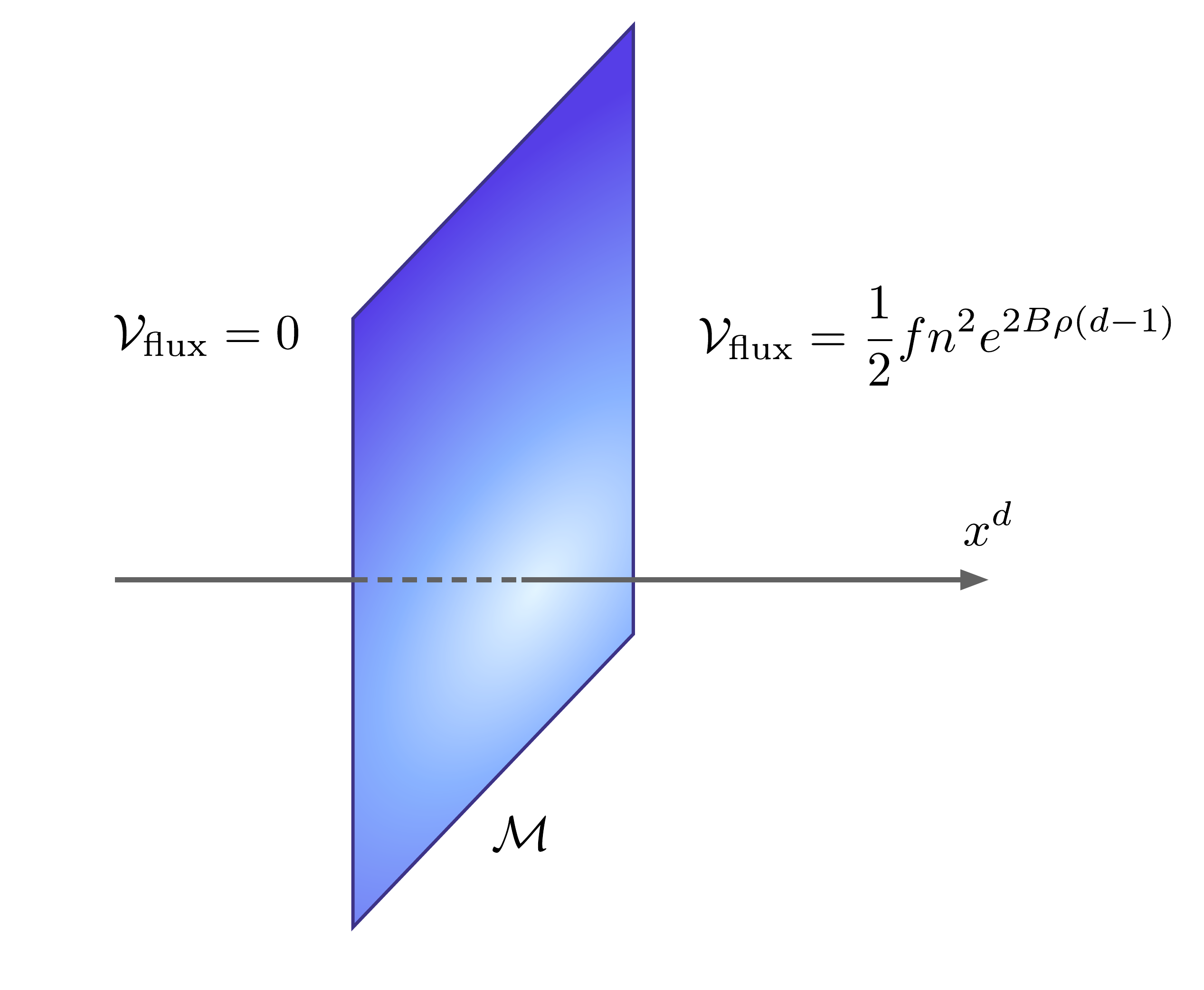}
		\caption{A membrane that interpolates between a configuration with null potential and one with potential $\mathcal{V} = \mathcal{V}_{\rm flux}$.\label{Fig:MemGen}}
	\end{figure}
\end{center}

We can now integrate out the $(d-1)$-form according to
\begin{equation}
	H^{\mu_1 \ldots \mu_d} = - \frac{C + \nu \, \Theta (x^d)}{\sqrt{-\widehat g_X}} \,  e^{\alpha \phi + 2{B} (d-1) \rho}  \, \varepsilon^{\mu_1 \ldots \mu_d} \, ,
\end{equation}
with $C$ an arbitrary real constant, and in the ensuing discussion, we shall take $C=0$. The action \eqref{eq:swamp_mem_S} then evaluates to
\begin{equation}
\begin{aligned}
\label{eq:swamp_mem_Sos}
S &= \int_X d^d x\, \sqrt{-\widehat g_X} \left(R -\frac12 (\partial \phi)^2  -\frac12 (\partial \rho)^2 - \mathcal{V}_{\rm gen}  \right) -\int_{\mathcal{M}} d^{d-1}\xi\, \sqrt{-h}\, \mathcal{T} \, ,
\end{aligned}
\end{equation}
where the potential generated by the membrane is
\begin{equation}
\label{eq:swamp_mem_Vgen0}
	\mathcal{V}_{\rm gen} = \Theta(x^d)\, \mathcal{V}_{\rm flux}\,.
\end{equation}
In other words, as depicted in Fig.~\ref{Fig:MemGen}, the membrane generates a potential for the scalar fields, and delimits a region where it is zero from one where it coincides with the flux-induced contribution of eq.~\eqref{eq:bound_redpotflux}. While the bare charge appears directly in the action in eq.~\eqref{eq:swamp_mem_S}, the \emph{physical} charge $\mathcal{Q}$ of the membrane can be most readily identified from the corresponding potential according to
\begin{equation}
\label{eq:swamp_mem_Vgen}
	\mathcal{V}_{\rm gen} = \Theta(x^d)\, \frac12 \, \mathcal{Q}^2\,.
\end{equation}

The scalar version of the membrane Weak Gravity conjecture (WGC) then predicts that there must exist at least a membrane satisfying
\begin{equation}
\label{eq:swamp_mem_WGC}
\frac{\mathcal{Q}}{\mathcal{T}} \geq \chi
\end{equation}
where $\chi$ is the extremality parameter. It is determined in terms of the gauge coupling $F$ according to
\begin{equation}
	\chi^2 = \frac14 \frac{G^{ij} \partial_i F \partial_j F}{F^2} - \frac{d-1}{d-2}
\end{equation}
and, for the cases at hand, it reads
\begin{equation}
\chi^2 = \frac14 \,\alpha^2 - \, \frac{(d-1)(d^2 - 11 d + 26)}{16(d-2)}\,.
\end{equation}
It is worthwhile mentioning that membranes obeying eq.~\eqref{eq:swamp_mem_WGC} also satisfy the Repulsive Force conjecture~\cite{Heidenreich:2019zkl,Herraez:2020tih,Lanza:2020qmt}: two identical membranes, with the same physical charge $\mathcal{Q}$ and same tension $\mathcal{T}$, are mutually repulsive whenever eq.~\eqref{eq:swamp_mem_WGC} holds, provided that they are sufficiently close to one another. From this viewpoint, saturating eq.~\eqref{eq:swamp_mem_WGC} translates into a balance of forces. 

For the heteretotic model $\chi_{\rm heterotic}^2 = \frac25$, while for the orientifold models $\chi_{\rm orientifold} = 0$. In the latter case, this implies that, whenever $\mathcal{Q}>0$, any membrane is self-repulsive and obeys the scalar WGC. In the former case, one can consider an extremal membrane whose tension is fixed by
\begin{equation}
\label{eq:swamp_ex_mem}
	\mathcal{Q} = \chi \mathcal{T}_{\rm ext}\,,
\end{equation}
with $\mathcal{Q}$ as in eq.~\eqref{eq:swamp_mem_Vgen}.
The tension of such a membrane, analogously to its supersymmetric counterparts~\cite{Bandos:2018gjp,Lanza:2020qmt}, has exactly the same field dependence as the potential in eq.~\eqref{eq:swamp_mem_Vgen}. Remarkably, in the region $x^d>0$, the potential generated by these extremal membranes, given by eq.~\eqref{eq:swamp_mem_Vgen}, satisfies the dS conjecture, since
\begin{equation}
\label{eq:swamp_mem_dS}
	\frac{|\nabla \mathcal{V}_{\rm flux}|}{\mathcal{V}_{\rm flux}} \geq {\rm min} \{ \alpha, 2{B}(d-1)\}\,.
\end{equation}
Furthermore, regardless of $\alpha$, the TCC in eq.~\eqref{eq:swamp_TCC} is also identically satisfied.

In addition, one can consider more general membranes which obey the strict WGC inequality in eq.~\eqref{eq:swamp_mem_WGC}. For instance, assuming still that the charge of the membrane is proportional to its tension,
\begin{equation}
\label{eq:swamp_non-ex_mem}
\mathcal{Q} = \varepsilon \, \chi \, \mathcal{T}_{\rm ext}\,,
\end{equation}
where the constant parameter $\varepsilon>1$. Such a membrane generates a potential of the type
\begin{equation}
\label{eq:swamp_non-ex_V}
\mathcal{V}_{\rm gen} = \Theta(x^d)\, \frac 12 \, \mathcal{Q}^2 = \varepsilon \, \mathcal{V}_{\rm flux} > \mathcal{V}_{\rm flux} 
\end{equation}
in the region $x^d>0$. Thus, interestingly, also non-extremal membranes of this type satisfy the de Sitter conjecture of eq.~\eqref{eq:swamp_mem_dS}.

However, in the preceding sections we have shown that the potential arising from non-supersymmetric string models is more general, as highlighted in eq.~\eqref{eq:bound_redpot}. In particular, in the action of eq.~\eqref{eq:swamp_mem_S} one ought to include the additional `spectator' potential
\begin{equation}
\label{eq:swamp_mem_hatV}
	\widehat{\mathcal{V}} = T \, e^{2 {B}\rho + \gamma \phi} - r \, e^{2({B}-{A})\rho}\,.
\end{equation}
Placing the $(d-1)$-form field on shell, the potential evaluates to
\begin{equation}
\label{eq:swamp_mem_Vgenb}
\mathcal{V} =  \mathcal{V}_{\rm gen} + \widehat{\mathcal{V}} \,.
\end{equation}
We can now inquire how the properties of the membrane, which generates the flux-induced contribution, affect the de Sitter conjecture. To this end, let us consider a generic membrane whose tension and charge are related according to eq.~\eqref{eq:swamp_non-ex_mem}, so that $\varepsilon \to 1$ corresponds to the extremal limit. Let us observe that the spectator potential can be recast in the form
\begin{equation}
	\widehat{\mathcal{V}} = u_\rho \partial_\rho \widehat{\mathcal{V}} +  u_\phi \partial_\phi \widehat{\mathcal{V}} \, ,
\end{equation}
with the same choice of $u_\phi$, $u_\rho$ of eq.~\eqref{eq:swamp_ineqV_u1}. Then, proceeding as in the preceding section, one arrives at
\begin{equation}
\label{eq:swamp_mem_CS}
|\nabla \mathcal{V}| \geq {\widetilde c}  \left(u_i \,G^{ij} \, \partial_j \mathcal{V}_{\rm gen} + \widehat{\mathcal{V}} \right)\,,
\end{equation}
valid in the region $x^d > 0$, with $\widetilde c$ chosen as in eq.~\eqref{eq:swamp_diseq0}. In conclusion, the above inequality leads to the de Sitter conjecture of eq.~\eqref{eq:swamp_no-dS} whenever 
\begin{equation}
\label{eq:swamp_mem_eps}
\varepsilon \geq \frac{1}{\alpha u_\phi + 2\beta (d -1) u_\rho } \, .
\end{equation}
For the heterotic model this would imply that $\varepsilon > \frac12$, and therefore also super-extremal membranes in the sense of eq.~\eqref{eq:swamp_non-ex_mem}, with $\varepsilon > 1$, would satisfy the de Sitter conjecture. On the other hand, sub-extremal membranes with $\varepsilon < 1$ might in principle violate it.

\subsection{The distance conjecture and the tower of states}

In any EFT, it is not expected that the (classical) moduli space can be explored completely. Indeed, in some corners of the moduli space the effective description is driven away from its regime of validity, since, for instance, quantum corrections are expected to be relevant. The distance conjecture~\cite{Ooguri:2006in} expresses such an obstacle. It states that, at certain points an (geodesic) infinite distance ${\rm d}$ away in field space, an infinite towers of state becomes massless according to
\begin{equation}
	m \sim e^{-\lambda {\rm d}}
\end{equation}
for some $\mathcal{O}(1)$ constant parameter $\lambda$. Thus, testing the conjecture requires firstly understanding infinite-distance loci in the moduli space, how to fields can approach them and, secondly, to identify the tower of states that become massless. While this conjecture has been thoroughly tested in supersymmetric settings~\cite{Grimm:2018ohb,Corvilain:2018lgw,Grimm:2018cpv}, as we shall now discuss it is expected to hold also in non-supersymmetric models.

To begin with, let us assume that the dilaton is fixed to a given value $\phi_0$ such that $g_{\rm s} = e^{\phi_0} \ll 1$. Then, $\rho \to -\infty$ is an infinite-distance limit, corresponding to a large internal volume. A natural candidate for the tower of states becoming massless in such a limit is thus Kaluza-Klein states. These arise from fluctuations of the dilaton, the graviton and the two-form around the background, and for $\ads$ solutions they were investigated in detail in~\cite{Basile:2018irz}. In the Einstein frame, and in terms of the $d$-dimensional Planck mass, their masses scale schematically according to
\begin{equation}
	m^2_{\rm KK} \sim \frac{M^2_{{\rm Pl},d}}{R^2 V} \sim \frac{M^2_{{\rm Pl},d}}{R^{2+d}}\,,
\end{equation}
where, in the last step, we have made assumed that the internal volume $V \sim R^d$. In the three-dimensional orientifold model, this would then lead to
\begin{equation}
m^2_{\rm KK,\, orientifold} \sim M^2_{{\rm Pl},3} e^{\frac{5}{4 \sqrt{7}} \rho} \, ,
\end{equation}
while, in the seven-dimensional heterotic models \, ,
\begin{equation}
m^2_{\rm KK,\, heterotic} \sim M^2_{{\rm Pl},7} e^{\frac{3 \sqrt{15}}{4} \rho} \, .
\end{equation}
Some crucial comments are now in order. As shown in~\cite{Basile:2018irz}, the inclusion of Kaluza-Klein modes in non-supersymmetric models can lead to perturbative instabilities. However, such instabilities are caused only by a \emph{finite} number of Kaluza-Klein modes. Thus, if the instabilities cannot be not removed, the dynamics drives the theory away from the original background, and to inquire whether an infinite tower of massless states in some corners of the moduli space becomes futile. On the other hand, if the instabilities can be removed, one expects to be able to explore the moduli space along the radion direction. The distance conjecture would then come into the picture, since an infinite tower of stable Kaluza-Klein modes would become massless as $\rho \to -\infty$.

It would be interesting to investigate whether the emergent tower of states predicted by the distance conjecture can be alternatively realized by particles or other extended objects, arising \textit{e.g.} wrapping branes around some internal cycles as in~\cite{Grimm:2018ohb,Font:2019cxq}. These further developments, however, are beyond the scope of this work and are left for future research.

\section{de Sitter on the brane-world}
\label{sec:braneworld}

\begin{center}
	\begin{figure}[t]
		\centering
		\includegraphics[width=7cm]{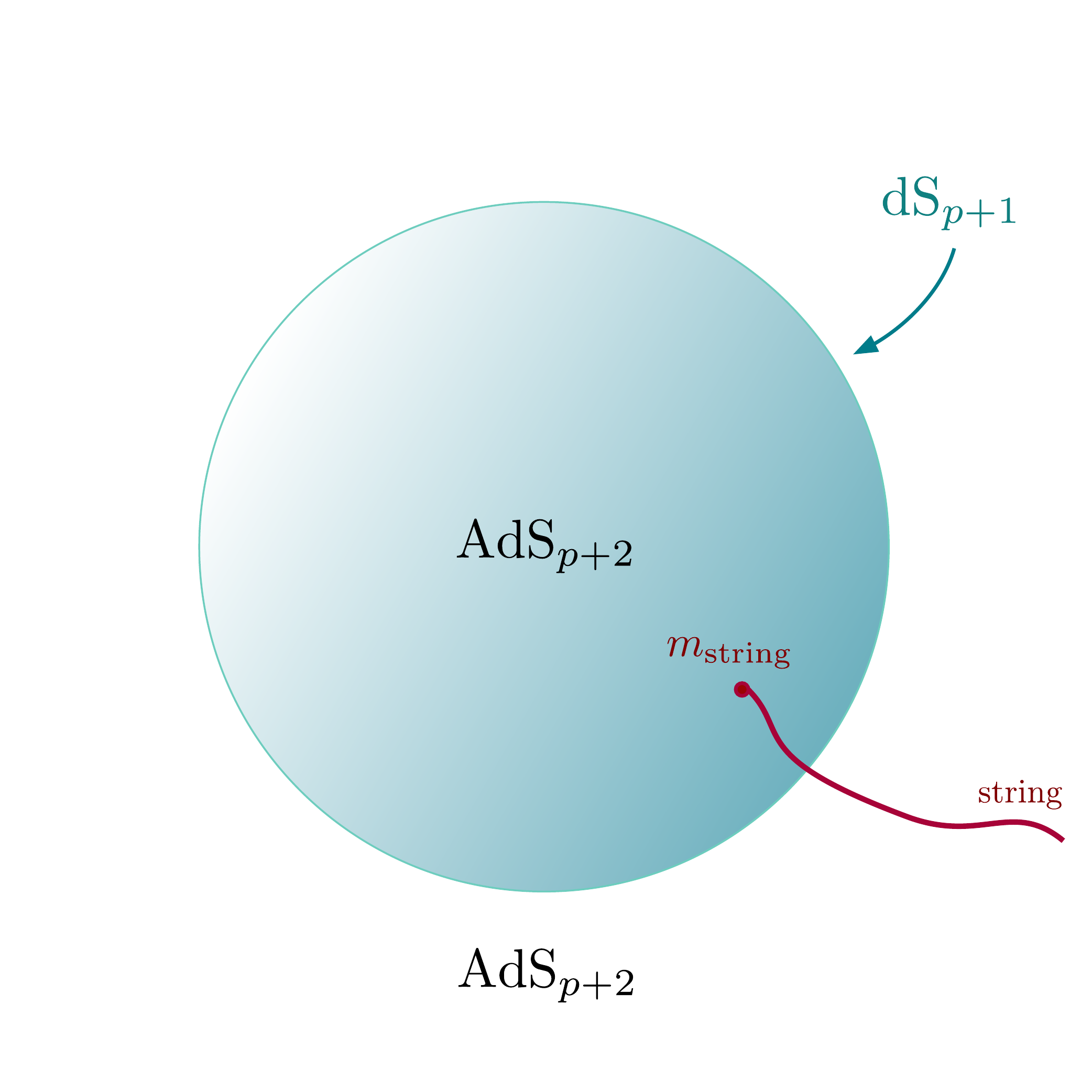}
		\caption{A bubble which interpolates between two ${\rm AdS}_{p+2}$ space-times, hosting a ${\rm dS}_{p+1}$ geometry on its world-volume. Open strings with a single endpoint attached to the bubble wall give rise to massive particles on the world-volume.\label{Fig:BW}}
	\end{figure}
\end{center}

According to the proposal of~\cite{Banerjee:2018qey, Banerjee:2019fzz, Banerjee:2020wix}, a thin-wall bubble nucleating between two $\ads_{p+2}$ space-times hosts a $\ds_{p+1}$ geometry on its wall\footnote{For some earlier works along these lines, see~\cite{Kaloper:1999sm, Shiromizu:1999wj, Vollick:1999uz, Gubser:1999vj, Hawking:2000kj}.}, as schematically depicted in Fig.~\ref{Fig:BW}. Here we make use of the results of~\cite{Antonelli:2019nar} to propose an embedding of scenarios of this type in string theory. Specifically, nucleation of $\text{D}1$-branes in the $\ads_3 \times \ess^7$ solution of~\cite{Mourad:2016xbk} and of $\text{NS}5$-branes in the $\ads_7 \times \ess^3$ solution of~\cite{Mourad:2016xbk} lead to a $\ds_2$ geometry and a $\ds_6$ geometry respectively\footnote{The analogous phenomenon in the case of $\text{D}3$-branes in the type $0'\text{B}$ model appears more elusive, since the corresponding bulk geometry is not $\ads_5 \times \ess^5$, and its large-flux behavior is not uniform~\cite{Dudas:2000sn, Angelantonj:1999qg, Angelantonj:2000kh}.}.

\subsection{The bulk setup}

The $\ads_{p+2} \times \ess^q$ solutions arise as special cases of the Freund-Rubin solutions, and we have discussed  them in Section~\ref{sec:FR_AdS}. In particular, in the orientifold models the $\ads_3 \times \ess^7$ solution is described by eq.~\eqref{eq:caseI_elbsb_sol}, where the flux is electric, while the heterotic $\ads_7 \times \ess^3$ solution is described by eq.~\eqref{eq:caseI_maghet_sol}, where the flux is magnetic in the `natural' frame.

\subsection{Generating bubbles}

The solutions described in the preceding section feature non-perturbative instabilities, whereby charged bubbles nucleate in $\ads$ via flux tunneling. In~\cite{Antonelli:2019nar} the associated decay rates were computed, and it was argued that these solutions arise as near-horizon limits of the gravitational back-reaction sourced by stacks of $\text{D}1$-branes and $\text{NS}5$-branes respectively. The corresponding bubbles are therefore the gravitational counterparts of these fundamental branes, that nucleate and expand due to an enhanced charge-to-tension ratio
\begin{eqaed}\label{eq:enhanced_ctt}
    \beta \equiv \left(\frac{\mathcal{T}}{\mathcal{Q}}\right)_{\!\text{eff}} = v_0 \left(\frac{\mathcal{T}}{\mathcal{Q}}\right)_{\!\text{bare}} \, , \qquad v_0 \equiv \sqrt{\frac{2(D-2)\gamma}{(p+1) ((q-1)\gamma-\alpha)}} \, .
\end{eqaed}
In the string models discussed in Section~\ref{sec:non-susy_strings} $v_0 > 1$, and thus also $\beta > 1$. In particular, one finds~\cite{Antonelli:2019nar}
\begin{eqaed}\label{eq:v0_orientifold}
    (v_0)_{\text{orientifold}} = \sqrt{\frac{3}{2}}
\end{eqaed}
for the orientifold models, while
\begin{eqaed}\label{eq:v0_heterotic}
    (v_0)_{\text{heterotic}} = \sqrt{\frac{5}{3}}
\end{eqaed}
for the heterotic model. This behavior resonates with considerations stemming from the WGC, since the presence of branes which are (effectively) lighter than their charge would usually imply a decay channel for extremal or near-extremal objects. While non-perturbative instabilities of non-supersymmetric $\ads$ due to brane nucleation have been thoroughly discussed in the literature~\cite{Maldacena:1998uz,Seiberg:1999xz,Ooguri:2016pdq}, we stress that in the present case this phenomenon arises from fundamental branes interacting in the absence of supersymmetry.

\subsection{The effective theory on the brane-world}

In the notation of~\cite{Antonelli:2019nar}, let us consider the landscape of $\ads_{p+2}$ space-times with curvature radii $\widetilde{L}$, expressed in the $(p+2)$-dimensional Einstein frame, specified by large flux numbers $n$. The equations of motion for a spherical brane (stack) of charge $\delta n \ll n$ that describe its expansion after nucleation involve the extrinsic curvature $\Theta$ of the world-volume, and stem from the Israel junction conditions~\cite{Israel1966, Barrabes:1991ng}
\begin{eqaed}\label{eq:junction_conditions_general}
	\kappa_{p+2}^2 \, \delta \left(\Theta \, (j^*g)_{ab} - \Theta_{ab} \right) = \widetilde{\tau}_p \, (j^*g)_{ab} \, ,
\end{eqaed}
where $\delta$ denotes the discontinuity across the brane and $\widetilde{\tau}_p$ is the (dressed) tension written in the $(p+2)$-dimensional Einstein frame. Writing the induced metric $j^*g$ on the brane, which is continuous, according to
\begin{eqaed}\label{eq:brane_metric}
	ds^2_{\text{brane}} = - \, dt^2 + a(t)^2 \, d\Omega_p^2 \, ,
\end{eqaed}
the junctions conditions read
\begin{eqaed}\label{eq:junction_eom}
	\delta \sqrt{\frac{1}{\widetilde{L}^2} + \frac{1 + \dot{a}^2}{a^2}} = \frac{\kappa_{p+2}^2 \, \widetilde{\tau}_p}{p} \, .
\end{eqaed}
In the thin-wall limit $\delta n \ll n$ eq.~\eqref{eq:junction_eom} reduces to
\begin{eqaed}\label{eq:thin-wall_junction}
	\sqrt{\frac{1}{\widetilde{L}^2} + \frac{1 + \dot{a}^2}{a^2}} & = \frac{p}{2\kappa_{p+2}^2 \, \widetilde{\tau}_p} \, \delta \left( \frac{1}{\widetilde{L}^2} \right) \\
	& = \frac{\epsilon}{\left( p+1 \right) \widetilde{\tau}_p} = \frac{\beta}{\widetilde{L}} \, ,
\end{eqaed}
where $\epsilon$ is the energy (density) carried by the brane. At the time of nucleation $\dot{a} = 0$, and $a(0) = \widetilde{\rho}$ gives the correct nucleation radius, while the time evolution of the scale factor $a$ is described by the Friedmann equation
\begin{eqaed}\label{eq:friedmann_eq_braneworld}
	\left(\frac{\dot{a}}{a}\right)^2 = - \, \frac{1}{a^2} + \frac{\beta^2 - 1}{\widetilde{L}^2} \, ,
\end{eqaed}
whence $a = \frac{1}{H} \cosh(H t)$ identifies the Hubble parameter
\begin{eqaed}\label{eq:hubble_constant_braneworld}
	H = \frac{1}{\widetilde{\rho}} = \frac{\sqrt{\beta^2 - 1}}{\widetilde{L}} \propto n^{- \, \frac{\gamma \left( 1 + \frac{q}{p} \right)}{(q-1)\gamma-\alpha}} \, .
\end{eqaed}
While the extremality parameter $\beta$ in the string models at stake is not close to unity, as in the near-extremal cases studied in~\cite{Banerjee:2018qey, Banerjee:2019fzz, Banerjee:2020wix}, the $\ads$ curvature is nevertheless parametrically small for large $n$, and therefore the curvature of the $\ds$ wall is also parametrically small.

Furthermore, it has been shown that the Einstein gravity propagating in the bulk induces, at large distances, lower-dimensional Einstein equations on the brane~\cite{Banerjee:2019fzz}, in a fashion reminiscent of Randall-Sundrum constructions~\cite{Randall:1999ee, Randall:1999vf, Giddings:2000mu, Dvali:2000hr, Karch:2000ct}\footnote{Despite some similarities, it is worth stressing that the present context is qualitatively different from scenarios of the Randall-Sundrum type.}. In order to elucidate this issue in the present case, where the branes deviate from extremality by the $\mathcal{O}\!\left(1\right)$ factor $v_0$, let us compare the on-shell action for the expanding brane, which takes the form
\begin{eqaed}\label{eq:on-shell_braneworld}
	S_p = \left( \beta - 1 \right) \widetilde{\tau} \int d^{p+1}\zeta \left( \frac{\widetilde{L}}{Z} \right)^{p+1}
\end{eqaed}
in Poincar\'e coordinates, with the corresponding Einstein-Hilbert action
\begin{eqaed}\label{eq:eh_braneworld}
	S_p^{\text{EH}} = \frac{1}{2\kappa_{p+1}^2} \int d^{p+1}\zeta \left( \frac{\widetilde{L}}{Z} \right)^{p+1} \left(R_{p+1} - 2\Lambda_{p+1}\right) \, ,
\end{eqaed}
since the resulting effective gravitational theory on the world-volume ought to reconstruct general covariance~\cite{Shiromizu:1999wj}\footnote{For a recent discussion in the context of entanglement islands, see~\cite{Chen:2020uac}.}. Since for $\ds_{p+1}$
\begin{eqaed}\label{eq:ds_lagrangian}
	R_{p+1} - 2\Lambda_{p+1} = 2p H^2 \, ,
\end{eqaed}
using eq.~\eqref{eq:hubble_constant_braneworld} and the defining relations
\begin{eqaed}
	\beta & \equiv \frac{\epsilon \, \widetilde{L}}{(p+1) \widetilde{\tau}} \, , \\
	\epsilon & \equiv \delta \widetilde{E}_0 = \frac{p(p+1)}{\kappa_{p+2}^2 \, \widetilde{L}^3} \, \delta \widetilde{L} \, ,
\end{eqaed}
introduced in~\cite{Antonelli:2019nar}, one finds the world-volume Newton constant
\begin{eqaed}\label{eq:action_matching}
	\kappa_{p+1}^2 = \beta \left(\beta + 1\right) \frac{\kappa_{p+2}^2}{\delta \widetilde{L}} \propto n^{1 - \, \frac{\gamma \left( 1 + \frac{q}{p} \right)}{(q-1)\gamma-\alpha}} \, ,
\end{eqaed}
which indeed reproduces the results of~\cite{Gubser:1999vj, Banerjee:2019fzz} in the near-extremal limit $\beta \; \to \; 1$. While for the orientifold models $p = 1$, and thus there would be no associated Planck mass $M_{\text{Pl}}^{1 - p} = \kappa_{p+1}^2$, in the heterotic model $p = 5$ and $\beta = \sqrt{\frac{5}{3}}$ for extremal $\text{NS}5$-branes, and thus the vacuum energy (density) in units of the $(p+1)$-dimensional Planck mass is given by
\begin{eqaed}\label{eq:cc_planck_het}
	\left(\frac{E_{p+1}}{M_{\text{Pl}}^{p+1}}\right)_{\!\text{heterotic}} = \frac{25}{18 \pi} \, \sqrt{\frac{5}{3}} \, \sqrt{1 + \sqrt{\frac{5}{3}}} \frac{\left(\kappa_{10} \, T^2 \right)}{\sqrt{T \, \delta n} \left(T \, n\right)^2} \, ,
\end{eqaed}
which is parametrically small for large $n$. This result actually holds whenever the bulk $\ads$ geometry exists, since
\begin{eqaed}\label{eq::cc_planck_gen}
	\frac{E_{p+1}}{M_{\text{Pl}}^{p+1}} \propto n^{- \frac{2 ((p + 1) \gamma + \alpha)}{(p - 1) ((q - 1) \gamma - \alpha)}} \, .
\end{eqaed}
It would be interesting to investigate in detail how world-volume matter and gauge fields couple the effective brane-world gravity, and whether the low-energy physics is constrained as a result. Furthermore, holographic considerations~\cite{Maxfield:2014wea, Antonelli:2018qwz} could shed some light on the late-time, strongly-coupled regime of these constructions.

\subsection{Massive particles}

It has been shown in~\cite{Banerjee:2019fzz, Banerjee:2020wix} that one can include radiation and matter densities in the Friedmann equation of eq.~\eqref{eq:friedmann_eq_braneworld} introducing black holes and `string clouds' respectively. While the former case appears problematic~\cite{Poletti:1994ww, Wiltshire:1994de, Chan:1995fr}, one can nevertheless reproduce the effect of introducing string clouds using probe open strings stretching between branes in $\ads$. In order to compute the mass $m_{\text{str}}$ of the point particle induced by an open string ending on a brane in more general settings, let us consider a bulk geometry with the symmetries corresponding to a flat (codimension one) brane, with transverse geodesic coordinate $\xi$, and thus a metric of the type
\begin{eqaed}\label{eq:brane_bulk_metric}
	ds^2 = d\xi^2 + \Omega(\xi)^2 \, \gamma_{\mu \nu}(x) \, dx^\mu \, dx^\nu \, .
\end{eqaed}
Let us further consider a string with tension $T$ stretched along $\xi$, attached to the brane at $\xi = \xi_b$, with longitudinal coordinates $x^\mu(\tau)$ in terms of the world-line of the induced particle. A suitable embedding with world-sheet coordinates $(\tau \, , \, \sigma)$ then takes the form
\begin{eqaed}\label{eq:embedding}
	X^\mu & = X^\mu(\tau, \sigma) \, , \qquad X^\mu(\tau, \sigma_b) \equiv x^\mu(\tau) \, , \\
	\xi & = \xi(\sigma) \, , \qquad \xi(\sigma_b) \equiv \xi_b \, ,
\end{eqaed}
with Neumann boundary conditions on the $X^\mu$, so that the induced metric determinant on the world-sheet yields the Nambu-Goto action
\begin{eqaed}\label{eq:induced_NG_action}
	S_{\text{NG}} = - \, T \int d\tau \, d\sigma \, \Omega \, \sqrt{ \Omega^2 \left( \dot{X} \cdot X' \right)^2 - \left( \xi'^2 + \Omega^2 \, X'^2 \right) \dot{X}^2} \, ,
\end{eqaed}
where $\dot{X}^2 \equiv \gamma_{\mu \nu}(X) \, \dot{X}^\mu \, \dot{X}^\nu$ and we have assumed that $\Omega > 0$ and $\xi' > 0$, since both $\xi$ and $\sigma$ parametrize the string stretching in the transverse direction. In turn, this implies that $\sigma_b < \sigma_f$, where $\xi(\sigma_f) \equiv \xi_f$ corresponds to the (conformal) boundary where $\Omega(\sigma_f) = 0$. Then, varying the action and integrating by parts gives the boundary term
\begin{eqaed}\label{eq:NG_action_variation}
	\delta S_{\text{NG}} = - \, T \int d\tau \, \Omega \, \delta \xi \, \sqrt{-\dot{X}^2} \,\bigg|_{\sigma_b}^{\sigma_f} \, ,
\end{eqaed}
up to terms that vanish on shell\footnote{Let us remark that, as usual, initial and final configurations are fixed in order that the Euler-Lagrange equations hold.}. Since the variation $\delta \xi_f = 0$, one can fix $X^\mu = X^\mu(\tau, \sigma_b) = x^\mu(\tau)$, and the resulting on-shell variation
\begin{eqaed}\label{eq:brane_variation}
	\delta S_{\text{NG}} = \delta \left( - \, T \int d\tau \int_{\xi_b}^{\xi_f} d\xi \, \Omega(\xi) \, \sqrt{-\dot{x}^2} \right)
\end{eqaed}
ought to be identified with the variation of the particle action
\begin{eqaed}\label{eq:particle_action}
	S_\text{particle} = - \, m_{\text{string}} \int d\tau \, \Omega(\xi_b) \, \sqrt{-\dot{x}^2} \, ,
\end{eqaed}
which one can also obtain evaluating eq.~\eqref{eq:induced_NG_action} for a rigid string. Hence,
\begin{eqaed}\label{eq:mass_fix}
	m_{\text{string}} = \frac{T}{\Omega(\xi_b)} \int_{\xi_b}^{\xi_f} d\xi \, \Omega(\xi) \, ,
\end{eqaed}
and for $\ads$, for which $\Omega(\xi) \propto e^{-\frac{\xi}{L}}$, eq.~\eqref{eq:mass_fix} reduces to $m_{\text{string}} = T \, L$, thus reproducing the results of~\cite{Banerjee:2019fzz, Banerjee:2020wix}. More generally, requiring that $\frac{\partial m_{\text{string}}}{\partial \xi_b} = 0$ gives the condition $\Omega'(\xi_b) = - \, \frac{m_{\text{string}}}{T} \, \Omega(\xi_b)$, \text{i.e.} the space-time is $\ads$ if the mass remains constant as the brane expands. Moreover, if the string stretches between $\xi_b$ and the position $\xi_{b'}$ of another brane, the endpoints of integration change, and if $\xi_b \sim \xi_{b'}$ one recovers the flat-space-time result $m_{\text{string}} \sim T \, \delta\xi$. While for fundamental strings stretching between $\text{D}1$-branes the resulting masses would be large, and would thus bring one outside the regime of validity of the present analysis, successive nucleation events would allow for arbitrarily light strings stretched between nearby branes, although the probability of such events is highly suppressed in the semi-classical limit. The resulting probability distribution of particle masses is correspondingly heavily skewed toward large values.

\subsection{de Sitter foliations from nothing}\label{sec:ds_nothing}

As a final comment, let us remark that the nucleation of bubbles of nothing~\cite{Witten:1981gj} offers another enticing possibility to construct $\ds$ configurations~\cite{Dibitetto:2020csn}. To our knowledge, realizations of this type of scenario in string theory have been mostly investigated breaking supersymmetry in lower-dimensional settings~\cite{Horowitz:2007pr}\footnote{Some lower-dimensional toy models offer flux landscapes where more explicit results can be obtained~\cite{BlancoPillado:2010df, Brown:2010mf, Brown:2011gt}.}. However, recent results indicate that nucleation of bubbles of nothing is quite generic, and occurs also in some supersymmetric cases~\cite{GarciaEtxebarria:2020xsr}. In particular, the supersymmetry-breaking $\mathbb{Z}_k$ orbifold of the type IIB $\ads_5 \times \ess^5$ solution, described in~\cite{Horowitz:2007pr}, appears to provide a calculable large-$N$ regime and a dual interpretation in terms of the corresponding orbifold of $\mathcal{N} = 4$ supersymmetric Yang-Mills theory in four dimensions, which is a $U(N)^k$ gauge theory that is expected to retain some of the properties of the parent theory~\cite{Kachru:1998ys, Lawrence:1998ja, Bershadsky:1998mb, Bershadsky:1998cb, Schmaltz:1998bg, Erlich:1998gb, Angelantonj:1999qg, Tong:2002vp}. For what concerns the $\adsts$ solutions discussed in~\cite{Antonelli:2019nar}, on the other hand, some evidence suggests that the decay rate per unit volume associated to the nucleation of bubble of nothing is subleading with respect to flux tunneling in single-flux landscapes~\cite{Brown:2010mf}, and thus in the $\adsts$ solutions of interest.

\section{Conclusions}\label{sec:conclusions}

In this paper we have explored a number of possibilities to realize $\ds$ configurations in the context of ten-dimensional string models where supersymmetry is broken at the string scale or is absent altogether. We have focused on the $USp(32)$ and type $0$'B orientifold models and on the heterotic $SO(16) \times SO(16)$ model. These models share the presence of an exponential runaway potential for the dilaton, a tantalizing feature that mirrors lower-dimensional anti-brane uplifts and compels one to look for $\ds$ vacua from a ten-dimensional vantage point.

To begin with, we have seeked stable $\ds$ warped flux compactifications on arbitrary internal manifolds. However, streamlining earlier results which hold for supergravity theories~\cite{Maldacena:2000mw, Giddings:2001yu}, we have formulated a general no-go theorem that excludes $\ds$ and Minkowski vacua in the dimensionally reduced theory within a region of parameter space. In particular, realizations of lower-dimensional $\ds$ vacua of this type are excluded for all the non-supersymmetric string models that we have considered. Furthermore, within bottom-up models where Freund-Rubin $\ds$ compactifications exist, they are always unstable: we have considered lower-dimensional EFTs described by eq.~\eqref{eq:swamp_redS}, which generalize those obtained compactifying non-supersymmetric string models. These depend on various parameters, including dilaton and gauge couplings. Consistently with the results of~\cite{Montero:2020rpl}, one can show that whenever $\ds$ solutions exist they are perturbatively unstable due to the universal dilaton-radion dynamics. 

A lower-dimensional, bottom-up perspective offers additional insights. The absence of classical $\ds$ vacua in non-supersymmetric string models resonates with some recently proposed Swampland conjectures, such as the de Sitter conjecture~\cite{Obied:2018sgi} and the `Transplanckian Censorship conjecture' (TCC)~\cite{Bedroya:2019snp}. We have shown that these hold for both the orientifold $USp(32)$ and type $0$'B models and for the heterotic $SO(16) \times SO(16)$ model, explicitly computing the relevant parameters and providing appropriate bounds. This result garners non-trivial evidence for the de Sitter conjecture and for the TCC in top-down non-supersymmetric settings. It would be interesting to further investigate additional Swampland conjectures within a non-supersymmetric context and the resulting constraints on non-supersymmetric string phenomenology~\cite{Abel:2015oxa,Abel:2017vos,March-Russell:2020lkq}. Here we have pointed out possible realizations of the `distance conjecture', identifying Kaluza-Klein states as the relevant tower of states that become massless at infinite distance in field space. A more detailed analysis would presumably require a deeper knowledge of the geometry of the moduli spaces which can arise in non-supersymmetric compactifications, albeit our arguments rest solely on the existence of the ubiquitous dilaton-radion sector. It would be also interesting to address whether the `Distant Axionic String conjecture'~\cite{Lanza:2020qmt}, which predicts the presence of axionic strings within any infinite-distance limit in field space, holds also in non-supersymmetric settings. 

Despite our preceding considerations, the absence of $\ds$ vacua does not necessarily preclude alternative realizations of $\ds$ cosmologies. According to a recently revisited proposal~\cite{Banerjee:2018qey, Banerjee:2019fzz, Banerjee:2020wix}, branes expanding within a bulk $\ads$ space-time host $\ds$ geometries on their world-volumes. In the non-supersymmetric string models that we have considered, the non-perturbative instabilities of the $\ads$ flux compactifications or~\cite{Mourad:2016xbk}, studied in~\cite{Antonelli:2019nar}, entail the nucleation of charged branes of codimension one in $\ads$, which mediate flux tunneling and separate $\ads$ regions with different flux numbers. Thus, it is natural to propose these branes as candidates to realize $\ds$ geometries on their world-volumes. However, the complete identification of the EFT living on the world-volume of such branes appears challenging and, although we have suggested some preliminary steps in this respect, further work is needed to make progress. For instance, the results of~\cite{Banerjee:2018qey,Banerjee:2019fzz,Banerjee:2020wix} suggest that Einstein gravity arises on the brane-world only at large distances, and is accompanied by corrections akin to those of more familiar scenarios of the Randall-Sundrum type. In this regard, it would be interesting to investigate whether world-volume theories of this kind constrain, \textit{e.g.}, which matter or gauge fields can be present. A detailed study of these promising scenarios might be a suitable starting point to shed some light on whether Swampland conjectures, which mostly concern bulk constructions, apply to brane-world models, and if so to which extent. 

\section*{Acknowledgements}

The authors would like to thank Suvendu Giri, Fernando Marchesano, Matt Reece, Savdeep Sethi and Irene Valenzuela for useful comments and discussions. In addition, we are deeply grateful to Augusto Sagnotti for his feedback on the manuscript, to Adam Brown and Alex Dahlen for having shared their Mathematica code and to Davide Bufalini for a thorough reading of the manuscript. IB is supported in part by Scuola Normale, by INFN (IS GSS-Pi)
and by the MIUR-PRIN contract 2017CC72MK$\_$003. SL is supported by a fellowship of Angelo Della Riccia Foundation, Florence and a fellowship of Aldo Gini Foundation, Padova.

\appendix

\section{Proof of the no-go theorem}
\label{app:no-go_proof}

In this appendix we provide the proof of the no-go theorems stated in Section~\ref{sec:no-go}, which proceeds along the same lines of~\cite{Maldacena:2000mw,Giddings:2001yu}. Let us consider a compactification of the ten-dimensional theory described by the action in eq.~\eqref{eq:action} over a closed, compact $q$-dimensional manifold $Y$. The metric ansatz for the reduction is
\begin{equation}
\label{no-go_gen_metr}
{d }s_{10}^2  = e^{- \frac{2q}{d-2}C(y)} \, \widehat g_{\mu\nu} (x) d x^\mu d x^\nu+ e^{2C(y)} \,  \widetilde g_{ab} (y) d y^a d y^b\, ,
\end{equation}
where $x^\mu$, $\mu = 0,1,\ldots,d-1$ denote the external coordinates and $y^a$,  $a=1,\ldots,q$ denote the internal coordinates over $Y$. Notice that the warp factor depends exclusively on the internal coordinates, and it is such that, in the reduced $d$-dimensional theory, the action is expressed in the Einstein frame. Furthermore, for the sake of generality we shall include the space-time-filling sources discussed in Section~\ref{sec:no-go_sources}, which are described by the action of eq.~\eqref{eq:sources_Sloc}. The complete ten-dimensional action is that of eq.~\eqref{eq:sources_Stot}. Accordingly, the trace-reversed stress-energy tensor is the sum of the bulk and source contributions,
\begin{equation}
	\widetilde{T}_{MN} = \widetilde{T}^{\rm bulk}_{MN}+\widetilde{T}^{\rm loc}_{MN} \, ,
\end{equation}
where
\begin{eqaed}\label{eq:souces_Tbulk}
	\widetilde{T}^{\rm bulk}_{MN} & = \frac12 \, \partial_M \phi \partial_N \phi + \frac{f(\phi)}{2(p+1)!} \, (H_{p+2}^2)_{MN}  + \frac{g_{MN}}{8} \left( V - \frac{p+1}{2(p+2)!} \, f(\phi) \, H_{p+2}^2 \right)
\end{eqaed}
and
\begin{eqaed}\label{eq:souces_Tloc}
	 {T}^{\rm loc}_{MN} & = - 
	\sigma_{\tau}  g_{MP} g_{NQ} \int d^{p+1}\xi\, h^{ij} \frac{\partial x^P}{\partial \xi^i} \frac{\partial x^Q}{\partial \xi^j} \tau_p(\phi) \delta^{(10)}(x^L - x^L(\xi)) 
	\\
	& = - 
	\sigma_{\tau}  \tau_p(\phi) \Pi_{MN} \delta^{(9-p)}(x_\perp^K - z^K) \, ,
\end{eqaed}
and, in passing from the first to the second line, we have employed the static gauge of eqs.~\eqref{eq:sources_xpar} and~\eqref{eq:sources_xperp}. The projector $\Pi_{MN}$ equals $g_{ij}$ whenever $M \, , N = i \, , j$, and vanishes otherwise. The ten-dimensional equations of motion obtained from the action in eq.~\eqref{eq:sources_Stot} are then
\begin{subequations}
	\label{eq:sources_eoms}
	\begin{align}
	\label{eq:sources_eoms_R}
	R_{MN} & =\widetilde{T}^{\rm bulk}_{MN} + \widetilde{T}^{\rm loc}_{MN} \, , 
	\\
	\label{eq:sources_eoms_phi}
	\Box \, \phi - V'(\phi) - \frac{f'(\phi)}{2(p+2)!} \, H_{p+2}^2 & = 2 \sigma \tau_p(\phi) \delta^{(9-p)}(x_\perp^K - z^K) \, ,
	\\
	\label{eq:sources_eoms_H}
	d \star (f(\phi) \, H_{p+2}) & =  - 2 q \delta^{(9-p)}(x_\perp^K - z^K) \, .
	\end{align}
\end{subequations}
To begin with, let us focus on eq.~\eqref{eq:sources_eoms_phi}, which one can recast as
\begin{equation}
\label{eq:no-go_gen_eomdil}
\widetilde \Box ( e^{-\frac{16}{d-2}B} \phi ) = e^{-2 \frac{10-d}{d-2} B } \left(V' + \frac{f'}{2 (p+2)!} H^2_{p+2}+ 2\sigma \tau_p \delta^{(9-p)}(x_\perp^K-z^K)\right)\,.
\end{equation}
A first constraint is obtained integrating eq.~\eqref{eq:no-go_gen_eomdil} over the internal manifold $Y$. Since $Y$ is a compact manifold without boundaries, the left-hand side of eq.~\eqref{eq:no-go_gen_eomdil} integrates to zero. Then one obtains
\begin{equation}
\label{eq:no-go_gen_constr1I}
\boxed{ \gamma\, \mathcal{I}_V + \frac{\alpha}{2} \mathcal{I}_H + 2 \sigma \sigma_\tau \mathcal{I}_{\rm loc} = 0 \, ,}
\end{equation}
where we have introduced
\begin{eqaed}
	\label{eq:no-go_gen_V_F_integrals}
	\mathcal{I}_V &\equiv \int d^q y \sqrt{\widetilde{g}(y)} \, e^{-2\frac{10-d}{d-2}C(y)} \, V \, ,
	\\
	\mathcal{I}_H &\equiv \int d^q y \sqrt{\widetilde{g}(y)} \, e^{-2\frac{10-d}{d-2}C(y)} \, \frac{f}{(p+2)!} \, H_{p+2}^2 \, ,
	\\
	\mathcal{I}_{\rm loc} &\equiv \int d^qy\, \sqrt{\widetilde{g}(y)}\,  e^{-2\frac{10-d}{d-2}C(y)} \,\tau_p(\phi) \,\delta (\Sigma) \, .
\end{eqaed}
A second constraint can be obtained from an appropriate integration of eq.~\eqref{eq:sources_eoms_R}. First, notice that splitting it into internal and external components with the metric ansatz in eq.~\eqref{no-go_gen_metr} yields the two independent equations
\begin{align}
\nonumber
\widehat R_{\mu \nu} & =  \bigg[\frac{1}{8} \, e^{-\frac{2q}{d-2}C} \left( V -\frac{p+1}{2(p+2)!} f  H^2_{p+2} \right)
\\
&\quad\, - \, e^{-\frac{16}{d-2} C} \left( \frac{q}{d-2} \widetilde \Box C -\frac{16}{(d-2)^2}(\widetilde \partial C)^2 \right) \bigg] \widehat g _{\mu\nu} + \widetilde T^{\rm loc}_{\mu\nu}\,,
\label{eq:sources_Rmunu}
\\
\nonumber
\widetilde R_{ab}  &+ \frac{16}{d-2} \, \nabla_a \partial_b C + \frac{8(4-2d-dq)}{(d-2)^2} \, \partial_a C \, \partial_b C - \left( \widetilde\Box C - \frac{16}{d-2}(\partial C)^2 \right) \widetilde g_{ab}  =
\\
&=\frac{1}{2}\, \partial_a \phi \, \partial_b \phi + \frac{f}{2(p+1)!} H^2_{ab} + \frac{1}{8} \, e^{2C} \left( V - \frac{p+1}{2(p+2)!} f H^2_{p+2}\right) \widetilde g_{ab} +  \widetilde T^{\rm loc}_{ab}\,, \label{eq:sources_Rab}
\end{align}
but for our purposes it is sufficient to consider their traces. The trace of eq.~\eqref{eq:sources_Rmunu} may be recast as
\begin{eqaed}
	\label{eq:no-go_gen_Rmunutr}
	\frac{qd}{(2-d)k} \widetilde\Box e^{kC} &= e^{ \chi C} \widehat R - \frac{qd}{d-2} (16-k(2-d))  e^{ \chi C} (\widetilde\partial C)^2
	\\
	&\quad-\frac{d}{16} e^{ \chi C + \frac{2q}{d-2}C} \left(2V - \frac{p+1}{(p+2)!} f H^2_{p+2}\right) - \widetilde T^{\rm loc}_{\rm ext}
\end{eqaed}
for any real $k$, where $\chi = \frac{16-2k+kd}{d-2}$. Here we have employed the following useful identity
\begin{equation}
\label{eq:no-go_ekb}
\widetilde \Box C = \frac{1}k e^{-k C}  \widetilde \Box e^{kC}-k (\widetilde \partial C)^2\,.
\end{equation}
In particular, for $k = \frac{16}{2-d}$, eq.~\eqref{eq:no-go_gen_Rmunutr} reduces to
\begin{eqaed}
	\label{eq:no-go_gen_Rmunutrb}
	\frac{qd}{16} \widetilde\Box e^{ \frac{16}{2-d}C} &= \widehat R -\frac{d}{16} e^{\frac{2q}{d-2}C} \left(2V - \frac{p+1}{(p+2)!} f H^2_{p+2}\right) -  \widetilde T^{\rm loc}_{\rm ext}\,.
\end{eqaed}
Although it is not strictly needed to prove of the no-go theorem, let us also provide the trace of eq.~\eqref{eq:sources_Rab} for completeness. It reduces to
\begin{eqaed}
	\label{eq:sources_Rintb}
	\frac{16-qd+2q}{l(d-2)}\widetilde\Box e^{lC} &= \left(l-\frac{8}{d-2}\right) \frac{qd - 2q -16}{d-2} e^{lC} (\widetilde\partial C)^2 -  e^{lC} \widetilde R  \,
	\\
	&\quad\,+  e^{lC}\left(\frac12(\widetilde \partial \phi)^2 + \frac{q}{8} \, e^{2C} \, V +  \frac{(8-q)p+16-q}{8(p+2)!}e^{2C} \, f H^2_{p+2} \right)\, 
	\\
	&\quad\,+e^{lC} \widetilde T^{\rm loc}_{\rm int}\,.
\end{eqaed}
The contributions to eqs.~\eqref{eq:no-go_gen_Rmunutrb} and~\eqref{eq:sources_Rintb} arising from the sources are
\begin{eqaed}
	\label{eq:sources_Ttrs}
	 \widetilde T^{\rm loc}_{\rm ext} &= -\sigma_\tau \frac{7-p}{2}  e^{-6C} \,\tau_p(\phi)\, \delta (\Sigma)     \,,
	\\
	 \widetilde T^{\rm loc}_{\rm int} &= -\sigma_\tau \frac{p-15}{4} e^{2C} \,\tau_p(\phi)\, \delta (\Sigma)   \,,
\end{eqaed}
where we have introduced the shorthand notation $\delta (\Sigma) = \delta^{(9-p)}(x_\perp^K - z^K)$. Integrating eq.~\eqref{eq:no-go_gen_Rmunutrb} over $Y$ then yields
\begin{equation}
\label{eq:no-go_gen_constr2I}
\boxed{ \frac{2}{d-2}  {\rm vol}_q \,\Lambda = \frac{1}{8}\,\mathcal{I}_V - \frac{1}{16} (p+1) \,\mathcal{I}_H - \frac{1}{8}\sigma_\tau  (7-p) \,\mathcal{I}_{\rm loc} \, ,}
\end{equation}
where ${\rm vol}_q$ denotes the volume of $Y$ and $\Lambda = \frac{d-2}{d} \widehat R$ denotes the space-time cosmological constant. Hence, eq.~\eqref{eq:no-go_gen_constr2I} is to be understood as a constraint on the allowed values of the $\Lambda$. The no-go theorems stated in Section~\ref{sec:no-go} are then obtained combining eq.~\eqref{eq:no-go_gen_constr2I} and eq.~\eqref{eq:no-go_gen_constr1I}, obtaining
\begin{equation}
\label{eq:no-go_gen_constr12I}
\frac{2}{d-2} {\rm vol}_q \Lambda = -\frac{1}{16}\left(\frac{\alpha}{\gamma}+(p+1)\right)\mathcal{I}_H + \sigma_\tau  \frac{1}{8} \left(p -7 -\frac{2\sigma}{\gamma}\right)\mathcal{I}_{\rm loc}\,.
\end{equation}
Thus, a sufficient condition in order \emph{not} to have dS or Minkoski vacua is
\begin{equation}
\label{eq:no-go_gen_s_constr1i}
\boxed{\frac{\alpha}{\gamma}+ (p+1) > 0 \quad {\rm and} \quad \sigma_\tau \left(p -7 -\frac{2 \sigma}{\gamma}  \right)< 0 , ,}
\end{equation}
as anticipated in eq.~\eqref{eq:no-go_gen_sources}, while the source-less case follows trivially.

\subsection{The dilaton potential as a \texorpdfstring{\hbox{$D$}}{D}-dimensional source}

As a byproduct of the preceding discussion, let us elaborate on the rôle of the dilaton potential in the non-supersymmetric models of Section~\ref{sec:non-susy_strings}. It is worthwhile noting that this contribution to eq.~\eqref{eq:action} may be understood as stemming from an extended object, filling the whole $D$-dimensional space-time, described by the action
\begin{equation}
\label{eq:V10D_S}
	S_{\rm loc} = - \int d \xi^D \sqrt{-h} {\mathcal{T}}_D(\phi)\,, \qquad {\rm with}\quad  \mathcal{T}_D \equiv \frac12 T e^{\gamma \phi}\,.
\end{equation}
In other words, in light of eq.~\eqref{eq:no-go_gen_constr2I}, in terms of eq.~\eqref{eq:no-go_gen_V_F_integrals} this can be recast as
\begin{equation}
	\label{eq:V10D_I}
	\mathcal{I}_V = 2 \, \mathcal{I}_{\rm loc} \, ,
\end{equation}
where $\mathcal{I}_{\rm loc}$ arises from the action in eq.~\eqref{eq:V10D_S}. This interpretation resonates with the orientifold models illustrated in Section~\ref{sec:non-susy_strings}, in which supersymmetry is broken spontaneously via space-time-filling branes and the endpoints of open strings are not constrained.

As noticed in~\cite{Polchinski:1995mt}, let us remark that the action in eq.~\eqref{eq:V10D_S}, unlike the more general eq.~\eqref{eq:sources_Sloc}, does not include a Wess-Zumino term of the form $q \int C_{10}$. Indeed, the degenerate ten-form $C_{10}$ is not dynamical, and thus integrating it out leads to $q = 0$, as required by anomaly cancellation.

\section{The radion-dilaton potential in the reduced theory}
\label{app:reduced_pot}

Following~\cite{Bremer:1998zp}, in this appendix we derive the effective radion-dilaton potential that we have employed in Section~\ref{sec:red_pot}. To begin with, let us consider the $D$-dimensional action
\begin{equation}
\label{eq:eft_10Daction}
S = \frac{1}{2}\int d^D x \, \sqrt{-g} \, \left( R - \frac{1}{2}\, (\partial \phi)^2 - V(\phi) - \frac{f(\phi)}{2}\, H_{p+2}|^2 \right) \, .
\end{equation}
Our aim is to reduce the above action on a compact, $d_Y$-dimensional manifold $Y$ in order to obtain a theory in $d_X = D-d_Y$ dimensions. We start from the metric ansatz~\cite{Bremer:1998zp,Montero:2020rpl}
\begin{equation}
\label{eq:eft_ds2}
d s^2 = e^{2B \rho (x)} d  \widehat s^2_X + e^{2 A \rho(x)} d  \widetilde s_Y^2 \, ,
\end{equation}
where $d \widehat s^2_X = \widehat g_{\mu\nu} d x^\mu d  x^\nu$, $\mu = 0,\ldots, d_X -1$ is the space-time metric and $d \widetilde s^2_Y = \widetilde g_{ab} d y^a d  y^b$, $a = 1,\ldots, d_Y$ is the internal metric. The warp factors encodes the dependence on the radion $\rho(x)$, and ${A}$ and ${B}$ are constants that we shall now determine. Using the ansatz in eq.~\eqref{eq:eft_ds2}, one can show that the ten-dimensional Ricci scalar reduces to
\begin{equation}
\label{eq:eft_Rred}
\begin{aligned}
R &= e^{-2A \rho} \widetilde R + e^{-2B \rho} \widehat R 
\\
&+e^{-2B\rho} a \Big[-2 d_Y \widehat\Box \rho - 2 (d_X -1) B \widehat\Box \rho - d_Y (d_Y+1) A^2 (\widehat\partial \rho)^2 
\\
&\quad\qquad- (d_X-1) (d_X-2) B^2(\widehat\partial \rho)^2 + 2(2-d_X) d_Y AB (\widehat\partial \rho)^2 \Big] \, .
\end{aligned}
\end{equation}
Furthermore, we assume that $\phi = \phi(x)$ depends only the external coordinates, and we require that the flux of $H_{p+2}$ be magnetic, threading the full internal manifold $Y$ according to
\begin{equation}
H_{d_Y} = n \, d \widetilde {\rm vol}_Y = n \,e^{-A d_Y} \,d  {\rm vol}_Y \,,
\end{equation}
supported by the quantization condition
\begin{equation}
\frac{1}{\widetilde{\rm vol}_Y} \int_{\widetilde Y} H_{d_Y} = n
\end{equation}
with $\widetilde{\rm vol}_Y = \int_{Y} \sqrt{\widetilde g_Y}$. The parameters ${A}$ and ${B}$ appearing in \eqref{eq:bound_alphabeta} can be then fixed requiring that the $d_X$-dimensional action be expressed in the Einstein frame, so that
\begin{equation}
\label{eq:eft_alpha}
{A} = \frac{2-d_X}{d_Y} {B}\,,
\end{equation}
and fixing the canonical normalization of the radion kinetic term, so that
\begin{equation}
\label{eq:eft_beta}
{B}^2 =-  \frac{d_Y}{2(2-d_X)(d_X+d_Y-2)}\,.
\end{equation}
Hence, reducing the action in eq.~\eqref{eq:eft_10Daction} using the metric in eq.~\eqref{eq:eft_ds2}, the curvature in eq.~\eqref{eq:eft_Rred} and eqs.~\eqref{eq:eft_alpha} and~\eqref{eq:eft_beta} one arrives at
\begin{equation}\label{eq:eft_redaction}
\begin{aligned}
S = \frac{\widetilde{\rm vol}_Y}{2} \int_X\, d^{d_X}x\, \sqrt{|\widehat g_X|} \left[\widehat R- \frac12 (\widehat\partial \rho)^2 -  \frac12 (\widehat\partial \phi)^2 - \mathcal{V}(\phi,\rho)\right] \, .
\end{aligned}
\end{equation}
The effective potential $\mathcal{V}(\phi,\rho)$ includes contributions from the ten-dimensional dilaton potential $V$, the magnetic flux and internal curvature, and it reads
\begin{equation}\label{eq:eft_redpot}
\begin{aligned}
\mathcal{V}(\phi,\rho) &= e^{2 B\rho} \, V(\phi) + \frac12 \, f \, n^2 \,  \sigma_Y \, e^{2 B \rho (d_X-1)} - r \, e^{2(B-A)\rho}\, ,
\end{aligned}    
\end{equation}
where we have introduced
\begin{eqaed}
r \equiv \frac{1}{\widetilde{\rm vol}_Y} \int_Y  d^{d_Y}x\, \sqrt{\widetilde g_Y}  \widetilde R(y)
\end{eqaed}
for convenience.


\begin{thebibliography}{100}
	
	\bibitem{Kachru:2003aw}
	S.~Kachru, R.~Kallosh, A.~D. Linde and S.~P. Trivedi, \emph{{De Sitter vacua in
			string theory}},
	\href{https://doi.org/10.1103/PhysRevD.68.046005}{\emph{Phys. Rev. D}
		{\bfseries 68} (2003) 046005}
	[\href{https://arxiv.org/abs/hep-th/0301240}{{\ttfamily hep-th/0301240}}].
	
	\bibitem{Danielsson:2009ff}
	U.~H. Danielsson, S.~S. Haque, G.~Shiu and T.~Van~Riet, \emph{{Towards
			Classical de Sitter Solutions in String Theory}},
	\href{https://doi.org/10.1088/1126-6708/2009/09/114}{\emph{JHEP} {\bfseries
			09} (2009) 114} [\href{https://arxiv.org/abs/0907.2041}{{\ttfamily
			0907.2041}}].
	
	\bibitem{Cordova:2018dbb}
	C.~C{\'o}rdova, G.~B. De~Luca and A.~Tomasiello, \emph{{Classical de Sitter
			Solutions of 10-Dimensional Supergravity}},
	\href{https://doi.org/10.1103/PhysRevLett.122.091601}{\emph{Phys. Rev. Lett.}
		{\bfseries 122} (2019) 091601}
	[\href{https://arxiv.org/abs/1812.04147}{{\ttfamily 1812.04147}}].
	
	\bibitem{Blaback:2019zig}
	J.~Bl{\r a}b{\"a}ck, U.~Danielsson, G.~Dibitetto and S.~Giri,
	\emph{{Constructing stable de Sitter in M-theory from higher curvature
			corrections}}, \href{https://doi.org/10.1007/JHEP09(2019)042}{\emph{JHEP}
		{\bfseries 09} (2019) 042}
	[\href{https://arxiv.org/abs/1902.04053}{{\ttfamily 1902.04053}}].
	
	\bibitem{Cordova:2019cvf}
	C.~C{\'o}rdova, G.~B. De~Luca and A.~Tomasiello, \emph{{New de Sitter Solutions
			in Ten Dimensions and Orientifold Singularities}},
	\href{https://arxiv.org/abs/1911.04498}{{\ttfamily 1911.04498}}.
	
	\bibitem{Koerber:2007xk}
	P.~Koerber and L.~Martucci, \emph{{From ten to four and back again: How to
			generalize the geometry}},
	\href{https://doi.org/10.1088/1126-6708/2007/08/059}{\emph{JHEP} {\bfseries
			08} (2007) 059} [\href{https://arxiv.org/abs/0707.1038}{{\ttfamily
			0707.1038}}].
	
	\bibitem{Moritz:2017xto}
	J.~Moritz, A.~Retolaza and A.~Westphal, \emph{{Toward de Sitter space from ten
			dimensions}}, \href{https://doi.org/10.1103/PhysRevD.97.046010}{\emph{Phys.
			Rev. D} {\bfseries 97} (2018) 046010}
	[\href{https://arxiv.org/abs/1707.08678}{{\ttfamily 1707.08678}}].
	
	\bibitem{Kallosh:2018nrk}
	R.~Kallosh and T.~Wrase, \emph{{dS Supergravity from 10d}},
	\href{https://doi.org/10.1002/prop.201800071}{\emph{Fortsch. Phys.}
		{\bfseries 67} (2019) 1800071}
	[\href{https://arxiv.org/abs/1808.09427}{{\ttfamily 1808.09427}}].
	
	\bibitem{Bena:2018fqc}
	I.~Bena, E.~Dudas, M.~Gra{\~ n}a and S.~L{\"u}st, \emph{{Uplifting Runaways}},
	\href{https://doi.org/10.1002/prop.201800100}{\emph{Fortsch. Phys.}
		{\bfseries 67} (2019) 1800100}
	[\href{https://arxiv.org/abs/1809.06861}{{\ttfamily 1809.06861}}].
	
	\bibitem{Gautason:2018gln}
	F.~Gautason, V.~Van~Hemelryck and T.~Van~Riet, \emph{{The Tension between 10D
			Supergravity and dS Uplifts}},
	\href{https://doi.org/10.1002/prop.201800091}{\emph{Fortsch. Phys.}
		{\bfseries 67} (2019) 1800091}
	[\href{https://arxiv.org/abs/1810.08518}{{\ttfamily 1810.08518}}].
	
	\bibitem{Danielsson:2018ztv}
	U.~H. Danielsson and T.~Van~Riet, \emph{{What if string theory has no de Sitter
			vacua?}}, \href{https://doi.org/10.1142/S0218271818300070}{\emph{Int. J. Mod.
			Phys. D} {\bfseries 27} (2018) 1830007}
	[\href{https://arxiv.org/abs/1804.01120}{{\ttfamily 1804.01120}}].
	
	\bibitem{Hamada:2019ack}
	Y.~Hamada, A.~Hebecker, G.~Shiu and P.~Soler, \emph{{Understanding KKLT from a
			10d perspective}}, \href{https://doi.org/10.1007/JHEP06(2019)019}{\emph{JHEP}
		{\bfseries 06} (2019) 019}
	[\href{https://arxiv.org/abs/1902.01410}{{\ttfamily 1902.01410}}].
	
	\bibitem{Carta:2019rhx}
	F.~Carta, J.~Moritz and A.~Westphal, \emph{{Gaugino condensation and small
			uplifts in KKLT}}, \href{https://doi.org/10.1007/JHEP08(2019)141}{\emph{JHEP}
		{\bfseries 08} (2019) 141}
	[\href{https://arxiv.org/abs/1902.01412}{{\ttfamily 1902.01412}}].
	
	\bibitem{Gautason:2019jwq}
	F.~Gautason, V.~Van~Hemelryck, T.~Van~Riet and G.~Venken, \emph{{A 10d view on
			the KKLT AdS vacuum and uplifting}},
	\href{https://doi.org/10.1007/JHEP06(2020)074}{\emph{JHEP} {\bfseries 06}
		(2020) 074} [\href{https://arxiv.org/abs/1902.01415}{{\ttfamily
			1902.01415}}].
	
	\bibitem{Vafa:2005ui}
	C.~Vafa, \emph{{The String landscape and the swampland}},
	\href{https://arxiv.org/abs/hep-th/0509212}{{\ttfamily hep-th/0509212}}.
	
	\bibitem{Brennan:2017rbf}
	T.~D. Brennan, F.~Carta and C.~Vafa, \emph{{The String Landscape, the
			Swampland, and the Missing Corner}},
	\href{https://doi.org/10.22323/1.305.0015}{\emph{PoS} {\bfseries TASI2017}
		(2017) 015} [\href{https://arxiv.org/abs/1711.00864}{{\ttfamily
			1711.00864}}].
	
	\bibitem{Palti:2019pca}
	E.~Palti, \emph{{The Swampland: Introduction and Review}},
	\href{https://doi.org/10.1002/prop.201900037}{\emph{Fortsch. Phys.}
		{\bfseries 67} (2019) 1900037}
	[\href{https://arxiv.org/abs/1903.06239}{{\ttfamily 1903.06239}}].
	
	\bibitem{Obied:2018sgi}
	G.~Obied, H.~Ooguri, L.~Spodyneiko and C.~Vafa, \emph{{De Sitter Space and the
			Swampland}},  \href{https://arxiv.org/abs/1806.08362}{{\ttfamily
			1806.08362}}.
	
	\bibitem{Maldacena:2000mw}
	J.~M. Maldacena and C.~Nunez, \emph{{Supergravity description of field theories
			on curved manifolds and a no go theorem}},
	\href{https://doi.org/10.1142/S0217751X01003937}{\emph{Int. J. Mod. Phys. A}
		{\bfseries 16} (2001) 822}
	[\href{https://arxiv.org/abs/hep-th/0007018}{{\ttfamily hep-th/0007018}}].
	
	\bibitem{Giddings:2001yu}
	S.~B. Giddings, S.~Kachru and J.~Polchinski, \emph{{Hierarchies from fluxes in
			string compactifications}},
	\href{https://doi.org/10.1103/PhysRevD.66.106006}{\emph{Phys. Rev. D}
		{\bfseries 66} (2002) 106006}
	[\href{https://arxiv.org/abs/hep-th/0105097}{{\ttfamily hep-th/0105097}}].
	
	\bibitem{Hertzberg:2007wc}
	M.~P. Hertzberg, S.~Kachru, W.~Taylor and M.~Tegmark, \emph{{Inflationary
			Constraints on Type IIA String Theory}},
	\href{https://doi.org/10.1088/1126-6708/2007/12/095}{\emph{JHEP} {\bfseries
			12} (2007) 095} [\href{https://arxiv.org/abs/0711.2512}{{\ttfamily
			0711.2512}}].
	
	\bibitem{AlvarezGaume:1986jb}
	L.~Alvarez-Gaume, P.~H. Ginsparg, G.~W. Moore and C.~Vafa, \emph{{An O(16) x
			O(16) Heterotic String}},
	\href{https://doi.org/10.1016/0370-2693(86)91524-8}{\emph{Phys. Lett.}
		{\bfseries B171} (1986) 155}.
	
	\bibitem{Dixon:1986iz}
	L.~J. Dixon and J.~A. Harvey, \emph{{String Theories in Ten-Dimensions Without
			Space-Time Supersymmetry}},
	\href{https://doi.org/10.1016/0550-3213(86)90619-X}{\emph{Nucl. Phys.}
		{\bfseries B274} (1986) 93}.
	
	\bibitem{Sagnotti:1995ga}
	A.~Sagnotti, \emph{{Some properties of open string theories}},  in
	\emph{{Supersymmetry and unification of fundamental interactions.
			Proceedings, International Workshop, SUSY 95, Palaiseau, France, May 15-19}},
	pp.~473--484, 1995, \href{https://arxiv.org/abs/hep-th/9509080}{{\ttfamily
			hep-th/9509080}}.
	
	\bibitem{Sagnotti:1996qj}
	A.~Sagnotti, \emph{{Surprises in open string perturbation theory}},
	\href{https://doi.org/10.1016/S0920-5632(97)00344-7}{\emph{Nucl. Phys. Proc.
			Suppl.} {\bfseries 56B} (1997) 332}
	[\href{https://arxiv.org/abs/hep-th/9702093}{{\ttfamily hep-th/9702093}}].
	
	\bibitem{Sugimoto:1999tx}
	S.~Sugimoto, \emph{{Anomaly cancellations in type I D-9 - anti-D-9 system and
			the USp(32) string theory}},
	\href{https://doi.org/10.1143/PTP.102.685}{\emph{Prog. Theor. Phys.}
		{\bfseries 102} (1999) 685}
	[\href{https://arxiv.org/abs/hep-th/9905159}{{\ttfamily hep-th/9905159}}].
	
	\bibitem{Antoniadis:1999xk}
	I.~Antoniadis, E.~Dudas and A.~Sagnotti, \emph{{Brane supersymmetry breaking}},
	\href{https://doi.org/10.1016/S0370-2693(99)01023-0}{\emph{Phys. Lett.}
		{\bfseries B464} (1999) 38}
	[\href{https://arxiv.org/abs/hep-th/9908023}{{\ttfamily hep-th/9908023}}].
	
	\bibitem{Angelantonj:1999jh}
	C.~Angelantonj, \emph{{Comments on open string orbifolds with a nonvanishing
			B(ab)}}, \href{https://doi.org/10.1016/S0550-3213(99)00662-8}{\emph{Nucl.
			Phys.} {\bfseries B566} (2000) 126}
	[\href{https://arxiv.org/abs/hep-th/9908064}{{\ttfamily hep-th/9908064}}].
	
	\bibitem{Aldazabal:1999jr}
	G.~Aldazabal and A.~M. Uranga, \emph{{Tachyon free nonsupersymmetric type IIB
			orientifolds via Brane - anti-brane systems}},
	\href{https://doi.org/10.1088/1126-6708/1999/10/024}{\emph{JHEP} {\bfseries
			10} (1999) 024} [\href{https://arxiv.org/abs/hep-th/9908072}{{\ttfamily
			hep-th/9908072}}].
	
	\bibitem{Angelantonj:1999ms}
	C.~Angelantonj, I.~Antoniadis, G.~D'Appollonio, E.~Dudas and A.~Sagnotti,
	\emph{{Type I vacua with brane supersymmetry breaking}},
	\href{https://doi.org/10.1016/S0550-3213(00)00052-3}{\emph{Nucl. Phys.}
		{\bfseries B572} (2000) 36}
	[\href{https://arxiv.org/abs/hep-th/9911081}{{\ttfamily hep-th/9911081}}].
	
	\bibitem{Antonelli:2019nar}
	R.~Antonelli and I.~Basile, \emph{{Brane annihilation in non-supersymmetric
			strings}}, \href{https://doi.org/10.1007/JHEP11(2019)021}{\emph{JHEP}
		{\bfseries 11} (2019) 021}
	[\href{https://arxiv.org/abs/1908.04352}{{\ttfamily 1908.04352}}].
	
	\bibitem{Montero:2020rpl}
	M.~Montero, T.~Van~Riet and G.~Venken, \emph{{A dS obstruction and its
			phenomenological consequences}},
	\href{https://doi.org/10.1007/JHEP05(2020)114}{\emph{JHEP} {\bfseries 05}
		(2020) 114} [\href{https://arxiv.org/abs/2001.11023}{{\ttfamily
			2001.11023}}].
	
	\bibitem{Bedroya:2019snp}
	A.~Bedroya and C.~Vafa, \emph{{Trans-Planckian Censorship and the Swampland}},
	\href{https://arxiv.org/abs/1909.11063}{{\ttfamily 1909.11063}}.
	
	\bibitem{Lanza:2020qmt}
	S.~Lanza, F.~Marchesano, L.~Martucci and I.~Valenzuela, \emph{{Swampland
			Conjectures for Strings and Membranes}},
	\href{https://arxiv.org/abs/2006.15154}{{\ttfamily 2006.15154}}.
	
	\bibitem{Banerjee:2018qey}
	S.~Banerjee, U.~Danielsson, G.~Dibitetto, S.~Giri and M.~Schillo,
	\emph{{Emergent de Sitter Cosmology from Decaying Anti--de Sitter Space}},
	\href{https://doi.org/10.1103/PhysRevLett.121.261301}{\emph{Phys. Rev. Lett.}
		{\bfseries 121} (2018) 261301}
	[\href{https://arxiv.org/abs/1807.01570}{{\ttfamily 1807.01570}}].
	
	\bibitem{Banerjee:2019fzz}
	S.~Banerjee, U.~Danielsson, G.~Dibitetto, S.~Giri and M.~Schillo, \emph{{de
			Sitter Cosmology on an expanding bubble}},
	\href{https://doi.org/10.1007/JHEP10(2019)164}{\emph{JHEP} {\bfseries 10}
		(2019) 164} [\href{https://arxiv.org/abs/1907.04268}{{\ttfamily
			1907.04268}}].
	
	\bibitem{Banerjee:2020wix}
	S.~Banerjee, U.~Danielsson and S.~Giri, \emph{{Dark bubbles: decorating the
			wall}}, \href{https://doi.org/10.1007/JHEP04(2020)085}{\emph{JHEP} {\bfseries
			20} (2020) 085} [\href{https://arxiv.org/abs/2001.07433}{{\ttfamily
			2001.07433}}].
	
	\bibitem{Dudas:2000bn}
	E.~Dudas, \emph{{Theory and phenomenology of type I strings and M theory}},
	\href{https://doi.org/10.1088/0264-9381/17/22/201}{\emph{Class. Quant. Grav.}
		{\bfseries 17} (2000) R41}
	[\href{https://arxiv.org/abs/hep-ph/0006190}{{\ttfamily hep-ph/0006190}}].
	
	\bibitem{Angelantonj:2002ct}
	C.~Angelantonj and A.~Sagnotti, \emph{{Open strings}},
	\href{https://doi.org/10.1016/S0370-1573(02)00273-9,
		10.1016/S0370-1573(03)00006-1}{\emph{Phys. Rept.} {\bfseries 371} (2002) 1}
	[\href{https://arxiv.org/abs/hep-th/0204089}{{\ttfamily hep-th/0204089}}].
	
	\bibitem{Mourad:2017rrl}
	J.~Mourad and A.~Sagnotti, \emph{{An Update on Brane Supersymmetry Breaking}},
	\href{https://arxiv.org/abs/1711.11494}{{\ttfamily 1711.11494}}.
	
	\bibitem{Sagnotti:1987tw}
	A.~Sagnotti, \emph{{Open Strings and their Symmetry Groups}},  in \emph{{NATO
			Advanced Summer Institute on Nonperturbative Quantum Field Theory (Cargese
			Summer Institute) Cargese, France, July 16-30, 1987}}, pp.~521--528, 1987,
	\href{https://arxiv.org/abs/hep-th/0208020}{{\ttfamily hep-th/0208020}}.
	
	\bibitem{Pradisi:1988xd}
	G.~Pradisi and A.~Sagnotti, \emph{{Open String Orbifolds}},
	\href{https://doi.org/10.1016/0370-2693(89)91369-5}{\emph{Phys. Lett.}
		{\bfseries B216} (1989) 59}.
	
	\bibitem{Horava:1989vt}
	P.~Horava, \emph{{Strings on World Sheet Orbifolds}},
	\href{https://doi.org/10.1016/0550-3213(89)90279-4}{\emph{Nucl. Phys.}
		{\bfseries B327} (1989) 461}.
	
	\bibitem{Horava:1989ga}
	P.~Horava, \emph{{Background Duality of Open String Models}},
	\href{https://doi.org/10.1016/0370-2693(89)90209-8}{\emph{Phys. Lett.}
		{\bfseries B231} (1989) 251}.
	
	\bibitem{Bianchi:1990yu}
	M.~Bianchi and A.~Sagnotti, \emph{{On the systematics of open string
			theories}}, \href{https://doi.org/10.1016/0370-2693(90)91894-H}{\emph{Phys.
			Lett.} {\bfseries B247} (1990) 517}.
	
	\bibitem{Bianchi:1990tb}
	M.~Bianchi and A.~Sagnotti, \emph{{Twist symmetry and open string Wilson
			lines}}, \href{https://doi.org/10.1016/0550-3213(91)90271-X}{\emph{Nucl.
			Phys.} {\bfseries B361} (1991) 519}.
	
	\bibitem{Bianchi:1991eu}
	M.~Bianchi, G.~Pradisi and A.~Sagnotti, \emph{{Toroidal compactification and
			symmetry breaking in open string theories}},
	\href{https://doi.org/10.1016/0550-3213(92)90129-Y}{\emph{Nucl. Phys.}
		{\bfseries B376} (1992) 365}.
	
	\bibitem{Sagnotti:1992qw}
	A.~Sagnotti, \emph{{A Note on the Green-Schwarz mechanism in open string
			theories}}, \href{https://doi.org/10.1016/0370-2693(92)90682-T}{\emph{Phys.
			Lett.} {\bfseries B294} (1992) 196}
	[\href{https://arxiv.org/abs/hep-th/9210127}{{\ttfamily hep-th/9210127}}].
	
	\bibitem{Dudas:2000nv}
	E.~Dudas and J.~Mourad, \emph{{Consistent gravitino couplings in
			nonsupersymmetric strings}},
	\href{https://doi.org/10.1016/S0370-2693(01)00777-8}{\emph{Phys. Lett.}
		{\bfseries B514} (2001) 173}
	[\href{https://arxiv.org/abs/hep-th/0012071}{{\ttfamily hep-th/0012071}}].
	
	\bibitem{Pradisi:2001yv}
	G.~Pradisi and F.~Riccioni, \emph{{Geometric couplings and brane supersymmetry
			breaking}}, \href{https://doi.org/10.1016/S0550-3213(01)00441-2}{\emph{Nucl.
			Phys.} {\bfseries B615} (2001) 33}
	[\href{https://arxiv.org/abs/hep-th/0107090}{{\ttfamily hep-th/0107090}}].
	
	\bibitem{Dienes:1990ij}
	K.~R. Dienes, \emph{{New string partition functions with vanishing cosmological
			constant}}, \href{https://doi.org/10.1103/PhysRevLett.65.1979}{\emph{Phys.
			Rev. Lett.} {\bfseries 65} (1990) 1979}.
	
	\bibitem{Dienes:1990qh}
	K.~R. Dienes, \emph{{GENERALIZED ATKIN-LEHNER SYMMETRY}},
	\href{https://doi.org/10.1103/PhysRevD.42.2004}{\emph{Phys. Rev. D}
		{\bfseries 42} (1990) 2004}.
	
	\bibitem{Kachru:1998hd}
	S.~Kachru, J.~Kumar and E.~Silverstein, \emph{{Vacuum energy cancellation in a
			nonsupersymmetric string}},
	\href{https://doi.org/10.1103/PhysRevD.59.106004}{\emph{Phys. Rev. D}
		{\bfseries 59} (1999) 106004}
	[\href{https://arxiv.org/abs/hep-th/9807076}{{\ttfamily hep-th/9807076}}].
	
	\bibitem{Angelantonj:2004cm}
	C.~Angelantonj and M.~Cardella, \emph{{Vanishing perturbative vacuum energy in
			nonsupersymmetric orientifolds}},
	\href{https://doi.org/10.1016/j.physletb.2004.06.058}{\emph{Phys. Lett. B}
		{\bfseries 595} (2004) 505}
	[\href{https://arxiv.org/abs/hep-th/0403107}{{\ttfamily hep-th/0403107}}].
	
	\bibitem{Abel:2017rch}
	S.~Abel and R.~J. Stewart, \emph{{Exponential suppression of the cosmological
			constant in nonsupersymmetric string vacua at two loops and beyond}},
	\href{https://doi.org/10.1103/PhysRevD.96.106013}{\emph{Phys. Rev. D}
		{\bfseries 96} (2017) 106013}
	[\href{https://arxiv.org/abs/1701.06629}{{\ttfamily 1701.06629}}].
	
	\bibitem{Mourad:2016xbk}
	J.~Mourad and A.~Sagnotti, \emph{{$AdS$ Vacua from Dilaton Tadpoles and Form
			Fluxes}}, \href{https://doi.org/10.1016/j.physletb.2017.02.053}{\emph{Phys.
			Lett.} {\bfseries B768} (2017) 92}
	[\href{https://arxiv.org/abs/1612.08566}{{\ttfamily 1612.08566}}].
	
	\bibitem{Gibbons:1984kp}
	G.~Gibbons, \emph{{ASPECTS OF SUPERGRAVITY THEORIES}},  in \emph{{XV GIFT
			Seminar on Supersymmetry and Supergravity}}, 6, 1984.
	
	\bibitem{Dasgupta:1999ss}
	K.~Dasgupta, G.~Rajesh and S.~Sethi, \emph{{M theory, orientifolds and G -
			flux}}, \href{https://doi.org/10.1088/1126-6708/1999/08/023}{\emph{JHEP}
		{\bfseries 08} (1999) 023}
	[\href{https://arxiv.org/abs/hep-th/9908088}{{\ttfamily hep-th/9908088}}].
	
	\bibitem{Green:2011cn}
	S.~R. Green, E.~J. Martinec, C.~Quigley and S.~Sethi, \emph{{Constraints on
			String Cosmology}},
	\href{https://doi.org/10.1088/0264-9381/29/7/075006}{\emph{Class. Quant.
			Grav.} {\bfseries 29} (2012) 075006}
	[\href{https://arxiv.org/abs/1110.0545}{{\ttfamily 1110.0545}}].
	
	\bibitem{Das:2019vnx}
	S.~Das, S.~S. Haque and B.~Underwood, \emph{{Constraints and horizons for de
			Sitter with extra dimensions}},
	\href{https://doi.org/10.1103/PhysRevD.100.046013}{\emph{Phys. Rev. D}
		{\bfseries 100} (2019) 046013}
	[\href{https://arxiv.org/abs/1905.05864}{{\ttfamily 1905.05864}}].
	
	\bibitem{Kutasov:2015eba}
	D.~Kutasov, T.~Maxfield, I.~Melnikov and S.~Sethi, \emph{{Constraining de
			Sitter Space in String Theory}},
	\href{https://doi.org/10.1103/PhysRevLett.115.071305}{\emph{Phys. Rev. Lett.}
		{\bfseries 115} (2015) 071305}
	[\href{https://arxiv.org/abs/1504.00056}{{\ttfamily 1504.00056}}].
	
	\bibitem{Gaiotto:2014kfa}
	D.~Gaiotto, A.~Kapustin, N.~Seiberg and B.~Willett, \emph{{Generalized Global
			Symmetries}}, \href{https://doi.org/10.1007/JHEP02(2015)172}{\emph{JHEP}
		{\bfseries 02} (2015) 172} [\href{https://arxiv.org/abs/1412.5148}{{\ttfamily
			1412.5148}}].
	
	\bibitem{Banks:2010zn}
	T.~Banks and N.~Seiberg, \emph{{Symmetries and Strings in Field Theory and
			Gravity}}, \href{https://doi.org/10.1103/PhysRevD.83.084019}{\emph{Phys.
			Rev.} {\bfseries D83} (2011) 084019}
	[\href{https://arxiv.org/abs/1011.5120}{{\ttfamily 1011.5120}}].
	
	\bibitem{Montero:2017yja}
	M.~Montero, A.~M. Uranga and I.~Valenzuela, \emph{{A Chern-Simons Pandemic}},
	\href{https://doi.org/10.1007/JHEP07(2017)123}{\emph{JHEP} {\bfseries 07}
		(2017) 123} [\href{https://arxiv.org/abs/1702.06147}{{\ttfamily
			1702.06147}}].
	
	\bibitem{McNamara:2019rup}
	J.~McNamara and C.~Vafa, \emph{{Cobordism Classes and the Swampland}},
	\href{https://arxiv.org/abs/1909.10355}{{\ttfamily 1909.10355}}.
	
	\bibitem{Basile:2018irz}
	I.~Basile, J.~Mourad and A.~Sagnotti, \emph{{On Classical Stability with Broken
			Supersymmetry}}, \href{https://doi.org/10.1007/JHEP01(2019)174}{\emph{JHEP}
		{\bfseries 01} (2019) 174}
	[\href{https://arxiv.org/abs/1811.11448}{{\ttfamily 1811.11448}}].
	
	\bibitem{Bonnefoy:2018tcp}
	Q.~Bonnefoy, E.~Dudas and S.~LÃŒst, \emph{{On the weak gravity conjecture in
			string theory with broken supersymmetry}},
	\href{https://doi.org/10.1016/j.nuclphysb.2019.114738}{\emph{Nucl. Phys. B}
		{\bfseries 947} (2019) 114738}
	[\href{https://arxiv.org/abs/1811.11199}{{\ttfamily 1811.11199}}].
	
	\bibitem{Bonnefoy:2020fwt}
	Q.~Bonnefoy, E.~Dudas and S.~L{\"u}st, \emph{{Weak gravity (and other
			conjectures) with broken supersymmetry}},  in \emph{{19th Hellenic School and
			Workshops on Elementary Particle Physics and Gravity}}, 3, 2020,
	\href{https://arxiv.org/abs/2003.14126}{{\ttfamily 2003.14126}}.
	
	\bibitem{Garg:2018reu}
	S.~K. Garg and C.~Krishnan, \emph{{Bounds on Slow Roll and the de Sitter
			Swampland}}, \href{https://doi.org/10.1007/JHEP11(2019)075}{\emph{JHEP}
		{\bfseries 11} (2019) 075}
	[\href{https://arxiv.org/abs/1807.05193}{{\ttfamily 1807.05193}}].
	
	\bibitem{Ooguri:2018wrx}
	H.~Ooguri, E.~Palti, G.~Shiu and C.~Vafa, \emph{{Distance and de Sitter
			Conjectures on the Swampland}},
	\href{https://doi.org/10.1016/j.physletb.2018.11.018}{\emph{Phys. Lett. B}
		{\bfseries 788} (2019) 180}
	[\href{https://arxiv.org/abs/1810.05506}{{\ttfamily 1810.05506}}].
	
	\bibitem{Dvali:2018fqu}
	G.~Dvali and C.~Gomez, \emph{{On Exclusion of Positive Cosmological Constant}},
	\href{https://doi.org/10.1002/prop.201800092}{\emph{Fortsch. Phys.}
		{\bfseries 67} (2019) 1800092}
	[\href{https://arxiv.org/abs/1806.10877}{{\ttfamily 1806.10877}}].
	
	\bibitem{Andriot:2018wzk}
	D.~Andriot, \emph{{On the de Sitter swampland criterion}},
	\href{https://doi.org/10.1016/j.physletb.2018.09.022}{\emph{Phys. Lett. B}
		{\bfseries 785} (2018) 570}
	[\href{https://arxiv.org/abs/1806.10999}{{\ttfamily 1806.10999}}].
	
	\bibitem{Grimm:2019ixq}
	T.~W. Grimm, C.~Li and I.~Valenzuela, \emph{{Asymptotic Flux Compactifications
			and the Swampland}},
	\href{https://doi.org/10.1007/JHEP06(2020)009}{\emph{JHEP} {\bfseries 06}
		(2020) 009} [\href{https://arxiv.org/abs/1910.09549}{{\ttfamily
			1910.09549}}].
	
	\bibitem{Bedroya:2019tba}
	A.~Bedroya, R.~Brandenberger, M.~Loverde and C.~Vafa, \emph{{Trans-Planckian
			Censorship and Inflationary Cosmology}},
	\href{https://doi.org/10.1103/PhysRevD.101.103502}{\emph{Phys. Rev. D}
		{\bfseries 101} (2020) 103502}
	[\href{https://arxiv.org/abs/1909.11106}{{\ttfamily 1909.11106}}].
	
	\bibitem{Kinney:2018nny}
	W.~H. Kinney, S.~Vagnozzi and L.~Visinelli, \emph{{The zoo plot meets the
			swampland: mutual (in)consistency of single-field inflation, string
			conjectures, and cosmological data}},
	\href{https://doi.org/10.1088/1361-6382/ab1d87}{\emph{Class. Quant. Grav.}
		{\bfseries 36} (2019) 117001}
	[\href{https://arxiv.org/abs/1808.06424}{{\ttfamily 1808.06424}}].
	
	\bibitem{Andriot:2018ept}
	D.~Andriot, \emph{{New constraints on classical de Sitter: flirting with the
			swampland}}, \href{https://doi.org/10.1002/prop.201800103}{\emph{Fortsch.
			Phys.} {\bfseries 67} (2019) 1800103}
	[\href{https://arxiv.org/abs/1807.09698}{{\ttfamily 1807.09698}}].
	
	\bibitem{Andriot:2019wrs}
	D.~Andriot, \emph{{Open problems on classical de Sitter solutions}},
	\href{https://doi.org/10.1002/prop.201900026}{\emph{Fortsch. Phys.}
		{\bfseries 67} (2019) 1900026}
	[\href{https://arxiv.org/abs/1902.10093}{{\ttfamily 1902.10093}}].
	
	\bibitem{Brown:1987dd}
	J.~D. Brown and C.~Teitelboim, \emph{{Dynamical Neutralization of the
			Cosmological Constant}},
	\href{https://doi.org/10.1016/0370-2693(87)91190-7}{\emph{Phys. Lett.}
		{\bfseries B195} (1987) 177}.
	
	\bibitem{Brown:1988kg}
	J.~D. Brown and C.~Teitelboim, \emph{{Neutralization of the Cosmological
			Constant by Membrane Creation}},
	\href{https://doi.org/10.1016/0550-3213(88)90559-7}{\emph{Nucl. Phys.}
		{\bfseries B297} (1988) 787}.
	
	\bibitem{Bousso:2000xa}
	R.~Bousso and J.~Polchinski, \emph{{Quantization of four form fluxes and
			dynamical neutralization of the cosmological constant}},
	\href{https://doi.org/10.1088/1126-6708/2000/06/006}{\emph{JHEP} {\bfseries
			06} (2000) 006} [\href{https://arxiv.org/abs/hep-th/0004134}{{\ttfamily
			hep-th/0004134}}].
	
	\bibitem{Lanza:2019nfa}
	S.~Lanza, \emph{{Exploring the Landscape of effective field theories}}, Ph.D.
	thesis, Padua U., 2019.
	\newblock \href{https://arxiv.org/abs/1912.08935}{{\ttfamily 1912.08935}}.
	
	\bibitem{Heidenreich:2019zkl}
	B.~Heidenreich, M.~Reece and T.~Rudelius, \emph{{Repulsive Forces and the Weak
			Gravity Conjecture}},
	\href{https://doi.org/10.1007/JHEP10(2019)055}{\emph{JHEP} {\bfseries 10}
		(2019) 055} [\href{https://arxiv.org/abs/1906.02206}{{\ttfamily
			1906.02206}}].
	
	\bibitem{Herraez:2020tih}
	A.~Herraez, \emph{{A Note on Membrane Interactions and the Scalar potential}},
	\href{https://arxiv.org/abs/2006.01160}{{\ttfamily 2006.01160}}.
	
	\bibitem{Bandos:2018gjp}
	I.~Bandos, F.~Farakos, S.~Lanza, L.~Martucci and D.~Sorokin,
	\emph{{Three-forms, dualities and membranes in four-dimensional
			supergravity}},  \href{https://arxiv.org/abs/1803.01405}{{\ttfamily
			1803.01405}}.
	
	\bibitem{Ooguri:2006in}
	H.~Ooguri and C.~Vafa, \emph{{On the Geometry of the String Landscape and the
			Swampland}},
	\href{https://doi.org/10.1016/j.nuclphysb.2006.10.033}{\emph{Nucl. Phys. B}
		{\bfseries 766} (2007) 21}
	[\href{https://arxiv.org/abs/hep-th/0605264}{{\ttfamily hep-th/0605264}}].
	
	\bibitem{Grimm:2018ohb}
	T.~W. Grimm, E.~Palti and I.~Valenzuela, \emph{{Infinite Distances in Field
			Space and Massless Towers of States}},
	\href{https://doi.org/10.1007/JHEP08(2018)143}{\emph{JHEP} {\bfseries 08}
		(2018) 143} [\href{https://arxiv.org/abs/1802.08264}{{\ttfamily
			1802.08264}}].
	
	\bibitem{Corvilain:2018lgw}
	P.~Corvilain, T.~W. Grimm and I.~Valenzuela, \emph{{The Swampland Distance
			Conjecture for K{\"a}hler moduli}},
	\href{https://doi.org/10.1007/JHEP08(2019)075}{\emph{JHEP} {\bfseries 08}
		(2019) 075} [\href{https://arxiv.org/abs/1812.07548}{{\ttfamily
			1812.07548}}].
	
	\bibitem{Grimm:2018cpv}
	T.~W. Grimm, C.~Li and E.~Palti, \emph{{Infinite Distance Networks in Field
			Space and Charge Orbits}},
	\href{https://doi.org/10.1007/JHEP03(2019)016}{\emph{JHEP} {\bfseries 03}
		(2019) 016} [\href{https://arxiv.org/abs/1811.02571}{{\ttfamily
			1811.02571}}].
	
	\bibitem{Font:2019cxq}
	A.~Font, A.~Herr{\'a}ez and L.~E. Ib{\'a}{\~n}ez, \emph{{The Swampland Distance
			Conjecture and Towers of Tensionless Branes}},
	\href{https://doi.org/10.1007/JHEP08(2019)044}{\emph{JHEP} {\bfseries 08}
		(2019) 044} [\href{https://arxiv.org/abs/1904.05379}{{\ttfamily
			1904.05379}}].
	
	\bibitem{Kaloper:1999sm}
	N.~Kaloper, \emph{{Bent domain walls as brane worlds}},
	\href{https://doi.org/10.1103/PhysRevD.60.123506}{\emph{Phys. Rev. D}
		{\bfseries 60} (1999) 123506}
	[\href{https://arxiv.org/abs/hep-th/9905210}{{\ttfamily hep-th/9905210}}].
	
	\bibitem{Shiromizu:1999wj}
	T.~Shiromizu, K.-i. Maeda and M.~Sasaki, \emph{{The Einstein equation on the
			3-brane world}},
	\href{https://doi.org/10.1103/PhysRevD.62.024012}{\emph{Phys. Rev. D}
		{\bfseries 62} (2000) 024012}
	[\href{https://arxiv.org/abs/gr-qc/9910076}{{\ttfamily gr-qc/9910076}}].
	
	\bibitem{Vollick:1999uz}
	D.~N. Vollick, \emph{{Cosmology on a three-brane}},
	\href{https://doi.org/10.1088/0264-9381/18/1/301}{\emph{Class. Quant. Grav.}
		{\bfseries 18} (2001) 1}
	[\href{https://arxiv.org/abs/hep-th/9911181}{{\ttfamily hep-th/9911181}}].
	
	\bibitem{Gubser:1999vj}
	S.~S. Gubser, \emph{{AdS / CFT and gravity}},
	\href{https://doi.org/10.1103/PhysRevD.63.084017}{\emph{Phys. Rev. D}
		{\bfseries 63} (2001) 084017}
	[\href{https://arxiv.org/abs/hep-th/9912001}{{\ttfamily hep-th/9912001}}].
	
	\bibitem{Hawking:2000kj}
	S.~Hawking, T.~Hertog and H.~Reall, \emph{{Brane new world}},
	\href{https://doi.org/10.1103/PhysRevD.62.043501}{\emph{Phys. Rev. D}
		{\bfseries 62} (2000) 043501}
	[\href{https://arxiv.org/abs/hep-th/0003052}{{\ttfamily hep-th/0003052}}].
	
	\bibitem{Dudas:2000sn}
	E.~Dudas and J.~Mourad, \emph{{D-branes in nontachyonic 0B orientifolds}},
	\href{https://doi.org/10.1016/S0550-3213(00)00781-1}{\emph{Nucl. Phys.}
		{\bfseries B598} (2001) 189}
	[\href{https://arxiv.org/abs/hep-th/0010179}{{\ttfamily hep-th/0010179}}].
	
	\bibitem{Angelantonj:1999qg}
	C.~Angelantonj and A.~Armoni, \emph{{Nontachyonic type 0B orientifolds,
			nonsupersymmetric gauge theories and cosmological RG flow}},
	\href{https://doi.org/10.1016/S0550-3213(00)00136-X}{\emph{Nucl. Phys.}
		{\bfseries B578} (2000) 239}
	[\href{https://arxiv.org/abs/hep-th/9912257}{{\ttfamily hep-th/9912257}}].
	
	\bibitem{Angelantonj:2000kh}
	C.~Angelantonj and A.~Armoni, \emph{{RG flow, Wilson loops and the dilaton
			tadpole}}, \href{https://doi.org/10.1016/S0370-2693(00)00475-5}{\emph{Phys.
			Lett.} {\bfseries B482} (2000) 329}
	[\href{https://arxiv.org/abs/hep-th/0003050}{{\ttfamily hep-th/0003050}}].
	
	\bibitem{Maldacena:1998uz}
	J.~M. Maldacena, J.~Michelson and A.~Strominger, \emph{{Anti-de Sitter
			fragmentation}},
	\href{https://doi.org/10.1088/1126-6708/1999/02/011}{\emph{JHEP} {\bfseries
			02} (1999) 011} [\href{https://arxiv.org/abs/hep-th/9812073}{{\ttfamily
			hep-th/9812073}}].
	
	\bibitem{Seiberg:1999xz}
	N.~Seiberg and E.~Witten, \emph{{The D1 / D5 system and singular CFT}},
	\href{https://doi.org/10.1088/1126-6708/1999/04/017}{\emph{JHEP} {\bfseries
			04} (1999) 017} [\href{https://arxiv.org/abs/hep-th/9903224}{{\ttfamily
			hep-th/9903224}}].
	
	\bibitem{Ooguri:2016pdq}
	H.~Ooguri and C.~Vafa, \emph{{Non-supersymmetric AdS and the Swampland}},
	\href{https://arxiv.org/abs/1610.01533}{{\ttfamily 1610.01533}}.
	
	\bibitem{Israel1966}
	W.~Israel, \emph{Singular hypersurfaces and thin shells in general relativity},
	\href{https://doi.org/10.1007/BF02710419}{\emph{Il Nuovo Cimento B
			(1965-1970)} {\bfseries 44} (1966) 1}.
	
	\bibitem{Barrabes:1991ng}
	C.~Barrabes and W.~Israel, \emph{{Thin shells in general relativity and
			cosmology: The Lightlike limit}},
	\href{https://doi.org/10.1103/PhysRevD.43.1129}{\emph{Phys. Rev.} {\bfseries
			D43} (1991) 1129}.
	
	\bibitem{Randall:1999ee}
	L.~Randall and R.~Sundrum, \emph{{A Large mass hierarchy from a small extra
			dimension}}, \href{https://doi.org/10.1103/PhysRevLett.83.3370}{\emph{Phys.
			Rev. Lett.} {\bfseries 83} (1999) 3370}
	[\href{https://arxiv.org/abs/hep-ph/9905221}{{\ttfamily hep-ph/9905221}}].
	
	\bibitem{Randall:1999vf}
	L.~Randall and R.~Sundrum, \emph{{An Alternative to compactification}},
	\href{https://doi.org/10.1103/PhysRevLett.83.4690}{\emph{Phys. Rev. Lett.}
		{\bfseries 83} (1999) 4690}
	[\href{https://arxiv.org/abs/hep-th/9906064}{{\ttfamily hep-th/9906064}}].
	
	\bibitem{Giddings:2000mu}
	S.~B. Giddings, E.~Katz and L.~Randall, \emph{{Linearized gravity in brane
			backgrounds}},
	\href{https://doi.org/10.1088/1126-6708/2000/03/023}{\emph{JHEP} {\bfseries
			03} (2000) 023} [\href{https://arxiv.org/abs/hep-th/0002091}{{\ttfamily
			hep-th/0002091}}].
	
	\bibitem{Dvali:2000hr}
	G.~Dvali, G.~Gabadadze and M.~Porrati, \emph{{4-D gravity on a brane in 5-D
			Minkowski space}},
	\href{https://doi.org/10.1016/S0370-2693(00)00669-9}{\emph{Phys. Lett. B}
		{\bfseries 485} (2000) 208}
	[\href{https://arxiv.org/abs/hep-th/0005016}{{\ttfamily hep-th/0005016}}].
	
	\bibitem{Karch:2000ct}
	A.~Karch and L.~Randall, \emph{{Locally localized gravity}},
	\href{https://doi.org/10.1088/1126-6708/2001/05/008}{\emph{JHEP} {\bfseries
			05} (2001) 008} [\href{https://arxiv.org/abs/hep-th/0011156}{{\ttfamily
			hep-th/0011156}}].
	
	\bibitem{Chen:2020uac}
	H.~Z. Chen, R.~C. Myers, D.~Neuenfeld, I.~A. Reyes and J.~Sandor,
	\emph{{Quantum Extremal Islands Made Easy, Part I: Entanglement on the
			Brane}},  \href{https://arxiv.org/abs/2006.04851}{{\ttfamily 2006.04851}}.
	
	\bibitem{Maxfield:2014wea}
	T.~Maxfield and S.~Sethi, \emph{{Domain Walls, Triples and Acceleration}},
	\href{https://doi.org/10.1007/JHEP08(2014)066}{\emph{JHEP} {\bfseries 08}
		(2014) 066} [\href{https://arxiv.org/abs/1404.2564}{{\ttfamily 1404.2564}}].
	
	\bibitem{Antonelli:2018qwz}
	R.~Antonelli, I.~Basile and A.~Bombini, \emph{{AdS Vacuum Bubbles, Holography
			and Dual RG Flows}},
	\href{https://doi.org/10.1088/1361-6382/aafef9}{\emph{Class. Quant. Grav.}
		{\bfseries 36} (2019) 045004}
	[\href{https://arxiv.org/abs/1806.02289}{{\ttfamily 1806.02289}}].
	
	\bibitem{Poletti:1994ww}
	S.~Poletti, J.~Twamley and D.~Wiltshire, \emph{{Charged dilaton black holes
			with a cosmological constant}},
	\href{https://doi.org/10.1103/PhysRevD.51.5720}{\emph{Phys. Rev. D}
		{\bfseries 51} (1995) 5720}
	[\href{https://arxiv.org/abs/hep-th/9412076}{{\ttfamily hep-th/9412076}}].
	
	\bibitem{Wiltshire:1994de}
	D.~L. Wiltshire, \emph{{Dilaton black holes with a cosmological term}},
	\href{https://doi.org/10.1017/S0334270000011164}{\emph{J. Austral. Math. Soc.
			B} {\bfseries 41} (1999) 198}
	[\href{https://arxiv.org/abs/gr-qc/9502038}{{\ttfamily gr-qc/9502038}}].
	
	\bibitem{Chan:1995fr}
	K.~C. Chan, J.~H. Horne and R.~B. Mann, \emph{{Charged dilaton black holes with
			unusual asymptotics}},
	\href{https://doi.org/10.1016/0550-3213(95)00205-7}{\emph{Nucl. Phys. B}
		{\bfseries 447} (1995) 441}
	[\href{https://arxiv.org/abs/gr-qc/9502042}{{\ttfamily gr-qc/9502042}}].
	
	\bibitem{Witten:1981gj}
	E.~Witten, \emph{{Instability of the Kaluza-Klein Vacuum}},
	\href{https://doi.org/10.1016/0550-3213(82)90007-4}{\emph{Nucl. Phys.}
		{\bfseries B195} (1982) 481}.
	
	\bibitem{Dibitetto:2020csn}
	G.~Dibitetto, N.~Petri and M.~Schillo, \emph{{Nothing really matters}},
	\href{https://arxiv.org/abs/2002.01764}{{\ttfamily 2002.01764}}.
	
	\bibitem{Horowitz:2007pr}
	G.~T. Horowitz, J.~Orgera and J.~Polchinski, \emph{{Nonperturbative Instability
			of AdS(5) x S**5/Z(k)}},
	\href{https://doi.org/10.1103/PhysRevD.77.024004}{\emph{Phys. Rev.}
		{\bfseries D77} (2008) 024004}
	[\href{https://arxiv.org/abs/0709.4262}{{\ttfamily 0709.4262}}].
	
	\bibitem{BlancoPillado:2010df}
	J.~J. Blanco-Pillado and B.~Shlaer, \emph{{Bubbles of Nothing in Flux
			Compactifications}},
	\href{https://doi.org/10.1103/PhysRevD.82.086015}{\emph{Phys. Rev. D}
		{\bfseries 82} (2010) 086015}
	[\href{https://arxiv.org/abs/1002.4408}{{\ttfamily 1002.4408}}].
	
	\bibitem{Brown:2010mf}
	A.~R. Brown and A.~Dahlen, \emph{{Bubbles of Nothing and the Fastest Decay in
			the Landscape}},
	\href{https://doi.org/10.1103/PhysRevD.84.043518}{\emph{Phys. Rev. D}
		{\bfseries 84} (2011) 043518}
	[\href{https://arxiv.org/abs/1010.5240}{{\ttfamily 1010.5240}}].
	
	\bibitem{Brown:2011gt}
	A.~R. Brown and A.~Dahlen, \emph{{On 'nothing' as an infinitely negatively
			curved spacetime}},
	\href{https://doi.org/10.1103/PhysRevD.85.104026}{\emph{Phys. Rev. D}
		{\bfseries 85} (2012) 104026}
	[\href{https://arxiv.org/abs/1111.0301}{{\ttfamily 1111.0301}}].
	
	\bibitem{GarciaEtxebarria:2020xsr}
	I.~GarcÃ­a~Etxebarria, M.~Montero, K.~Sousa and I.~Valenzuela, \emph{{Nothing
			is certain in string compactifications}},
	\href{https://arxiv.org/abs/2005.06494}{{\ttfamily 2005.06494}}.
	
	\bibitem{Kachru:1998ys}
	S.~Kachru and E.~Silverstein, \emph{{4-D conformal theories and strings on
			orbifolds}}, \href{https://doi.org/10.1103/PhysRevLett.80.4855}{\emph{Phys.
			Rev. Lett.} {\bfseries 80} (1998) 4855}
	[\href{https://arxiv.org/abs/hep-th/9802183}{{\ttfamily hep-th/9802183}}].
	
	\bibitem{Lawrence:1998ja}
	A.~E. Lawrence, N.~Nekrasov and C.~Vafa, \emph{{On conformal field theories in
			four-dimensions}},
	\href{https://doi.org/10.1016/S0550-3213(98)00495-7}{\emph{Nucl. Phys.}
		{\bfseries B533} (1998) 199}
	[\href{https://arxiv.org/abs/hep-th/9803015}{{\ttfamily hep-th/9803015}}].
	
	\bibitem{Bershadsky:1998mb}
	M.~Bershadsky, Z.~Kakushadze and C.~Vafa, \emph{{String expansion as large N
			expansion of gauge theories}},
	\href{https://doi.org/10.1016/S0550-3213(98)00272-7}{\emph{Nucl. Phys.}
		{\bfseries B523} (1998) 59}
	[\href{https://arxiv.org/abs/hep-th/9803076}{{\ttfamily hep-th/9803076}}].
	
	\bibitem{Bershadsky:1998cb}
	M.~Bershadsky and A.~Johansen, \emph{{Large N limit of orbifold field
			theories}}, \href{https://doi.org/10.1016/S0550-3213(98)00526-4}{\emph{Nucl.
			Phys.} {\bfseries B536} (1998) 141}
	[\href{https://arxiv.org/abs/hep-th/9803249}{{\ttfamily hep-th/9803249}}].
	
	\bibitem{Schmaltz:1998bg}
	M.~Schmaltz, \emph{{Duality of nonsupersymmetric large N gauge theories}},
	\href{https://doi.org/10.1103/PhysRevD.59.105018}{\emph{Phys. Rev.}
		{\bfseries D59} (1999) 105018}
	[\href{https://arxiv.org/abs/hep-th/9805218}{{\ttfamily hep-th/9805218}}].
	
	\bibitem{Erlich:1998gb}
	J.~Erlich and A.~Naqvi, \emph{{Nonperturbative tests of the parent / orbifold
			correspondence in supersymmetric gauge theories}},
	\href{https://doi.org/10.1088/1126-6708/2002/12/047}{\emph{JHEP} {\bfseries
			12} (2002) 047} [\href{https://arxiv.org/abs/hep-th/9808026}{{\ttfamily
			hep-th/9808026}}].
	
	\bibitem{Tong:2002vp}
	D.~Tong, \emph{{Comments on condensates in nonsupersymmetric orbifold field
			theories}}, \href{https://doi.org/10.1088/1126-6708/2003/03/022}{\emph{JHEP}
		{\bfseries 03} (2003) 022}
	[\href{https://arxiv.org/abs/hep-th/0212235}{{\ttfamily hep-th/0212235}}].
	
	\bibitem{Abel:2015oxa}
	S.~Abel, K.~R. Dienes and E.~Mavroudi, \emph{{Towards a nonsupersymmetric
			string phenomenology}},
	\href{https://doi.org/10.1103/PhysRevD.91.126014}{\emph{Phys. Rev.}
		{\bfseries D91} (2015) 126014}
	[\href{https://arxiv.org/abs/1502.03087}{{\ttfamily 1502.03087}}].
	
	\bibitem{Abel:2017vos}
	S.~Abel, K.~R. Dienes and E.~Mavroudi, \emph{{GUT precursors and entwined SUSY:
			The phenomenology of stable nonsupersymmetric strings}},
	\href{https://doi.org/10.1103/PhysRevD.97.126017}{\emph{Phys. Rev. D}
		{\bfseries 97} (2018) 126017}
	[\href{https://arxiv.org/abs/1712.06894}{{\ttfamily 1712.06894}}].
	
	\bibitem{March-Russell:2020lkq}
	J.~March-Russell and R.~Petrossian-Byrne, \emph{{QCD, Flavor, and the de Sitter
			Swampland}},  \href{https://arxiv.org/abs/2006.01144}{{\ttfamily
			2006.01144}}.
	
	\bibitem{Polchinski:1995mt}
	J.~Polchinski, \emph{{Dirichlet Branes and Ramond-Ramond charges}},
	\href{https://doi.org/10.1103/PhysRevLett.75.4724}{\emph{Phys. Rev. Lett.}
		{\bfseries 75} (1995) 4724}
	[\href{https://arxiv.org/abs/hep-th/9510017}{{\ttfamily hep-th/9510017}}].
	
	\bibitem{Bremer:1998zp}
	M.~S. Bremer, M.~J. Duff, H.~Lu, C.~N. Pope and K.~S. Stelle, \emph{{Instanton
			cosmology and domain walls from M theory and string theory}},
	\href{https://doi.org/10.1016/S0550-3213(98)00764-0}{\emph{Nucl. Phys.}
		{\bfseries B543} (1999) 321}
	[\href{https://arxiv.org/abs/hep-th/9807051}{{\ttfamily hep-th/9807051}}].
	
\end{thebibliography}

\providecommand{\href}[2]{#2}\begingroup\raggedright\endgroup

\end{document}